\documentclass[aps,prc,twocolumn,superscriptaddress]{revtex4-2}
\usepackage{graphicx}
\usepackage{dcolumn}
\usepackage{color} 
\RequirePackage{booktabs}
\usepackage{longtable}

\usepackage{amsfonts}
\usepackage{amssymb}
\usepackage{amsmath}
\usepackage{amsxtra}

\begin{document}

\title{Improved \boldmath$sd$ shell effective interactions from Daejeon16}

\date{\today}

\author{Ik~Jae~Shin}
\email[]{geniean@ibs.re.kr}
\affiliation{Rare Isotope Science Project, Institute for Basis Science, Daejeon 34000, Republic of Korea}
\author{Nadezda~A.~Smirnova} 
\email[]{nadezda.smirnova@u-bordeaux.fr}
\affiliation{LP2IB (CNRS/IN2P3 -- Universit\'e de Bordeaux), 33175 Gradignan cedex, France }
\author{Andrey~M.~Shirokov}
\email[]{shirokov@nucl-th.sinp.msu.ru}
\affiliation{Skobeltsyn Institute of Nuclear Physics, Lomonosov Moscow State University, 
Moscow 119991, Russia}
\author{Zuxing~Yang}
\email[]{zuxing.yang@riken.jp}
\affiliation{Institute of Modern Physics, Chinese Academy of Science, Lanzhou, Gansu  730000, People's Republic of China}
\affiliation{RIKEN Nishina Center, Wako, Saitama 351-0198, Japan}
\author{Bruce~R.~Barrett}
\affiliation{Department of Physics, University of Arizona, Tucson, Arizona 85721, USA}
\author{Zhen~Li} 
\email[]{lizhen@cenbg.in2p3.fr}
\affiliation{LP2IB (CNRS/IN2P3 -- Universit\'e de Bordeaux), 33175 Gradignan cedex, France }
\author{Youngman~Kim}
\email[]{ykim@ibs.re.kr}
\affiliation{Center for Exotic Nuclear Studies, Institute for Basic Science, Daejeon 34126, Republic of Korea}
\author{Pieter~Maris}
\email[]{pmaris@iastate.edu}
\affiliation{Department of Physics and Astronomy, Iowa State University, Ames, Iowa 50011, USA}
\author{James~P.~Vary}
\email[]{jvary@iastate.edu}
\affiliation{Department of Physics and Astronomy, Iowa State University, Ames, Iowa 50011, USA}

\begin{abstract}
We present new microscopic effective shell-model interactions in the valence $sd$ shell,
obtained from the modern Daejeon16 nucleon-nucleon potential
using no-core shell-model (NCSM) wave functions of $^{18}$F at $N_{\rm max} =6$ (total oscillator quanta of excitation) 
model space and the Okubo--Lee--Suzuki transformation. 
First, we explore the convergence properties of our calculations and show that the excitation energies of
states in $^{18}$F, characterized by the largest valence-like configurations, are reasonably converged and the lowest states are 
in sensible agreement with experiment.
Then, we investigate the monopole properties of that interaction in comparison with the phenomenological
universal $sd$-shell interaction, USDB, and with the previously derived interaction at $N_{\rm max} =4$. 
Theoretical binding energies and low-energy spectra of the O isotopes, as well as low-energy spectra of 
a selection of $sd$-shell nuclei, are presented. 
We conclude that the use of larger-space NCSM wave functions leads to a noticeable improvement in the quality of 
the derived effective interaction.
We propose monopole modifications of the Daejeon16 centroids 
which further improve the agreement with experiment throughout the $sd$ shell, as demonstrated by a compilation
of spectra contained in Supplemental Material.
\end{abstract}


\maketitle

\section{Introduction}

The last decade has seen new approaches to the long-standing nuclear-structure problem of 
the construction of effective model-space interactions. Among them are advanced applications
of the many-body perturbation theory~\cite{MHJ95,Otsuka3N,Holt12,Fukui18}, 
as well as recently developed non-perturbative approaches based
on the similarity-renormalization group (SRG) concept and known as in-medium SRG (IMSRG)~\cite{Bogner2014,Stroberg2016,StrPRL118}, 
coupled-cluster theory~\cite{Jansen2014} and an approach based  on the Okubo-Lee-Suzuki (OLS) transformation of no-core shell-model
(NCSM) wave functions~\cite{Lisetskiy_2008,Lisetskiy_2009,Dikmen2015,SmBa2019}.
Many of these approaches have introduced explicitly 
the three-nucleon forces~\cite{Otsuka3N,Holt12,Fukui18,Bogner2014,Stroberg2016,StrPRL118,Jansen2014},
improving the monopole term of the corresponding interactions to get closer to the successful phenomenological interactions
such as USD and USDB~\cite{USD,USDab} in the $sd$ shell or KB3G~\cite{KB3G} and GXPF1A~\cite{GXPF1A} in the $pf$ shell. 
All these are important steps towards {\em ab-initio} theory of atomic nuclei.

In this article we present our progress along the lines reported in Ref.~\cite{SmBa2019}, where an effective
interaction for the $sd$ shell-model space was constructed based on the NCSM results using various nucleon-nucleon
potentials.
In that work the NCSM calculations were performed for $^{18}$F, as well as for $^{16,17}$O and $^{17}$F
in the model space defined by maximal total oscillator quanta of excitation  $N_{\rm max}=4$. 
For the oscillator parameter of the single-particle basis, $\hbar \Omega $=14 MeV was chosen.
The unitary OLS transformation was exploited to reduce the Hamiltonian to a block-diagonal form, separating
the model space ($sd$ shell) from the rest. Subtraction of the NCSM core energy and one-body contributions generated
a set of 63 two-body matrix elements (TBMEs), defining a residual two-body interaction in the $sd$-shell valence space.
Among different high-precision potentials employed, such as 
chiral N3LO (from Ref.~\cite{EnMa03}), JISP16~\cite{JISP16} and Daejeon16~\cite{DJ16}, 
the best agreement with experiment for selected $sd$-shell nuclei was achieved for Daejeon16.
It was also shown that the effective valence-space interaction derived from the Daejeon16 
with the correction to the monopole part resulted in a very good description of O binding energies and showed improved description of spectra.
This conclusion was supported by the study of upper
$sd$-shell nuclei which have been investigated in Ref.~\cite{ChSr2022}, using the same interactions.
In addition, in Ref.~\cite{ArimaMemorial} we have initiated construction of effective electromagnetic operators for valence-space calculations
with effective interactions originating from Daejeon16.

In the present study, we again employ the Daejeon16 interaction and construct an $sd$-shell effective interaction now based on $N_{\rm max}=6$ NCSM results. 
The results are compared with the previously derived interaction at $N_{\rm max}=4$ and with the phenomenological USDB interaction.

The paper is organized as follows. After a brief introduction to the method of deriving the effective valence-space interactions, 
we discuss the convergence properties of our calculations, as well as explain the challenges in the state selection, 
which provides support for our choice of the oscillator parameter. 
Then, we present the resulting valence space interaction, DJ16$_6$, obtained from the Daejeon16 $NN$ potential at $N_{\rm max} =6$ and 
compare it with the previously derived interaction at $N_{\rm max} =4$, DJ16$_4$.
In particular, we discuss the properties of two-body centroids which describe a spherical mean-field and are important
bellwethers for nuclear spectroscopy.
The main part of the article demonstrates the comparison of 
the low-lying spectra and binding energies of the O isotopes to illustrate the $T=1$ component. 
Then, we discuss the 
spectra of the odd-$A$ F isotopes and $^{39}$K to understand the main features of the proton-neutron centroids.
We also test the interaction on some selected $sd$ shell nuclei.
We then perform some minimal modifications of the mainly $T=1$ centroids of the effective interaction DJ16$_6$ 
in order to improve the agreement with the experimental data.
The extensive calculation of spectra of $sd$ shell nuclei are presented in Supplemental Material.
We conclude with our summary and outlook.

\section{Microscopic two-body interactions: formalism, convergence and state selection}

The derivation of the effective valence-space interaction has been discussed in detail 
in Refs.~\cite{Lisetskiy_2008,Lisetskiy_2009,Dikmen2015,SmBa2019}.
We start with a translationally-invariant Hamiltonian for $A$ point-like nucleons
interacting via a realistic $NN$ interaction
\begin{equation}
\label{H_intrinsic}
H =\sum_{i}^{A}\frac{\vec{p_i}^2}{2m}- \frac{\vec{P}^2}{2mA}
         +\sum_{i<j}^{A}V_{ij}^{NN},
\end{equation}
where $m$ is the nucleon mass  (approximated here as the average of the neutron and proton mass), 
$\vec{p}_i$ are nucleonic momenta, $\vec{P}=\sum_{i=1}^A \vec{p}_i$,
and $V_{ij}^{NN}$ denotes the $NN$ interaction. 
The two-body Coulomb interaction is included between the protons.

Within the NCSM, the eigenproblem for $H$ is solved by diagonalizing the
Hamiltonian matrix in a many-body spherical harmonic-oscillator (HO) basis. The model space
is defined by two parameters: (i) by a given HO energy quantum, $\hbar \Omega $, and 
(i) by a cut-off in the total number of the HO excitation quanta, $N_{\rm max}$. 
This means that we retain only many-body configurations, satisfying the condition
$N_{\rm min} \le \sum_{i=1}^A (2n_i+l_i) = N_{\rm min} + N$$\le N_{\rm min} + N_{\rm max}$,
where $n_i$ is the single-particle radial HO quantum number, 
$l_i$ is the single-particle orbital angular momentum quantum number, while
$N_{\rm min}$ is the minimum of the summation that satisfies the Pauli principle for the chosen $A$-nucleon system.
Consequently, $N = 0, \ldots, N_{\rm max}$.

One of the important advantages of the HO basis is that it allows one to separate 
the spurious center-of-mass motion. 
%
{
In practice, a truncation by the total number of HO quanta in the many body system leads to an exact factorization of the center-of-mass wave function and the intrinsic wave function; 
adding a center-of-mass term $\beta (H_{CM}-\frac32 \hbar \Omega )$ 
to the Hamiltonian~\eqref{H_intrinsic} shifts states with center-of-mass excitations by $\beta \hbar \Omega $~\cite{Lawson}}.

In the present study, we perform the NCSM calculations using the bare Daejeon16 $NN$ 
potential.  That is, we omit the first OLS renormalization, previously performed at the 2-nucleon cluster level~\cite{SmBa2019}, 
since Daejeon16 is reasonably soft and calculations in $N_{\rm max} = 6$ basis spaces are close enough 
to convergence for our purposes.
As a starting point, we calculate the ground state energy and the spectra of $^{16}$O, $^{17}$O, $^{17}$F and $^{18}$F
at a set of $\hbar \Omega $ values ranging from 12~MeV to 26~MeV, using a highly parallelized 
Multi-Fermion Dynamics for nuclei (MFDn) code~\cite{MFDn1,MFDn2,MFDn3,MFDn4,MFDn5}.

\begin{figure*}[tbp!]
 \centering
  \includegraphics[width=0.32\textwidth]{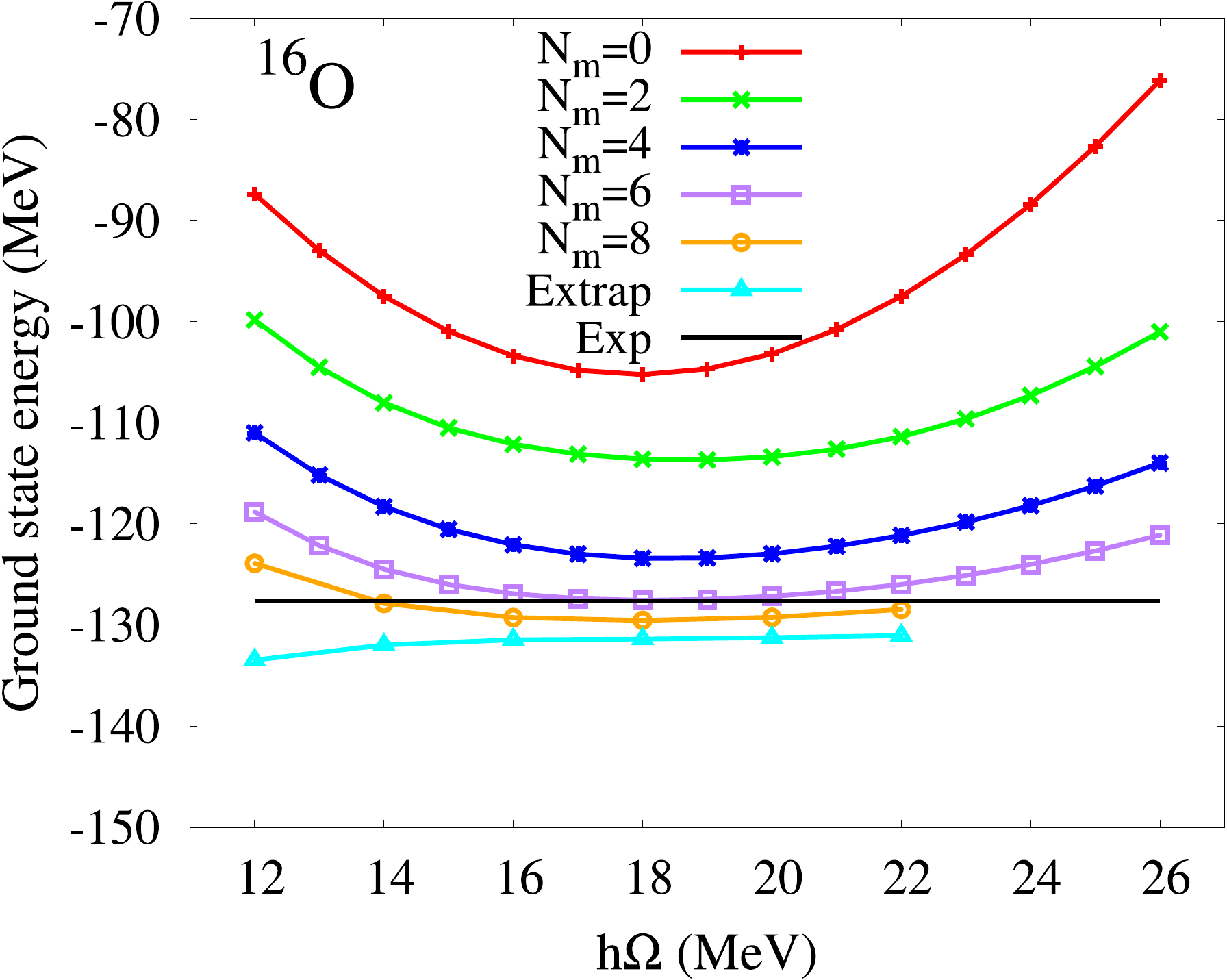}
  \includegraphics[width=0.32\textwidth]{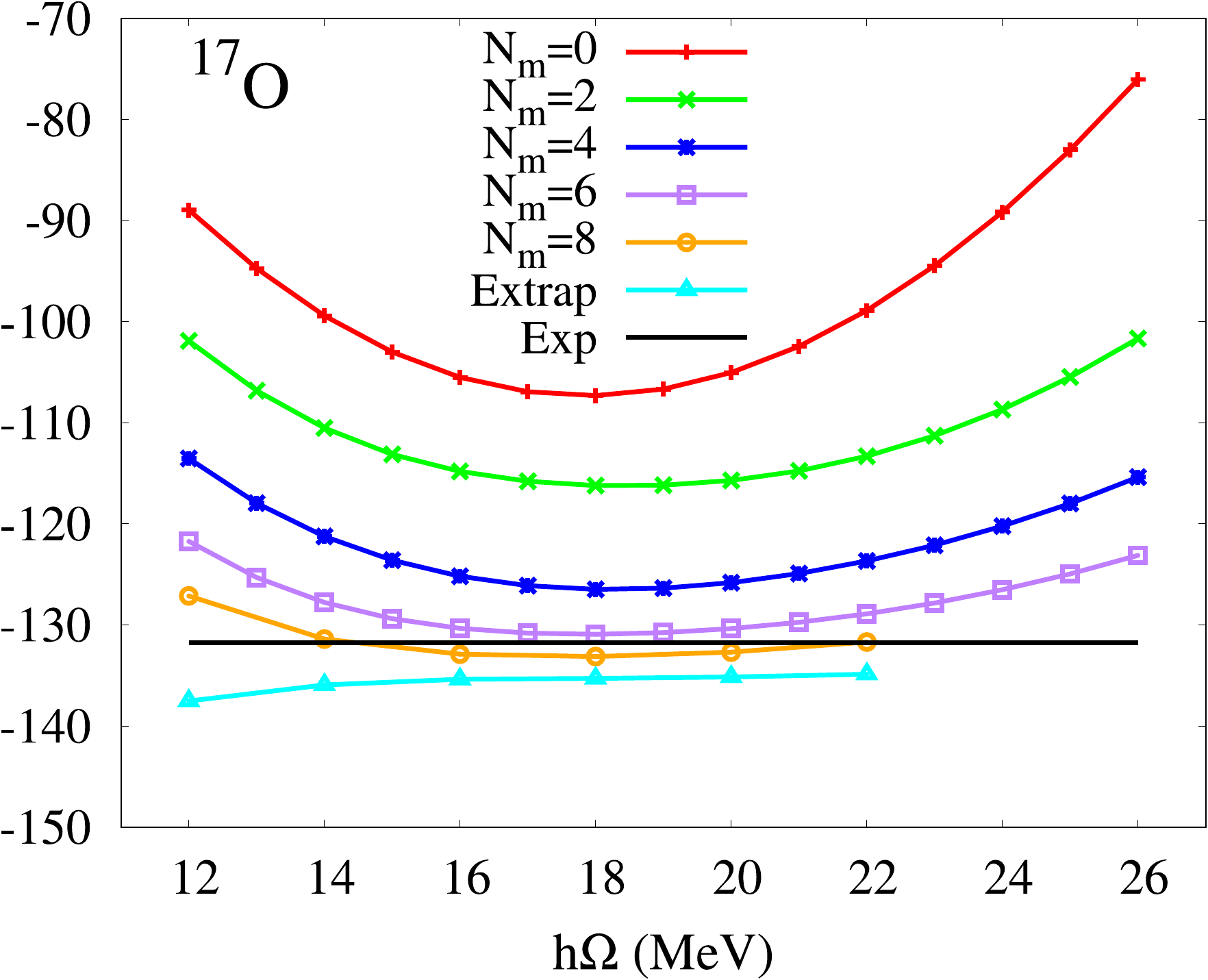} 
  \includegraphics[width=0.32\textwidth]{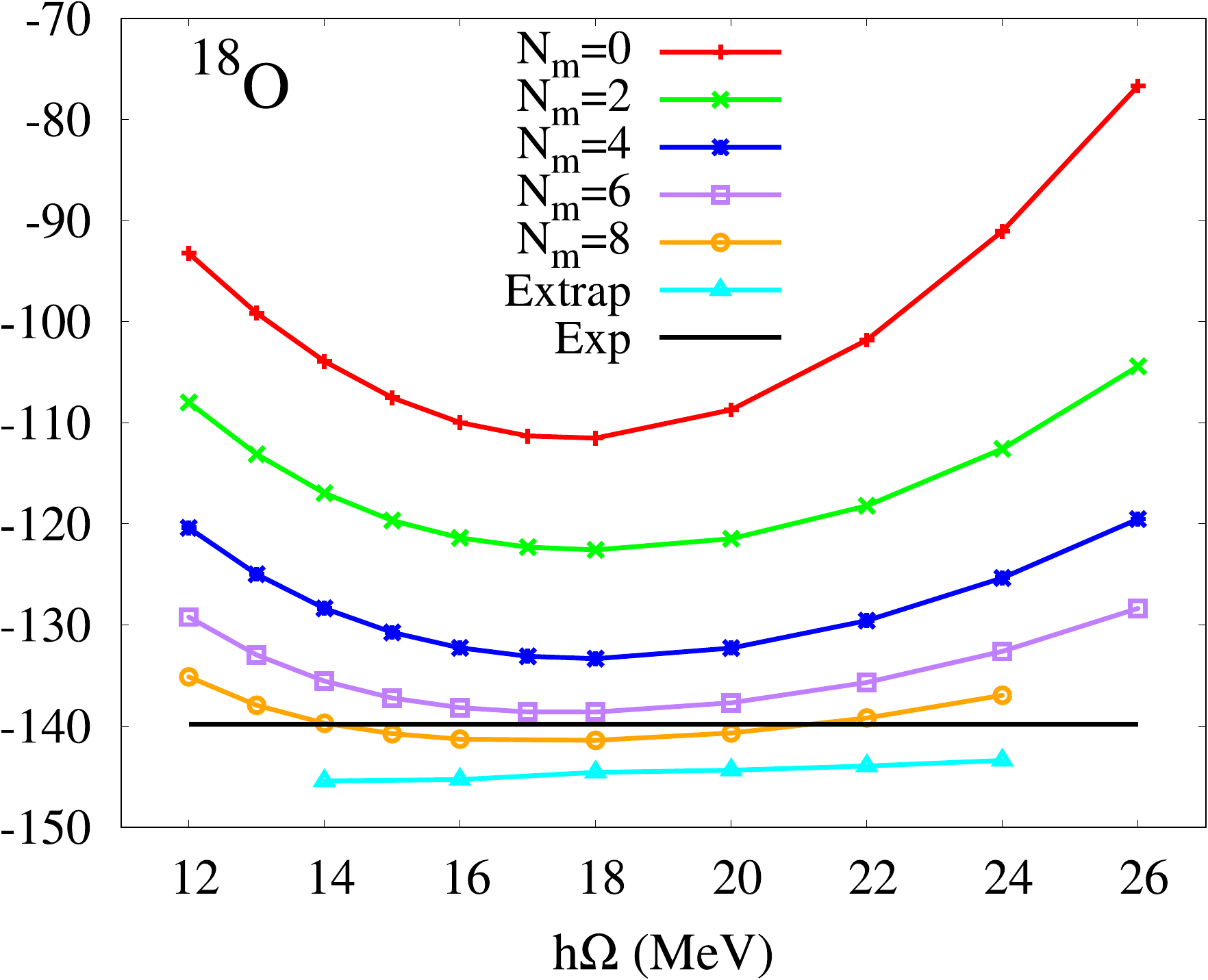} 
  \caption{\label{fig:conv} (Color online) Ground-state energies of $^{16,18}$O and the lowest $5/2^+$ state energy of $^{17}$O
obtained within the NCSM from the Daejeon16 $NN$ potential in comparison with experiment.  } 
\end{figure*}

\begin{figure*}[!tp]
 \centering
  \includegraphics[width=0.3\textwidth]{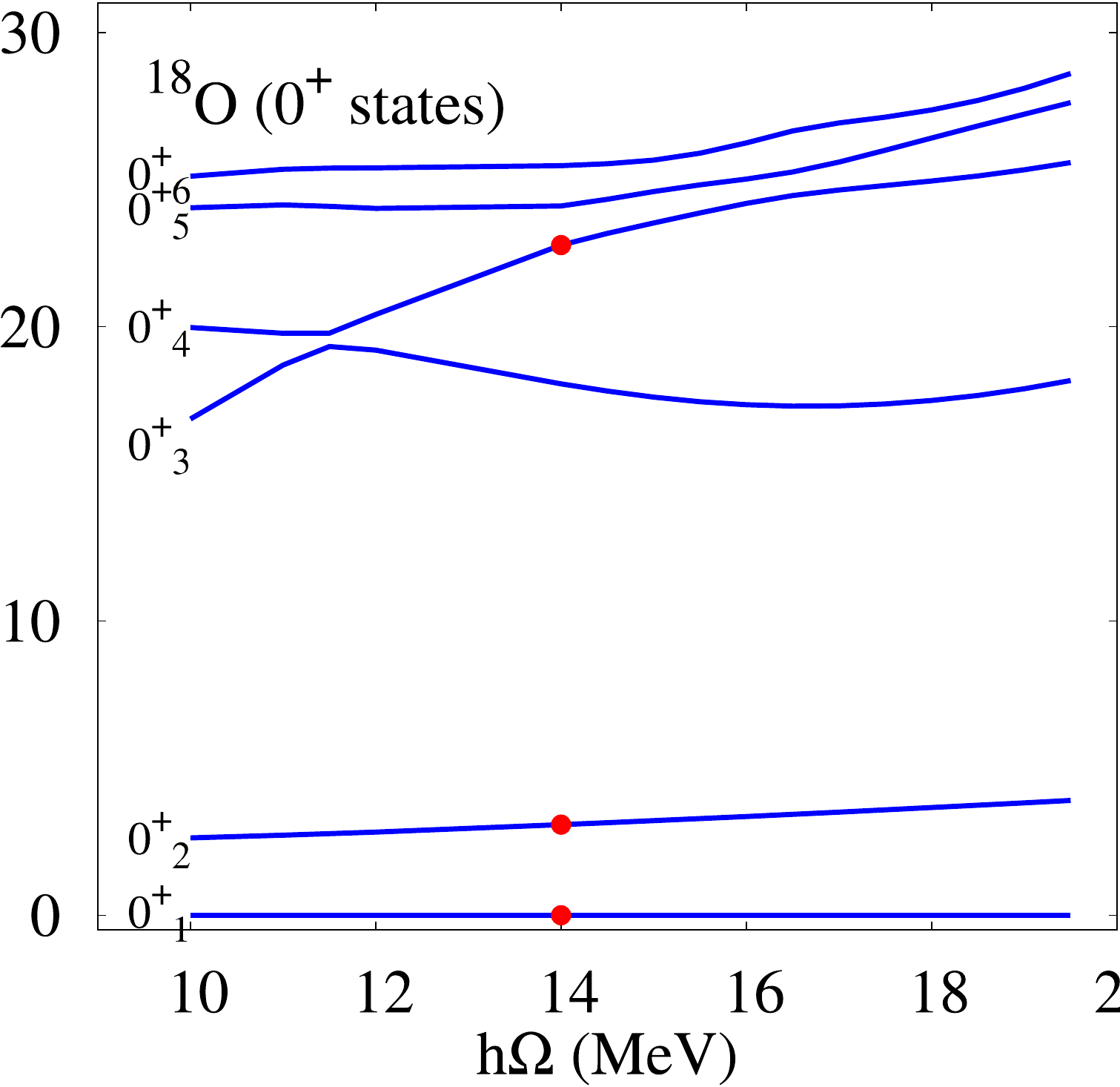} 
  \includegraphics[width=0.3\textwidth]{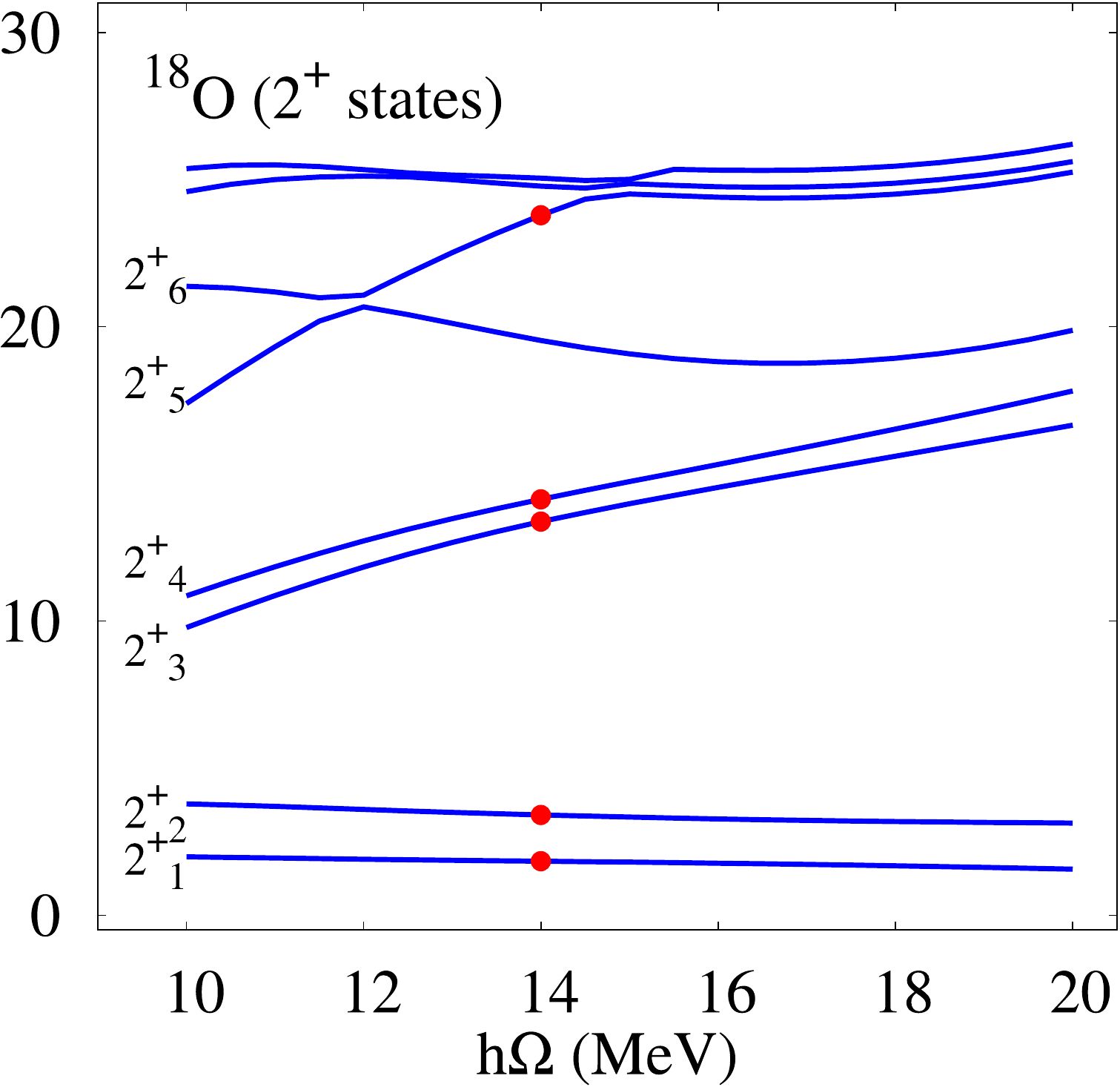} 
  \includegraphics[width=0.3\textwidth]{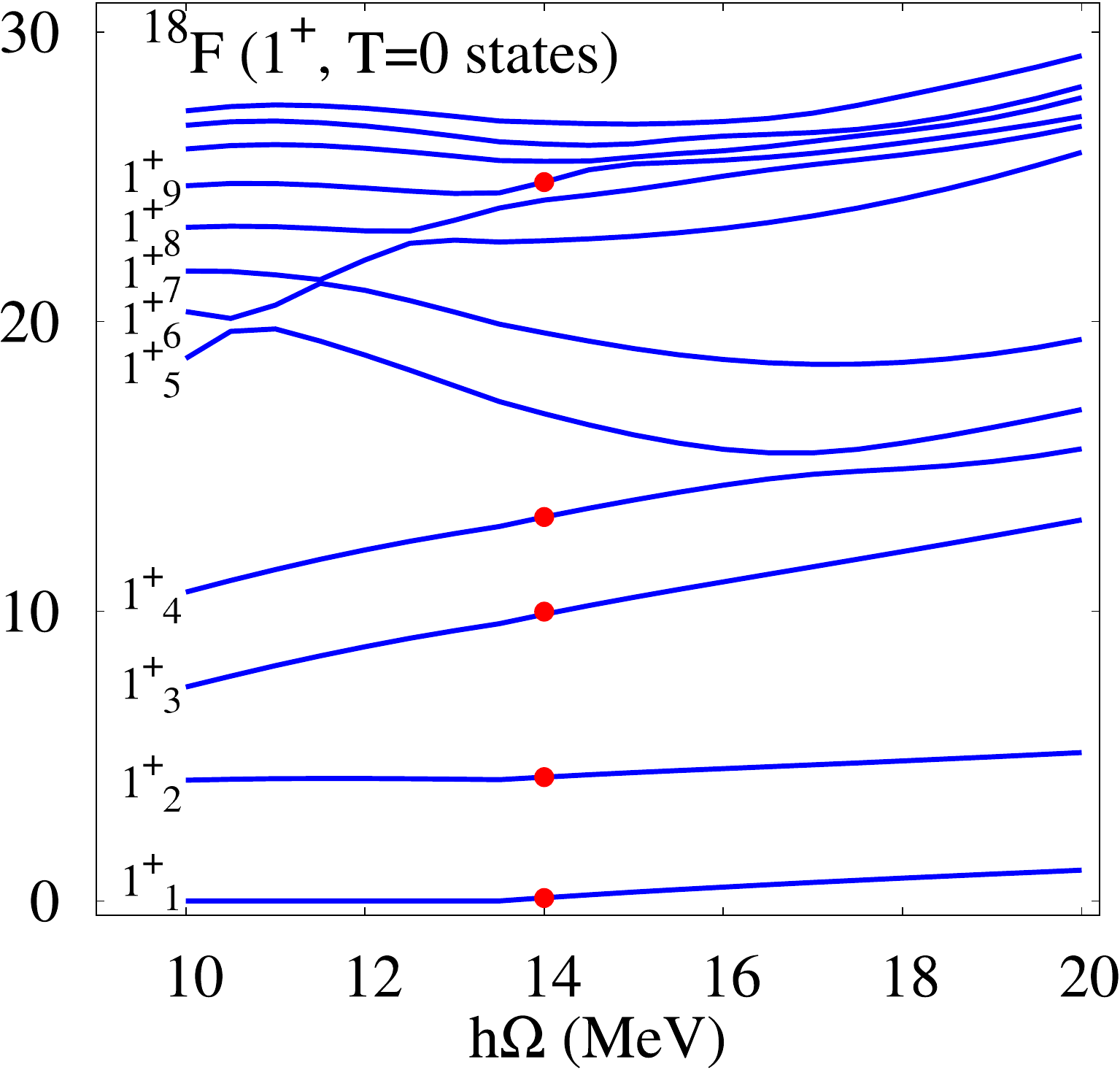} 
  \caption{\label{fig:state_selection} (Color online) Evolution of $0^+$ and $2^+$ states in $^{18}$O and $1^+, T=0$ states in $^{18}$F,
obtained within the NCSM with the Daejeon16 $NN$ potential at $N_{\rm max}=4$ as a function of $\hbar \Omega $. 
For illustration purpose, the calculation of $^{18}$F was done without Coulomb to avoid isospin mixing.
The states, labeled by red dots, are those having largest $N=0$ components at $\hbar \Omega =14$~MeV. See text for detail.} 
\end{figure*}

\begin{figure}[tbp!]
 \centering
  \includegraphics[width=\columnwidth]{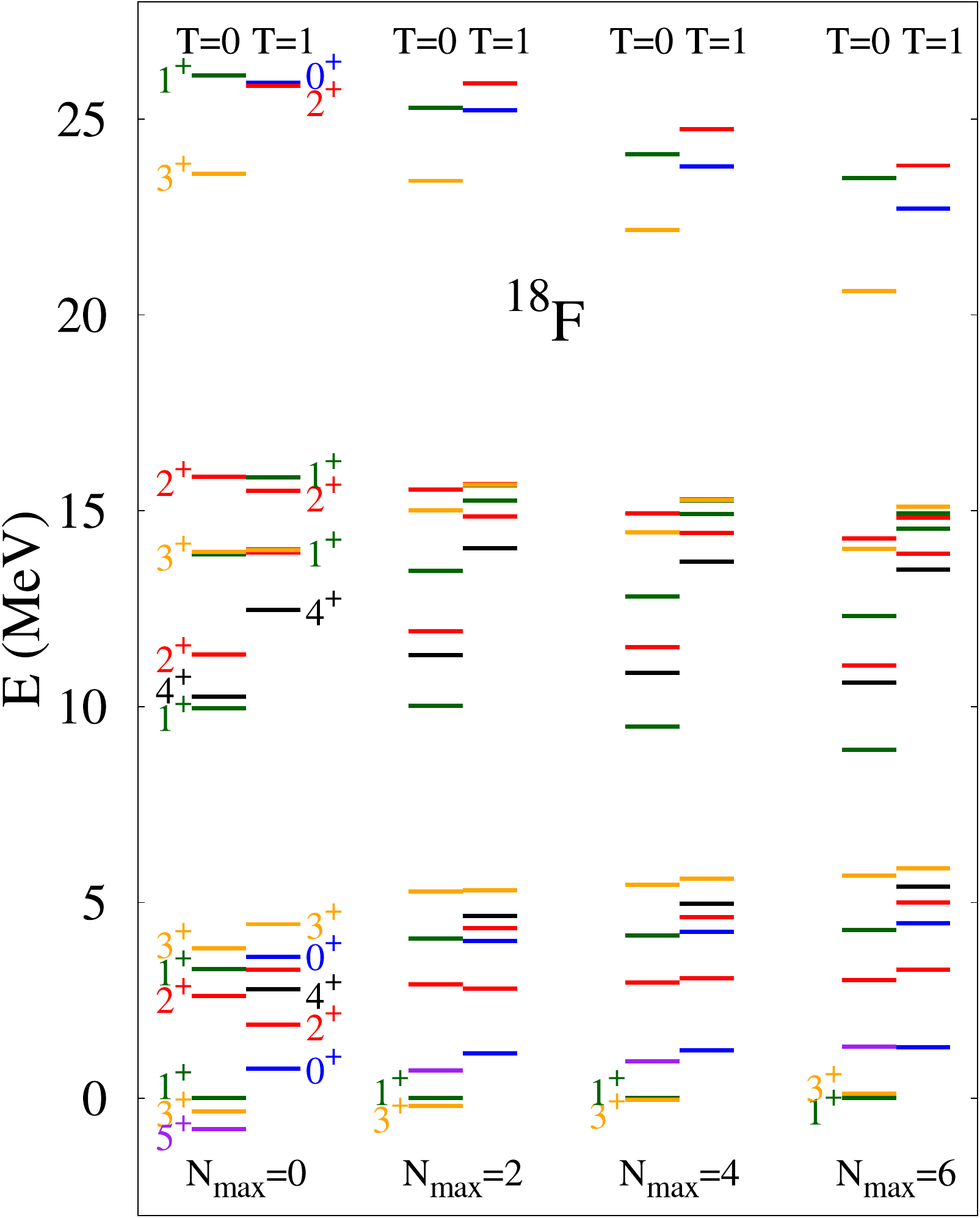} 
  \caption{\label{fig:conv_exc} (Color online) Excitation spectrum of twenty-eight states having the largest $N=0$ component
in $^{18}$F as obtained within the NCSM from  the Daejeon16 $NN$ potential at $\hbar \Omega =14$~MeV.
The energies are indicated relative to the position of the lowest ($1^+$, $T=0$) state, since it is
the known ground state spin and parity of $^{18}$F.
$T=0$ states are shown on the left, while the $T=1$ spectrum is on the right of each $N_{\rm max}$ column.} 
\end{figure}

The microscopic effective interactions obtained from NCSM depend on $N_{\rm max}$ and 
$\hbar \Omega $. 
On the other hand, the eigenvalues of the bound states of the NCSM results 
converge as $N_{\rm max}$ increases
towards results independent of $\hbar \Omega $ and $N_{\rm max}$~\cite{Barrett_2013}.
The ground state energies of oxygen isotopes with $A=16, 18$ and the energy of
the lowest $5/2^+$ state of $^{17}$O are shown in Fig.~\ref{fig:conv}
as functions of $\hbar \Omega $ for $N_{\rm max}=0,2,4,6,8$.
We also present the extrapolated values obtained following
the ``extrapolation B'' procedure of Ref.~\cite{Maris2009}.
First, we remark that Daejeon16, being a soft potential, results in a rather rapid convergence for these states.
Second, we note that for all nuclei we obtain a slight overbinding as compared to experiment.
The minimum energy is attained near $\hbar \Omega =18$~MeV, which is also the case for excited states (not shown in this figure).


Although the largest model space for which NCSM calculations are feasible for $A=18$ 
is the one defined by $N_{\rm max}=8$, to construct an effective interaction we need to identify
the required 28 NCSM eigenstates. 
It turns out that the states $0^+_3$, $2^+_5$, $T=1$ and $1^+_5$, $3^+_2$, $T=0$ are shifted higher in energy and appear in a region of dense NCSM solutions, with increasing $\hbar \Omega $.
This phenomenon is illustrated in Fig.~\ref{fig:state_selection} for $N_{\rm max} =4$ calculations for $0^+$ and $2^+$ in $^{18}$O and 
for $1^+,T=0$ states in $^{18}$F (without Coulomb for clarity of presentation).
For example, at $\hbar \Omega =12$~MeV, the three lowest $0^+$, $T=1$ states and five lowest $2^+$, $T=1$ states are the states
which carry the largest $N=0$ components among all the states of the same $J^{\pi },T$ quantum numbers.
However, already at $\hbar \Omega =14$ MeV, because of the (avoided) level crossing, $0^+_3$ becomes an intruder state and it is $0^+_4$ 
which is selected for the OLS transformation (see left panel of  Fig.~\ref{fig:state_selection}).
Similarly, at $\hbar \Omega =14$~MeV we chose $2^+_6$ as a fifth $2^+$, $T=1$ state for the OLS transformation as evident from the middle panel of Fig.~\ref{fig:state_selection}.
The situation with $1^+$ and $3^+$, $T=0$ states is even more complicated: these states rapidly go up in energy and penetrate into a region
of the high level density. The twelve lowest $1^+$, $T=0$ states in $^{18}$F are shown in the right panel of Fig.~\ref{fig:state_selection}.
Besides the four lowest states, it is the 9th $1^+$, $T=0$ state which is included for the construction of the $sd$-shell interaction.

The above-described situation becomes more complicated at $N_{\rm max}=6$. 
Thus, we require a large number of converged states. 
Their identification becomes more difficult due to numerous (avoided) level crossings and fractioning of the states.
In general, there is a good continuity of the values of TBMEs derived as a function of $N_{\rm max}$ and of $\hbar \Omega $
up to $\hbar \Omega =18$ MeV, where more serious ambiguities in the state selection enter.
Aiming at stable results, we thus limit ourselves to the $N_{\rm max}=6$ model space and, 
following the previous studies~\cite{Dikmen2015,SmBa2019}, we derive the effective interaction at $\hbar \Omega =14$~MeV. 
This value is close to the empirical shell-model value for $^{16}$O~\cite{Kirson07}.
The selected 28 eigenstates of $^{18}$F characterized by the largest contribution of $N=0$ components 
at  $\hbar \Omega =14$ MeV and $N_{\rm max}=6$ (see Table~\ref{tab:F18_Ex_ncsm_Nmax6}, left part) in comparison with the states selected
in our previous work at $N_{\rm max}=4$ (right part of the Table).
These states have been used to set up the OLS transformation to the $sd$-shell valence space.
Although the $N=0$ component of the 
$0^+_3$, $2^+_5$, $T=1$ and $1^+_5$, $3^+_2$, $T=0$ states in $^{18}$F is falling below 30\%,
we are able to clearly identify them among a set of states with even smaller components in the valence space.
{
The probability of the $N=0$ component, denoted as $\alpha^2_{N=0}$, is given for each selected state in Table~\ref{tab:F18_Ex_ncsm_Nmax6}}.
We checked that the use of  $\hbar \Omega =16$ and 18~MeV does not bring any qualitative improvement to the interactions,
but complicates the construction because of the problems indicated above.

To illustrate the convergence of our calculations, we show in  Fig.~\ref{fig:conv_exc}
the excited states of $^{18}$F, characterized by the largest $N=0$ components, as a function of $N_{\rm max}$. 
Indeed, the majority of the states stay relatively constant in energy. Only the highest $0^+$ and $2^+$ $T=1$ states
and the highest $1^+$ and $3^+$ $T=0$ states, containing lower contribution of $N=0$ components, 
show a decrease in energy when moving from $N_{\rm max}=4$ to $N_{\rm max}=6$.
The other states are reasonably converged.
We also remark that the absolute values of $N=0$ components reduce at  $N_{\rm max}=6$ as compared to those
obtained at to $N_{\rm max}=4$. This is a consequence of the increase of the model space.

\begin{table}[!t]
\vspace{-2ex}
\caption{\label{tab:F18_Ex_ncsm_Nmax6}The NCSM energies (in MeV) of the lowest
28 states $J^\pi_i$ of $^{18}$F calculated in the $N_{\rm max}=6$ model space (left)
and $N_{\rm max}=4$ model space (right) using the Daejeon16 $NN$ interaction 
with $\hbar \Omega=14$ MeV. 
These are states which contain the largest $N=0$ components (reported in the column $\alpha_{N=0}^2$),
and therefore are chosen for the OLS transformation to the valence space ($N'_{\rm max}=0$).}
\begin{ruledtabular}
\begin{tabular}{rrrr|rrrr}
$J^\pi_i$ & T & $N_{\rm max}=6$ & $\alpha_{N=0}^2$ & $J^\pi_i$ & T & $N_{\rm max}=4$  & $\alpha_{N=0}^2$ \\ \hline
$1^+_1$ & 0  & $-133.311$ & 0.583 & $3^+_1$ & 0 & $-126.069$ & 0.675 \\
$3^+_1$ & 0  & $-133.191$ & 0.600 & $1^+_1$ & 0 & $-126.032$ & 0.668 \\
$0^+_1$ & 1  & $-132.017$ & 0.583 & $5^+_1$ & 0 & $-125.087$ & 0.694 \\
$5^+_1$ & 0  & $-131.991$ & 0.619 & $0^+_1$ & 1 & $-124.817$ & 0.670 \\
$2^+_1$ & 0  & $-130.288$ & 0.588 & $2^+_1$ & 0 & $-123.081$ & 0.672 \\
$2^+_2$ & 1  & $-130.028$ & 0.597 & $2^+_2$ & 1 & $-122.965$ & 0.679 \\
$1^+_2$ & 0  & $-129.016$ & 0.602 & $1^+_2$ & 0 & $-121.884$ & 0.684 \\
$0^+_2$ & 1  & $-128.837$ & 0.602 & $0^+_2$ & 1 & $-121.778$ & 0.682 \\
$2^+_3$ & 1  & $-128.316$ & 0.616 & $2^+_3$ & 1 & $-121.402$ & 0.691 \\
$4^+_1$ & 1  & $-127.909$ & 0.625 & $4^+_1$ & 1 & $-121.071$ & 0.700 \\
$3^+_2$ & 0  & $-127.624$ & 0.609 & $3^+_2$ & 0 & $-120.591$ & 0.690 \\
$3^+_3$ & 1  & $-127.443$ & 0.604 & $3^+_3$ & 1 & $-120.421$ & 0.684 \\
$1^+_3$ & 0  & $-124.407$ & 0.528 & $1^+_3$ & 0 & $-116.545$ & 0.640 \\
$4^+_2$ & 0  & $-122.693$ & 0.572 & $4^+_2$ & 0 & $-115.164$ & 0.668 \\
$2^+_4$ & 0  & $-122.255$ & 0.481 & $2^+_4$ & 0 & $-114.521$ & 0.656 \\
$1^+_4$ & 0  & $-121.000$ & 0.508 & $1^+_4$ & 0 & $-113.214$ & 0.632 \\
$4^+_3$ & 1  & $-119.817$ & 0.568 & $4^+_3$ & 1 & $-112.337$ & 0.671 \\
$2^+_5$ & 1  & $-119.401$ & 0.510 & $2^+_5$ & 1 & $-111.594$ & 0.651 \\
$3^+_4$ & 0  & $-119.281$ & 0.546 & $3^+_4$ & 0 & $-111.579$ & 0.658 \\
$2^+_6$ & 0  & $-119.018$ & 0.526 & $1^+_5$ & 1 & $-111.112$ & 0.660 \\
$1^+_5$ & 1  & $-118.766$ & 0.551 & $2^+_6$ & 0 & $-111.092$ & 0.638 \\
$2^+_7$ & 1  & $-118.483$ & 0.551 & $2^+_7$ & 1 & $-110.803$ & 0.651 \\
$1^+_6$ & 1  & $-118.381$ & 0.560 & $1^+_6$ & 1 & $-110.779$ & 0.658 \\
$3^+_5$ & 1  & $-118.205$ & 0.572 & $3^+_5$ & 1 & $-110.748$ & 0.673 \\
$3^+_6$ & 0  & $-112.705$ & 0.280 & $3^+_6$ & 0 & $-103.869$ & 0.588 \\
$0^+_3$ & 1  & $-110.598$ & 0.172 & $0^+_3$ & 1 & $-102.246$ & 0.454 \\
$1^+_7$ & 0  & $-109.812$ & 0.193 & $1^+_7$ & 0 & $-101.928$ & 0.330 \\
$2^+_8$ & 1  & $-109.495$ & 0.279 & $2^+_8$ & 1 & $-101.291$ & 0.598 \\
\end{tabular}
\end{ruledtabular}
\end{table}

Performing the OLS transformation,
the effective valence Hamiltonian, $H_{18}^{P'}$, an effective one- and two-body operator for $N'_{\rm max}=0$, is derived 
(here prime is used to denote the valence space as in the previous works~\cite{Dikmen2015,SmBa2019}).
By construction, the energies of $H_{18}^{P'}$ for 2 valence nucleons in the $sd$ shell exactly coincide
with the selected eigenvalues of NCSM Hamiltonian for $^{18}$F in the full $N_{\rm max}=6$ oscillator space. 
See Ref.~\cite{Dikmen2015} for more details.

A NCSM calculation with the same $N_{\rm max}=6$ and  $\hbar \Omega =14$~MeV
is performed for $^{16}$O to get the core energy and for $^{17}$O and $^{17}$F.
Subtracting the core energy from the latter calculations, one obtains effective neutron
and proton one-body terms.
Subtraction of the core energy plus the one-body terms from the effective Hamiltonian for $^{18}$F, 
allows one to obtain the residual TBMEs to be used in the valence-space shell-model calculations.
The core energy and the single-particle energies obtained from the Daejeon16 $NN$ potential
are given in Table~\ref{tab:spe_DJ16} ($N_{\rm max}=6$ results in comparison with $N_{\rm max}=4$ results), 
while the $N_{\rm max}=6$ TBMEs  are summarized in Table~I of the Supplemental Material. 
The files with TBMEs are also available online~\cite{git}.

\begin{table}[t] 
\vspace{-2ex}
\caption{\label{tab:spe_DJ16}Neutron (``$\nu$'') and proton (``$\pi$'') single-particle energies (in MeV) 
obtained from the bare Daejeon16 potential for $A=17$ at $N_{\rm max}=6$ (left) and $N_{\rm max}=4$ (right)
for $\hbar \Omega = 14$~MeV.}
\begin{ruledtabular}
\begin{tabular}{c|ccc|ccc|}
 & \multicolumn{3}{c|}{$N_{\rm max}=6$} & \multicolumn{3}{c|}{$N_{\rm max}=4$} \\
\hline
 & \multicolumn{3}{c|}{$E_{\rm core}=-124.485$} & \multicolumn{3}{c|}{$E_{\rm core}=-118.307$} \\[1mm] 
\hline
$(nlj)$ & $1s_{1/2}$ &$0d_{5/2}$ & $0d_{3/2}$ & $1s_{1/2}$ &$0d_{5/2}$ & $0d_{3/2}$  \\[1mm] 
\hline
$\epsilon_{\nu }(nlj)$ & $-3.697$ & $-3.299$ & $5.823$ & $-3.115$ & $-2.953$ &  $6.889$ \\
$\epsilon_{\pi }(nlj)$ & $-0.253$ & $ 0.290$ & $9.063$ & $ 0.362$ & $ 0.621$ & $10.174$ \\
\end{tabular}
\end{ruledtabular}
\end{table} 

\begin{figure*}[!t]
\vspace{1ex}
 \centering
  \includegraphics[width=0.8\textwidth]{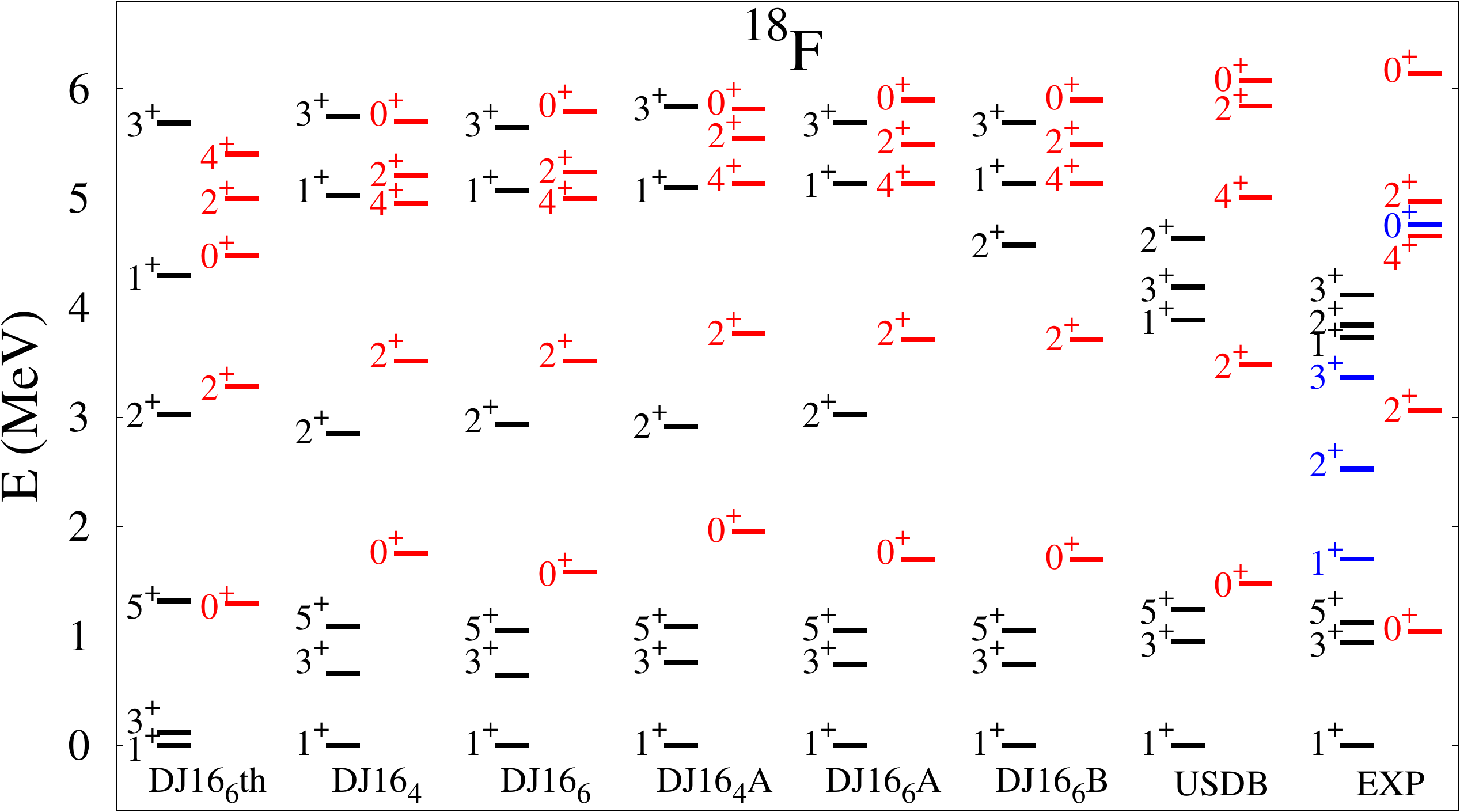}  
  \caption{\label{fig:F18} (Color online) Low-energy spectra (6 lowest $T=0$ states in black and 5 lowest $T=1$ states in red) 
of $^{18}$F obtained from USDB and from the microscopic effective interactions derived using Daejeon16 
in comparison with experiment. For DJ16$_6$th, the isospin symmetry is approximate. 
In the experimental spectrum we indicate additional possible low-energy intruder states (in blue color):
($1^+_2$, $2^+_1$, $3^+_2$, $T=0$) and   ($0^+_2$, $T=1$), which should not be compaired with results from $sd$-shell theory. 
The USDB single-particle energies have been used in all calculations except for DJ16$_6$th
which is the NCSM spectrum at $N_{\rm max}=6$ and $\hbar\Omega $= 14~MeV.
DJ16$_4$A is a monopole-modified version of DJ16$_4$, while DJ16$_6$A and DJ16$_4$B are modified versions of 
DJ16$_6$ (see following sections). } 
\vspace{-1ex}
\end{figure*}

As is evident from Table~\ref{tab:spe_DJ16}, theoretical single-particle
energies obtained at $N_{\rm max}=6$ are qualitatively similar to those obtained at $N_{\rm max}=4$ 
to within an overall shift, and
therefore, they are also very different from the best empirical values as set by the USDB Hamiltonian:
$\varepsilon (0d_{5/2})=-3.9257$~MeV, $\varepsilon (1s_{1/2})=-3.2079$~MeV and 
$\varepsilon (0d_{3/2})=2.1117$~MeV. 
First, the $d_{5/2}$ and $s_{1/2}$ orbitals remain inverted.
Second, the spin-orbit splitting between $d_{3/2}$ and $d_{5/2}$, being about 9~MeV at $N_{\rm max}=6$,
is still much larger than the empirical value of about 6.0~MeV.
These deficiencies in single-particle energies lead to similar problems with the description of nuclear spectra and binding energies
 as was discussed in Ref.~\cite{SmBa2019}.
While we plan to further investigate the theoretical single-particle energies in the future,
in the present study we limit ourselves to the comparison of the TBMEs and, therefore, we adopt the USDB single-particle
energies which are the same for neutrons and protons for our valence space calculations that compare theory with experiment.

Fig.~\ref{fig:F18} shows the low-energy spectrum of $^{18}$F obtained from USDB and from various
microscopic effective interactions obtained from Daejeon16 as well as the experimental spectrum. 
The first column, DJ16$_6$th, shows the theoretical spectrum
from the NCSM at $N_{\rm max}=6$  (Table~\ref{tab:F18_Ex_ncsm_Nmax6}).
The same spectrum is reproduced  from the effective $sd$-shell interaction with theoretical single-particle energies
given in Table~\ref{tab:spe_DJ16}.
The interactions named DJ16$_4$ and DJ16$_4$A are those obtained from the NCSM calculations at $N_{\rm max}=4$ -- 
the original and the monopole-modified, respectively, which have been thoroughly investigated in Ref.~\cite{SmBa2019} 
(in that work they were denoted as DJ16 and DJ16A, respectively).
We recall that DJ16$_4$ was obtained from the NCSM calculation with OLS-transformed two-nucleon Hamiltonian.
The resulting TBMEs, as well as O binding energies and excitation spectra of studied nuclei 
are very close to those obtained from the NCSM with the bare Daejeon16 potentials 
without the OLS renormalization.
Therefore, it is reasonable to use the published results with DJ16$_4$ for comparison in the present study.
The interaction named DJ16$_6$ is obtained 
by the NCSM calculations with Daejeon16 at $N_{\rm max}=6$ in the present work, 
while DJ16$_6$A and DJ16$_6$B are two phenomenologically modified versions of DJ16$_6$, as will be explained in the following sections.
All calculations, except for DJ16$_6$th, employ the USDB single-particle energies and, therefore, the states
are characterized by a given total isospin quantum number. In the case of DJ16$_6$th, the isospin symmetry is approximate.

In general, we notice a reasonable agreement between all theoretical spectra shown in Fig.~\ref{fig:F18} and 
the experimental data.
All interactions predict correct spin and parities of the ground and the first excited state.
We notice that the $T=1$ band lies higher than experiment. 
Also, in the five theoretical spectra (counting from the left), $2^+_1$, $T=0$ appears to be too low, 
while $1^+_2$ and $3^+_2$, $T=0$, appear to be too high. 
The former feature is improved by DJ16$_6$B, since we have modified a few non-diagonal ($J=2, T=0$) TBMEs.
We did not make any attempts to correct for the position of the $1^+_2$ or $3^+_2$, $T=0$ states.
The phenomenological USDB Hamiltonian well describes these particular spectral features.

In the present study, to adopt the derived TBMEs through the $sd$ shell, we use the same phenomenological
$A$ dependence as previously, namely, the $(A/A_0)^{-0.3}$ scaling ($A_0=18$) used also with USDB~\cite{USDab}.
In theory, $A$-dependence would arise from the many-body effective interactions derivable within the OLS procedures, 
which is beyond the scope of the present work.
The single-particle energies are kept constant for all calculations.



\begin{figure*}
 \centering
  \includegraphics[height=.18\textwidth]{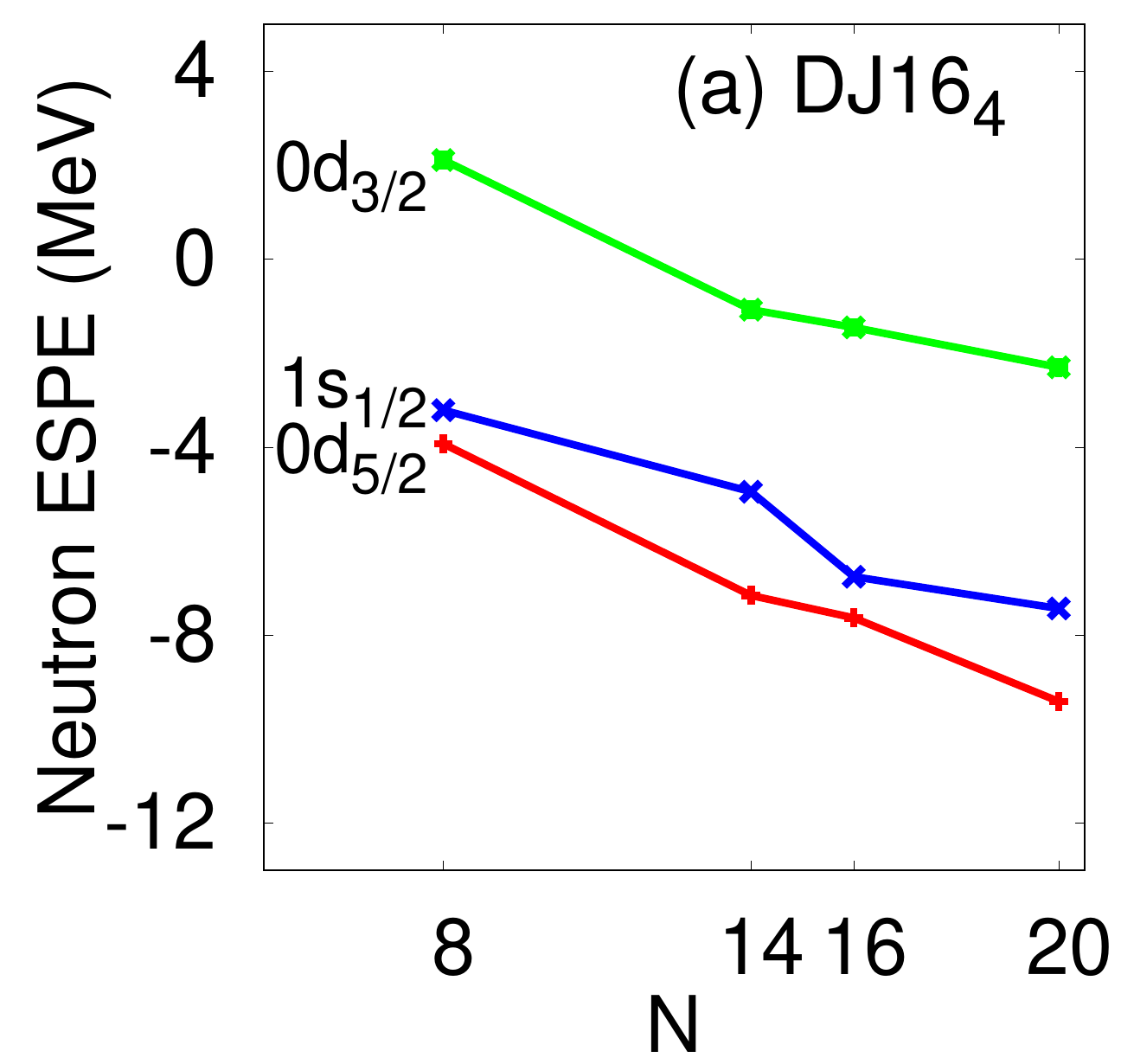}  
  \includegraphics[height=.18\textwidth]{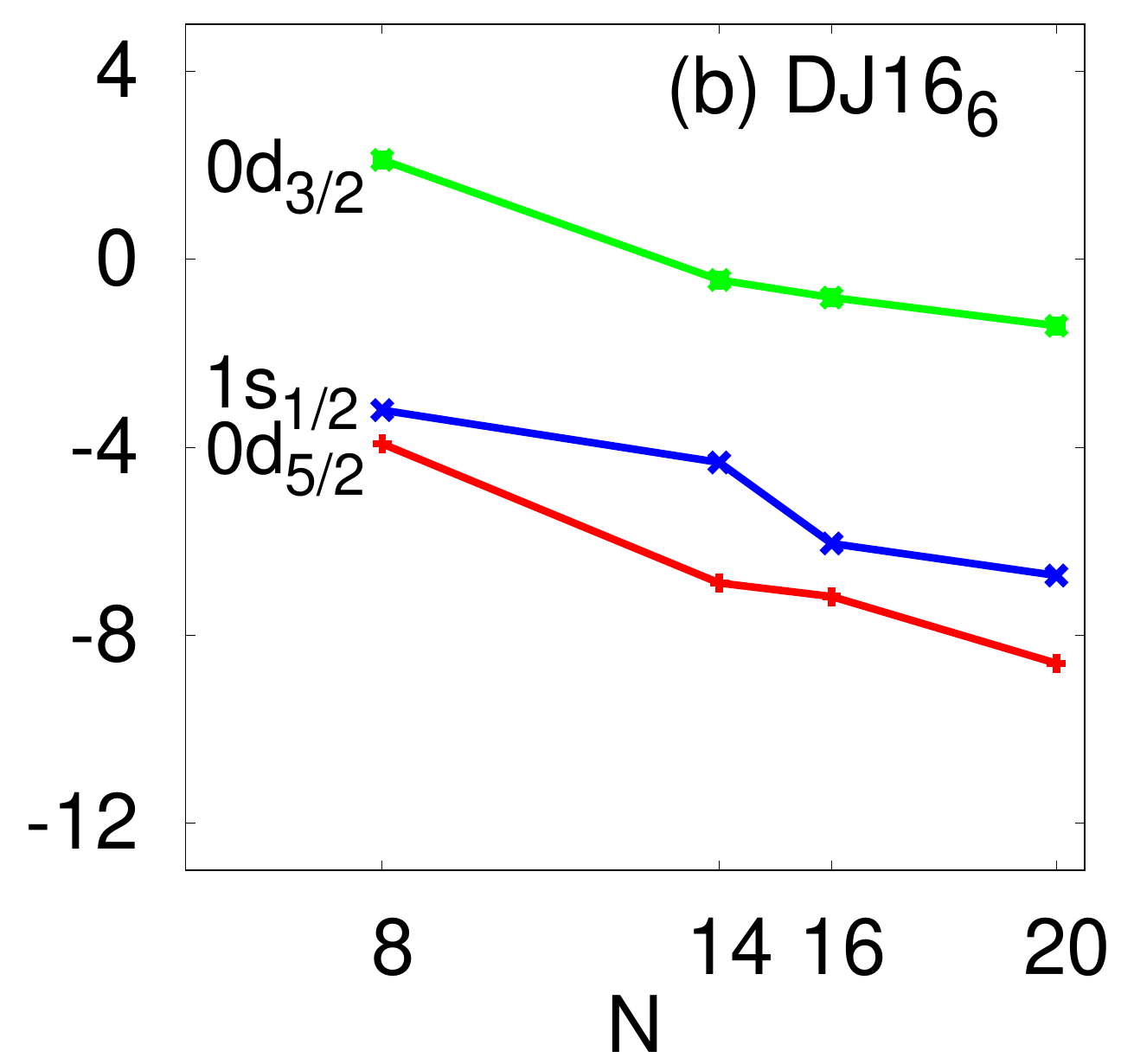}  
  \includegraphics[height=.18\textwidth]{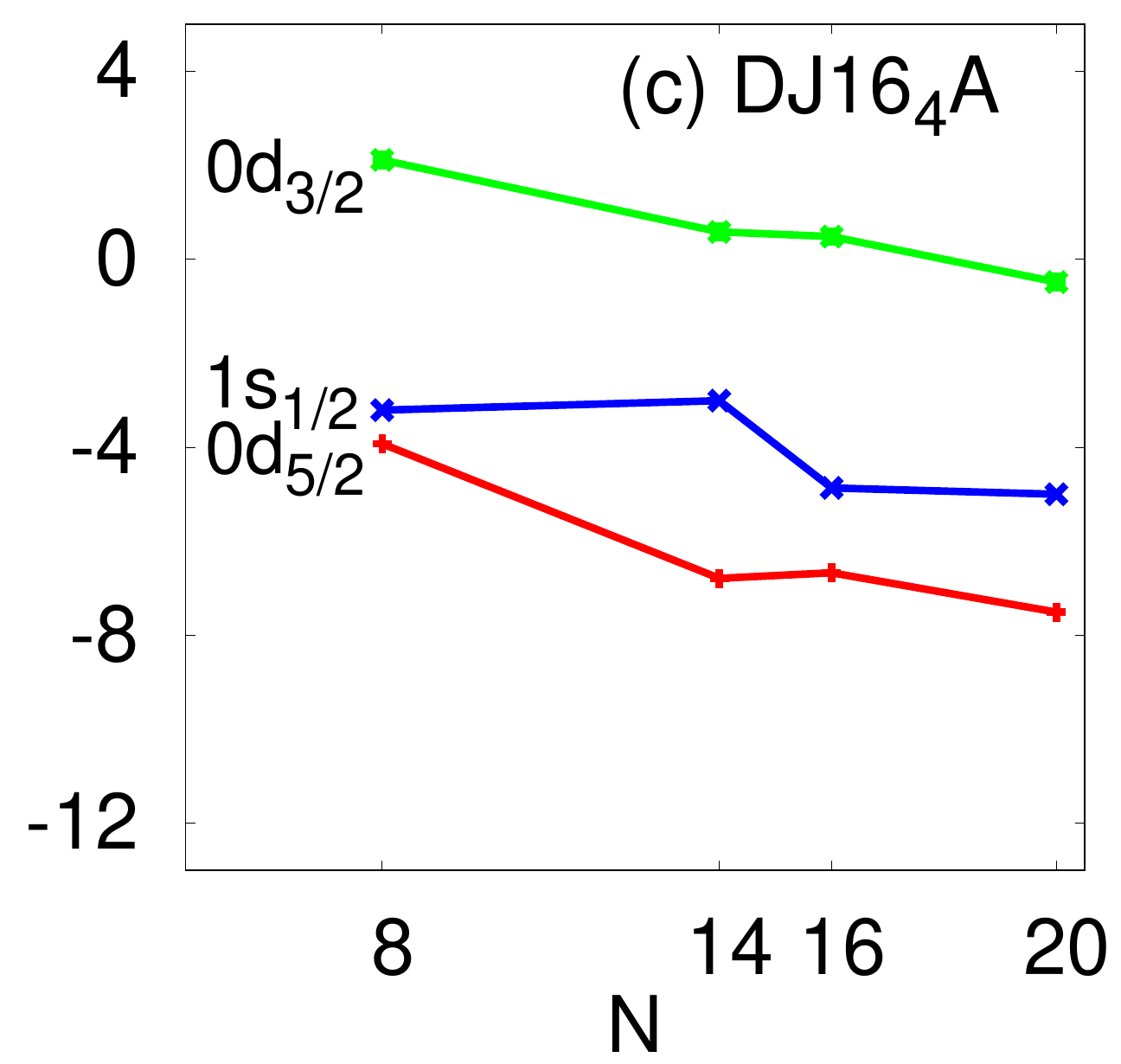}  
  \includegraphics[height=.18\textwidth]{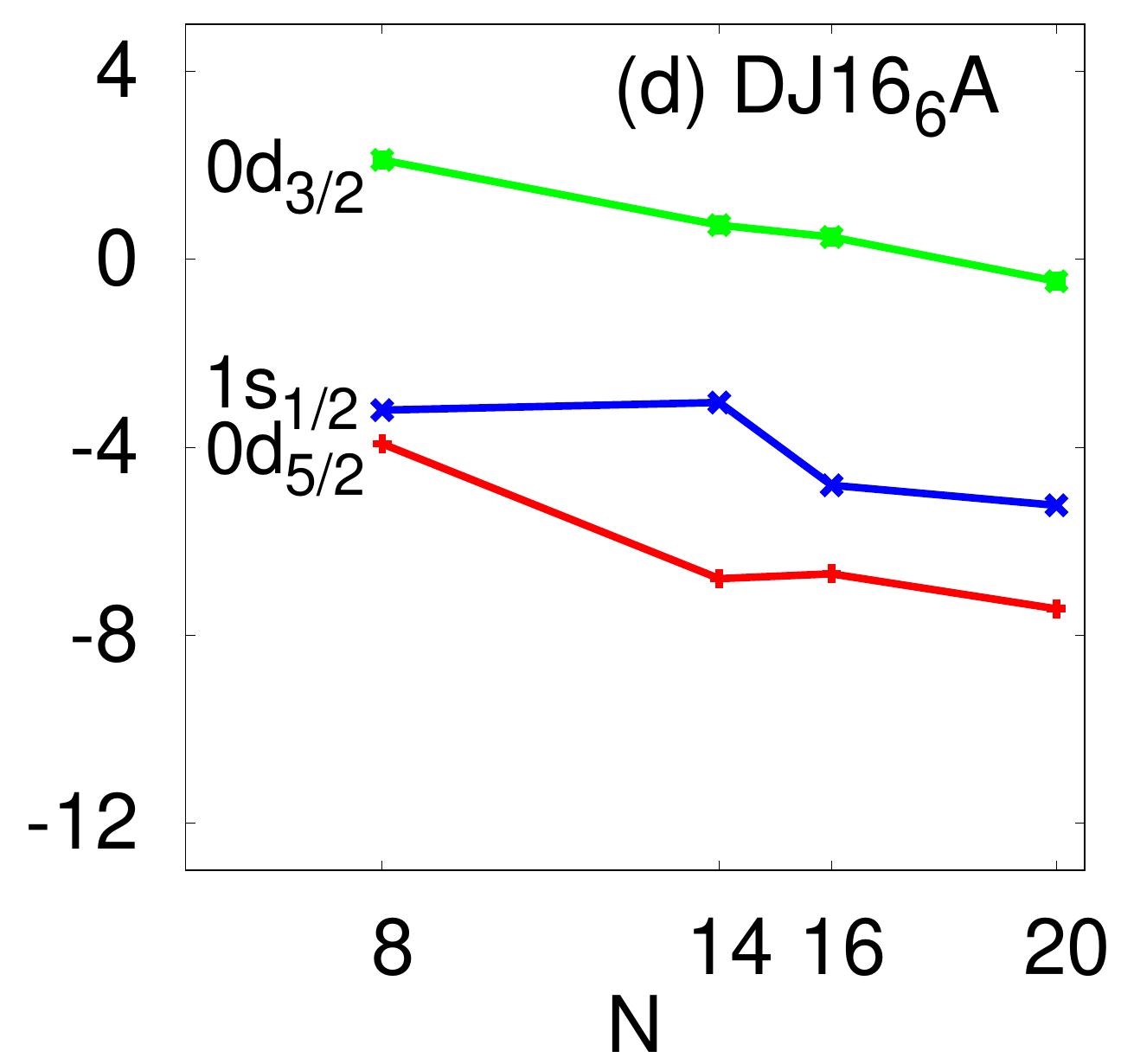} 
  \includegraphics[height=.18\textwidth]{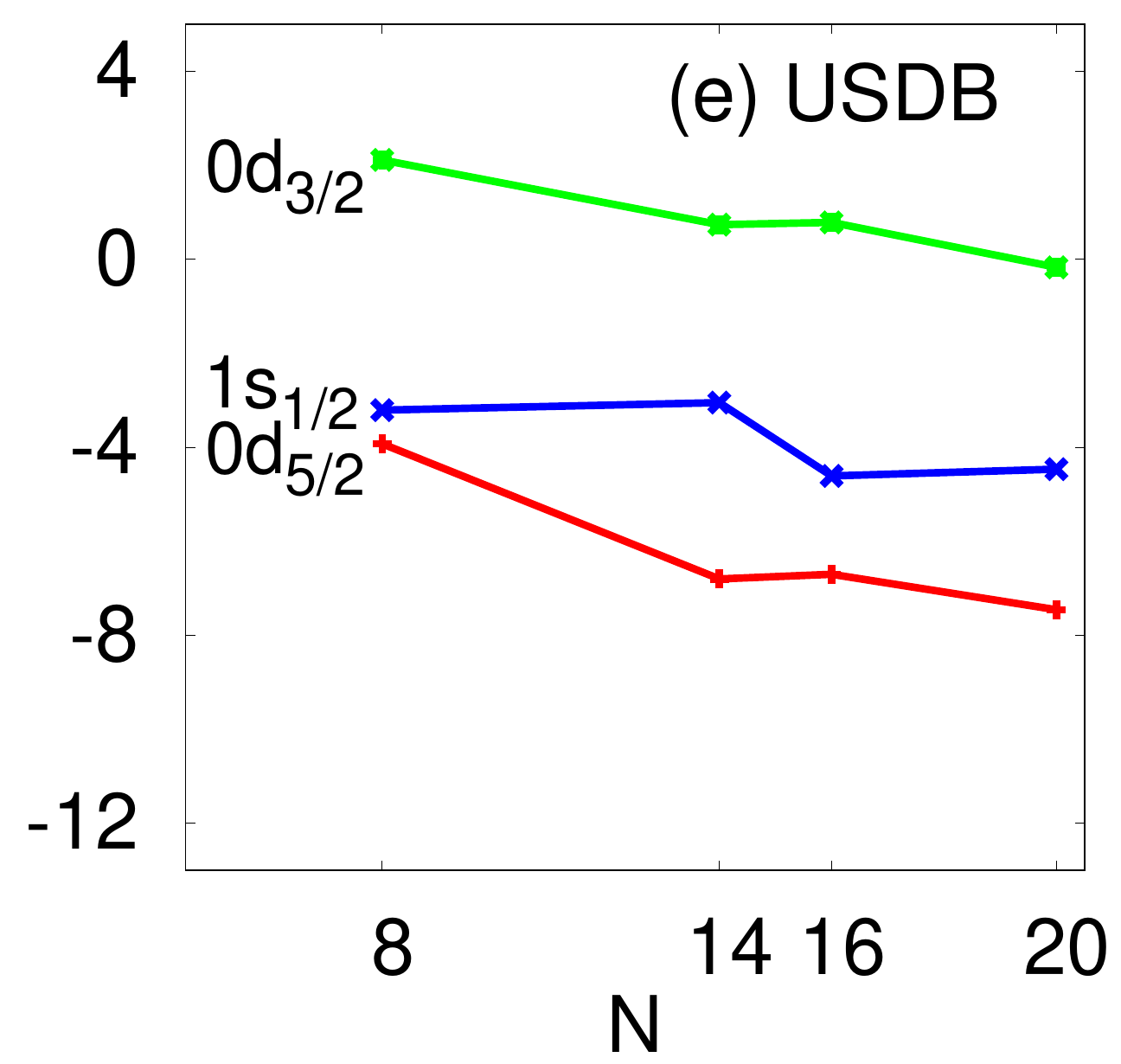} \\
  \includegraphics[height=.18\textwidth]{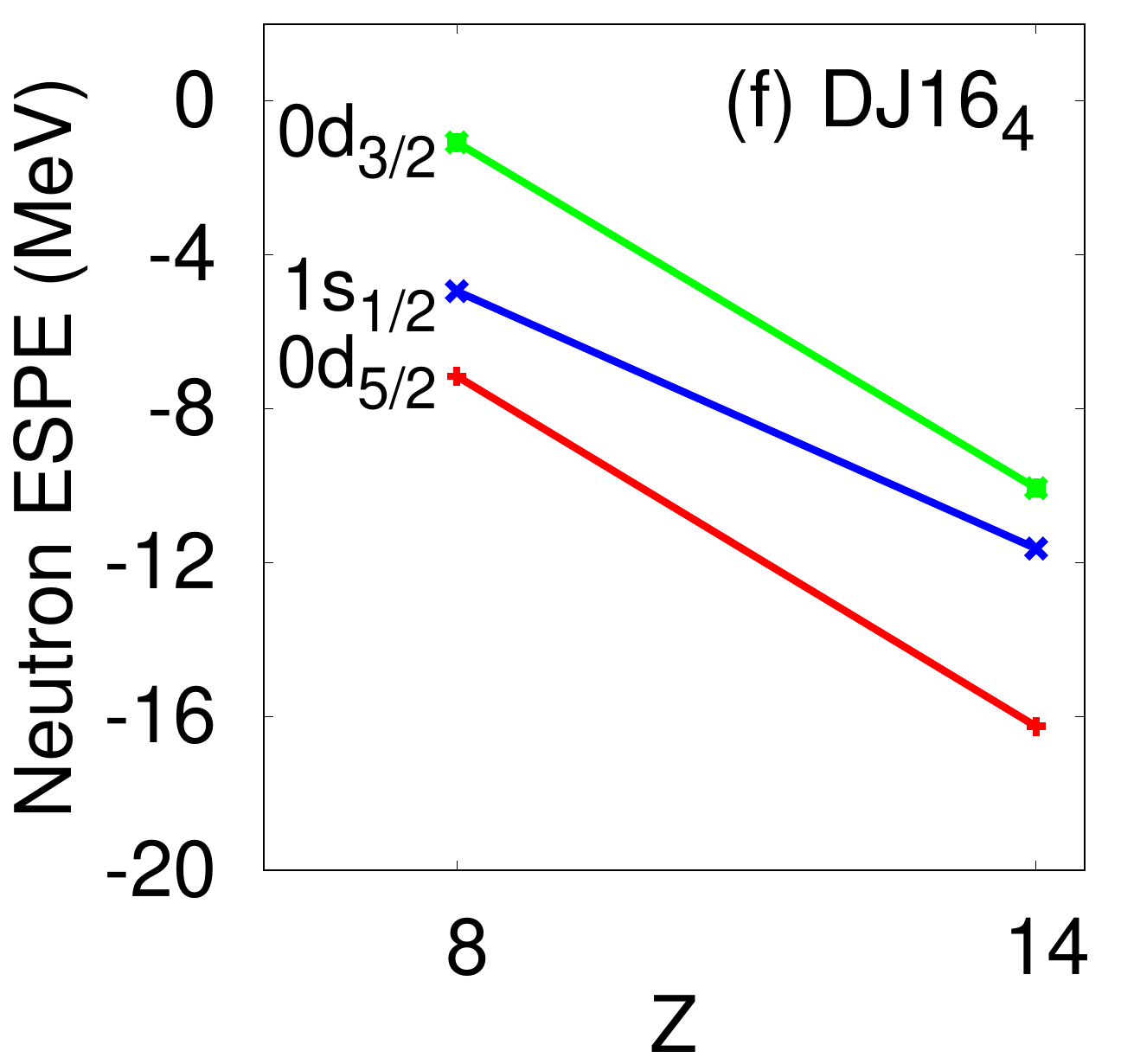}  
  \includegraphics[height=.18\textwidth]{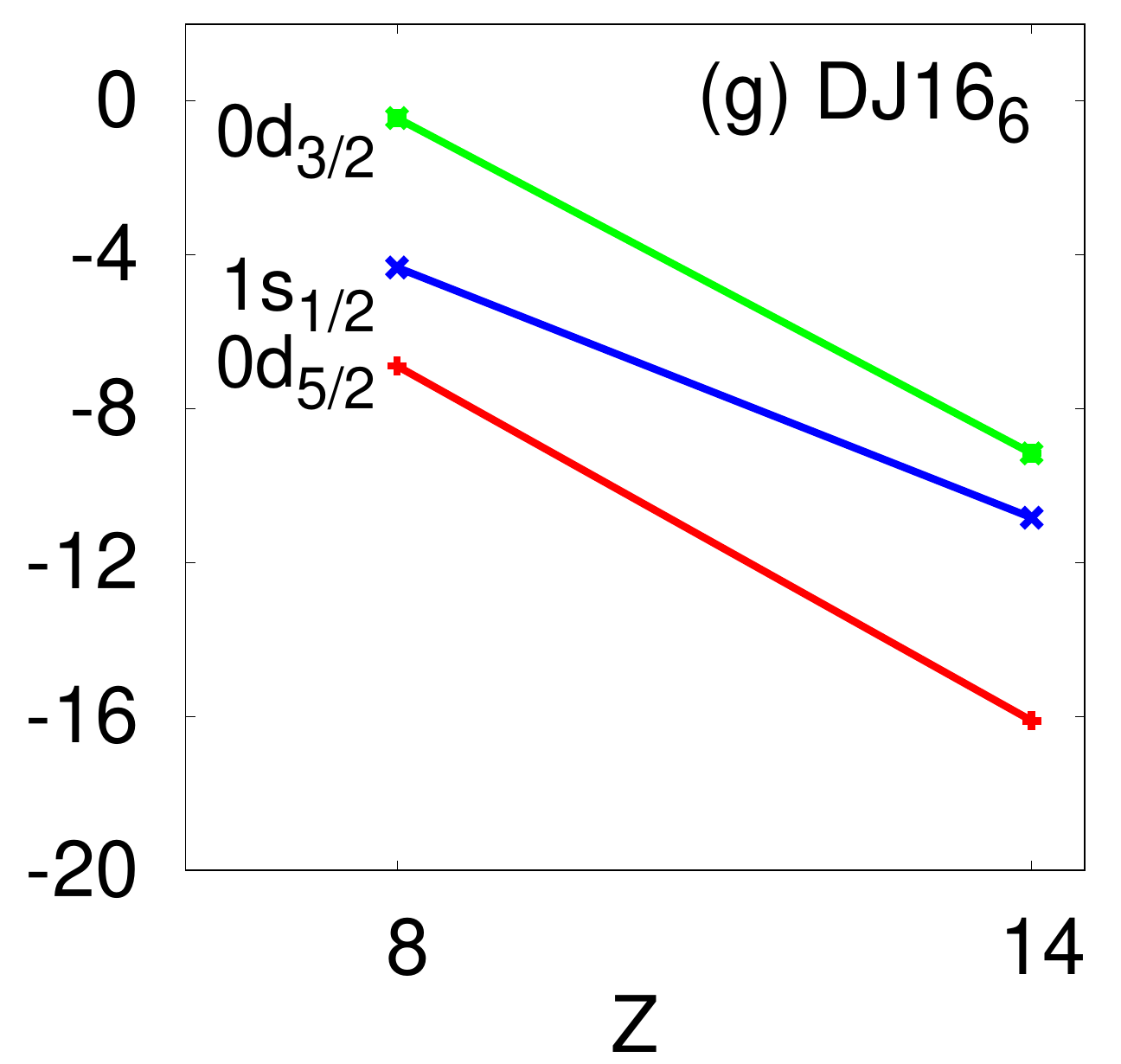}  
  \includegraphics[height=.18\textwidth]{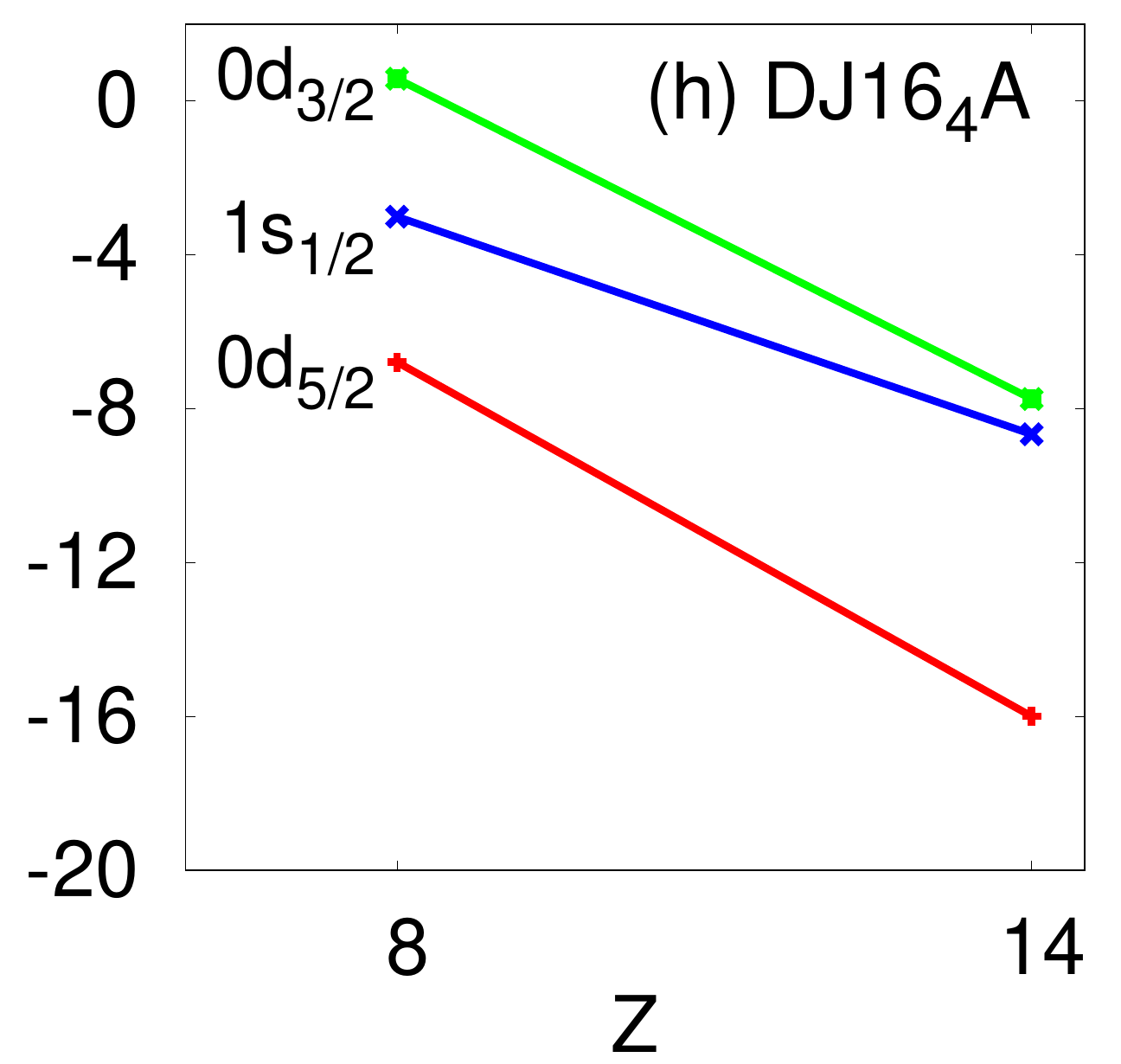}  
  \includegraphics[height=.18\textwidth]{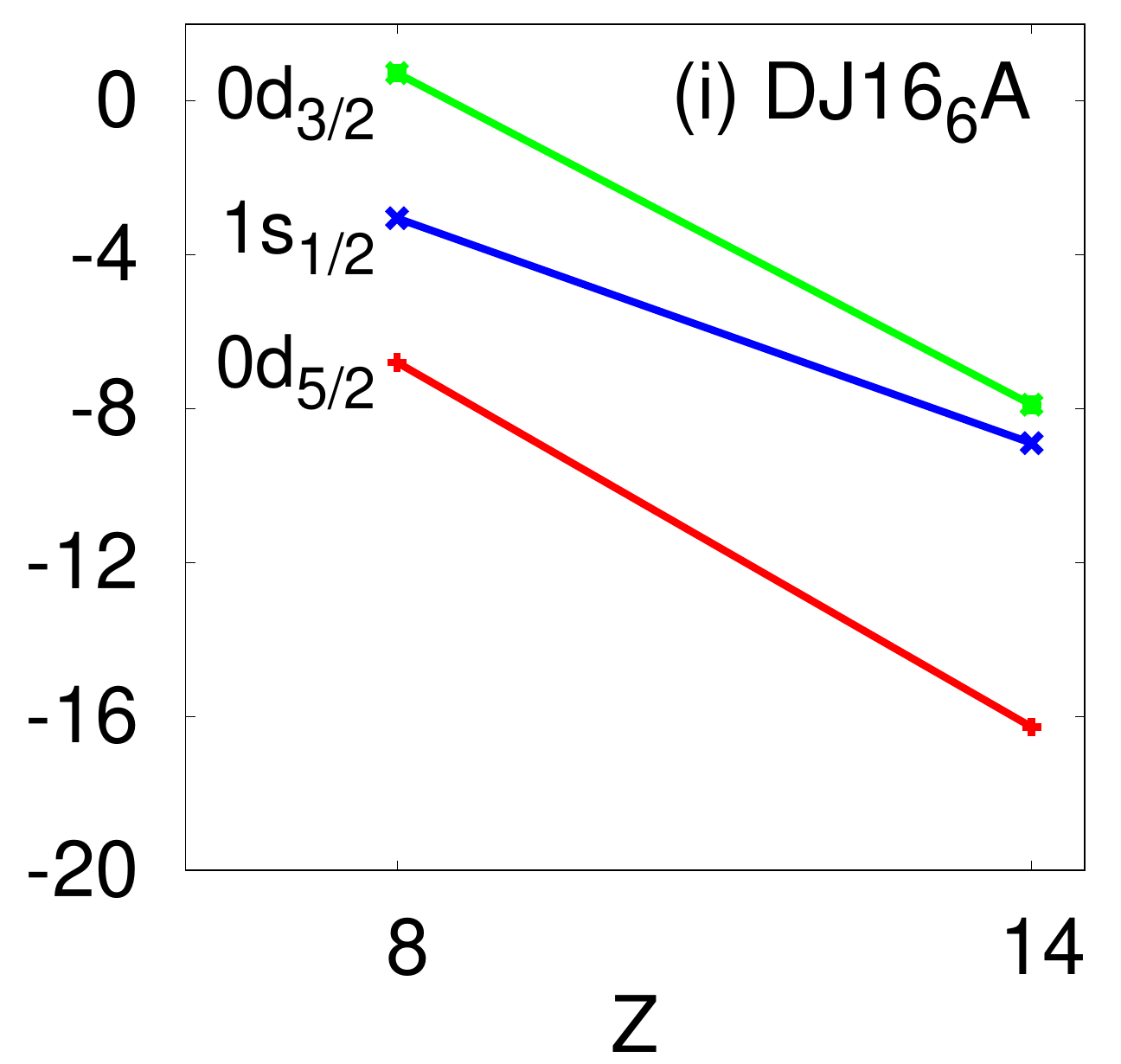}
  \includegraphics[height=.18\textwidth]{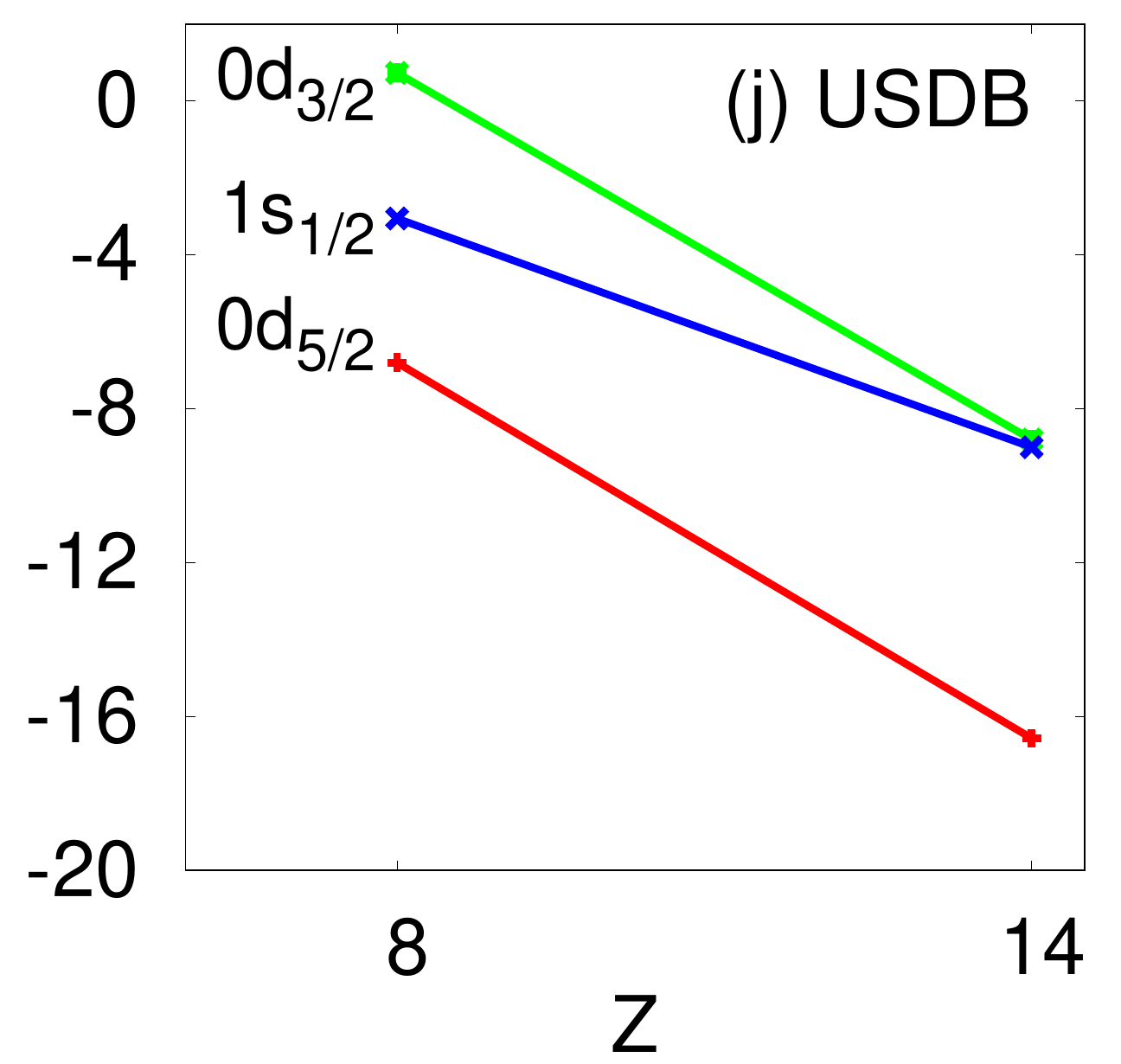} 
  \caption{\label{fig:ESPEs} (Color online) Variation of neutron ESPEs 
(a-e) in O isotopes with neutron number $N$ (upper row) and
(f-j) in $N=14$ isotones (lower row) calculated 
using the microscopic effective interactions 
obtained from Daejeon16 and the phenomenological one (USDB).
}
\vspace{-1.5ex}
\end{figure*}

\section{Monopole properties}

We initiate our analysis of the quality of our derived valence-space interactions from a study of their monopole component.
The monopole part~\cite{BaFr64} of the valence-space shell-model Hamiltonian plays an important role for \mbox{spectroscopic} properties
since it encapsulates the robust evolution of the spherical nuclear mean field 
as a function of valence nucleons~\cite{PoZu81,ZuDu95}.
A very useful insight is provided by the so-called {\it effective single-particle energies (ESPEs)}~\cite{OtHo01,Kumar2019}.
Assuming a normal filling of the single-particle orbitals as could be obtained from a pure monopole Hamiltonian,
we evaluate ESPEs for a closed sub-shell nucleus ($A$)
with respect to a reference nucleus ($A_r$) according to the expression:
\begin{equation}
\label{ESPE}
\tilde \varepsilon^{\rho }_k (A) = \varepsilon^{\rho }_k (A_r) + 
\sum\limits_{k', \rho' }V^{\rho \rho '}_{kk'}(A) \, n^{\rho '}_{k'} \, .
\end{equation}
Here, $k$ $(k')$ refer to
a complete set of quantum numbers of a harmonic oscillator orbital, e.\:g., $k\equiv (n_k l_k j_k)$,
$k'$ runs over valence space orbitals beyond $A_r$ and $n^{\rho '}_{k'}$ is the occupation
number of the orbital $k'$ for nucleons of the type $\rho '$. The quantity
$V^{\rho \rho' }_{kk'}$ are centroids of the two-body interaction,
\begin{equation}
\label{centroid}
V^{\rho \rho' }_{kk'} = \frac{ \sum\limits_J 
\langle k^{\phantom{'}}_{\rho } k'_{\rho'} |V| k^{\phantom{'}}_{\rho } k'_{\rho'}\rangle^{\phantom{'}}_J (2J+1)}
{ \sum\limits_J (2J+1)}  ,
\end{equation}
where $\rho $, $\rho' $ denote protons ($\pi $) or neutrons ($\nu $) and  
the total angular momentum of a two-body state $J$ 
runs over all values allowed by the Pauli principle.
{
Since the TBMEs $\langle k^{\phantom{'}}_{\rho } k'_{\rho'} |V| k^{\phantom{'}}_{\rho } k'_{\rho'}\rangle^{\phantom{'}}_J$
do not depend on the total angular momentum projection 
$M$, 
we skip it 
in the notation.}
The $T=1$ and $T=0$ centroids ($V^{T=0,1}_{kk'}$) can be obtained via the same equation, 
but using either $T=1$ or $T=0$ TBMEs in the summation.

The monopole part of the Hamiltonian describes a {\it spherical nuclear mean field}, 
which plays a lead role in the filling of orbitals and establishing (sub)shell gaps.
Its single-particle states, or ESPEs, provide an important ingredient for the arrangement
of shells and the interplay between spherical and deformed configurations in nuclei.
The higher multipole part of the interaction provides the so-called {\it correlation energy} 
for particle-hole excitations across the shell gap.
Large shell gaps 
are a prerequisite for magic numbers. 
A reduction of the spherical shell gaps may lead to 
a deformed ground state, if the correlation energy of a given excited (intruder) configuration 
is large enough to overcome the naive cost in energy for producing the excited configuration.

It has been recognized~\cite{PoZu81} that the main defect of the traditional microscopic
effective interactions derived from two-body $NN$ potentials is
an unsatisfactory monopole term, resulting in the absence of sufficiently large sub-shell gaps and
providing overbinding of systems beyond traditional closed-shell nuclei.
This in turn leads to the lack of sphericity in closed sub-shell nuclei and failures in the description
of open-shell nuclei.
Given the importance of the monopole component of an effective interaction, 
we now provide a detailed analysis of the ESPEs of the valence-space interactions under consideration here.

To illustrate the properties of the $T=1$ centroids of the microscopic effective interactions
in comparison with those of the phenomenological USDB interaction,
we show in Fig.~\ref{fig:ESPEs}(a-e) the neutron ESPEs in O isotopes with closed neutron sub-shells 
($^{16}$O, $^{22}$O, $^{24}$O and $^{28}$O)  as a function of the neutron number.
We  assume a normal filling of the orbitals with the order
determined by single-particle energies with respect to the core nucleus
(a Hartree--Fock approximation).
The ESPEs are thus represented by straight segments, whose slopes are given by the corresponding centroids
of the two-body interaction, as seen from Eq.~(\ref{ESPE}).
 
In all cases, the starting point is the $A$-independent
single-particle energies from the USDB Hamiltonian, quoted in the previous section.
The TBMEs of USDB and of the microscopic effective interactions are scaled as stated above.

While the neutrons fill $0d_{5/2}$, $1s_{1/2}$ and $0d_{3/2}$ orbitals 
(from $^{16}$O to $^{22}$O, then to $^{24}$O and on to $^{28}$O), the ESPEs acquire shifts 
due to additional increments provided by the respective centroids.
Two important features of the phenomenological USDB interaction
are easily seen. First, there is the appearance of a relatively large $N=14$ sub-shell closure in $^{22}$O. 
Similarly, there is a clear 
$N=16$ shell gap in $^{24}$O, resulting in the corresponding magic structure of that nucleus.

As seen from Fig.~\ref{fig:ESPEs}(a), the neutron ESPEs obtained from DJ16$_4$ do
show the $N=14$ subshell in $^{22}$O, 
but its $d_{5/2}$-$s_{1/2}$ gap is not as large as that from USDB (Fig.~\ref{fig:ESPEs}(e)).
This feature was discussed in detail in our previous work~\cite{SmBa2019}, where DJ16$_4$ was selected as a potential
providing the maximum $N=14$ subshell gap among a selected set of  microscopic interactions.  
At the same time, we notice that the new interaction DJ16$_6$, obtained in the present work from Daejeon16 at $N_{\rm max}=6$,
is characterized by a larger $N=14$ subshell gap (see Fig.~\ref{fig:ESPEs}(b)).
The numerical values of all the $N=14$ shell gaps in the O isotopes as obtained from various interactions shown in Fig.~\ref{fig:ESPEs}
are summarized in Table~\ref{tab:O_shell_gaps}.

\begin{table}[!b]
\vspace{-2.5ex}
\centering
\caption{Evolution of the $N=14$ sub-shell gap and neutron $d_{5/2}{-}d_{3/2}$ spin-orbit  splitting in the O isotopes 
from $^{16}$O to $^{28}$O as obtained from the ESPEs of different Hamiltonians.}\vspace{1ex}
\label{tab:O_shell_gaps}
\begin{ruledtabular}
\begin{tabular}{l|r|r|r|r|r|r|r|r} 
& \multicolumn{4}{c|}{Gap $\nu (s_{1/2}{-}d_{5/2})$} & \multicolumn{4}{c}{Gap $\nu (d_{3/2}{-}d_{5/2})$} \\[1mm]
\hline
& \multicolumn{4}{c|}{MeV} & \multicolumn{4}{c}{MeV} \\[1mm]
\hline
 & $^{16}$O & $^{22}$O & $^{24}$O & $^{28}$O &  $^{16}$O & $^{22}$O & $^{24}$O & $^{28}$O \\[1mm]  
\hline
DJ16$_4$        & 0.72 & 2.21 &  0.88 & 1.98 & 6.04 & 6.08 & 6.18  & 7.11 \\
DJ16$_6$        & 0.72 & 2.56 &  1.13 & 1.87 & 6.04 & 6.45 & 6.36  & 7.18 \\
DJ16$_4$A       & 0.72 & 3.78 &  1.80 & 2.51 & 6.04 & 7.37 & 7.15  & 7.02 \\
DJ16$_6$A       & 0.72 & 3.75 &  1.89 & 2.20 & 6.04 & 7.52 & 7.16  & 6.97 \\
USDB            & 0.72 & 3.75 &  2.09 & 2.99 & 6.04 & 7.53 & 7.49  & 7.28 \\
\end{tabular}
\end{ruledtabular}
\end{table}

\begin{table*}[t!]
\vspace{-2ex}
\centering
\caption{Evolution of the neutron sub-shell shell gaps in $N=14$ isotones from $^{22}$O to $^{28}$Si as obtained from
the ESPEs of different Hamiltonians.}
\label{tab:N14_shell_gaps}\vspace{1ex}
\begin{ruledtabular}
\begin{tabular}{l|r|r|r|r|r|r|r|r|r} 
& \multicolumn{3}{c|}{Gap $\nu (s_{1/2}{-}d_{5/2})$} & \multicolumn{3}{c|}{Gap $\nu (d_{3/2}{-}d_{5/2})$} 
& \multicolumn{3}{c}{Gap $\nu (d_{3/2}{-}s_{1/2})$}  \\
\hline
& \multicolumn{3}{c|}{MeV} & \multicolumn{3}{c|}{MeV} & \multicolumn{3}{c}{MeV} \\[1mm]
\hline
& $^{22}$O & $^{28}$Si & Diff & $^{22}$O &  $^{28}$Si & Diff & $^{22}$O & $^{28}$Si & Diff \\[1mm]  
\hline
DJ16$_4$        & 2.21 & 4.62 & {\bf 2.41} & 6.08 & 6.19 & {\bf 0.11}       & 3.87 & 1.57    & $\mathbf{-2.30}$ \\
DJ16$_6$        & 2.56 & 5.28 & {\bf 2.72} & 6.45 & 6.95 & {\bf 0.50}       & 3.88 & 1.67    & $\mathbf{-2.21}$ \\
DJ16$_4$A       & 3.78 & 7.33 & {\bf 3.55} & 7.37 & 8.25 & {\bf 0.88}       & 3.59 & 0.92    & $\mathbf{-2.67}$ \\
DJ16$_6$A       & 3.75 & 7.38 & {\bf 3.63} & 7.52 & 8.24 & {\bf 0.72}       & 3.77 & 0.86    & $\mathbf{-2.91}$ \\
USDB            & 3.75 & 7.57 & {\bf 3.82} & 7.53 & 7.77 & {\bf 0.23}       & 3.78 & 0.20    & $\mathbf{-3.58}$  \\
\end{tabular}
\end{ruledtabular}
\end{table*}

The evolution of the $N=14$ shell gap is governed by the difference between
$V^{T=1}_{d_{5/2} d_{5/2}}$ and $V^{T=1}_{d_{5/2} s_{1/2}}$ centroids of the TBMEs.
The detailed spin-tensor structure of those centroids will be discussed below.

We also note that the spin-orbit splitting between $0d_{3/2}$ and $0d_{5/2}$ ESPEs
is somewhat smaller for DJ16$_4$ and DJ16$_6$ compared with USDB (see also Table~\ref{tab:O_shell_gaps}).
Specifically, at $N=14$ and $N=16$, this spin-orbit splitting for DJ16$_6$ is only about 15\% smaller than that provided by USDB.

The second visible difference between USDB and both DJ16$_4$ and DJ16$_6$ is
related to the magnitude of all slopes of ESPEs in O isotopes: 
these two DJ16's typically provide a larger negative increment of the centroids with $N$ than those of USDB
(see Fig.~\ref{fig:ESPEs}). 
Since the monopole Hamiltonian provides the major contribution to the nuclear binding, 
we can immediately conclude that DJ16$_4$ and DJ16$_6$ will result in the overbinding of O isotopes, which we will discuss later. 
An encouraging trend is seen in that the inclusion of additional correlations arising 
at  $N_{\rm max}=6$ improves the monopoles compared with USDB which will reduce the resulting overbinding of O isotopes.

In Ref.~\cite{Otsuka3N}, the behavior of the ESPEs obtained from microscopic effective interactions,
based on a $NN$ potential, was ascribed to the missing $3N$ forces (see Fig.~2 of that reference).
Indeed, the centroids of the microscopic interactions obtained on the basis of $NN$ plus $3N$ forces show a much better
agreement with the centroids of phenomenological interactions~\cite{StrPRL118,Otsuka3N,Holt12,Fukui18}.
Daejeon16 is obtained from phase-equivalent transformations of SRG-evolved chiral $NN$ potential at N3LO, 
which tends to incorporate effects of many-body forces.
Thus, it may be that increasing the model space, when progressing from DJ16$_4$ to DJ16$_6$, more completely incorporates the influence of those many-body forces effectively included in Daejeon16 and results in the  improved monopole behaviors.

The proton-neutron centroids can be analyzed using the neutron ESPEs in 
$N=14$ isotones from $^{22}$O to $^{28}$Si, i.\:e. when protons fill the $d_{5/2}$ orbital.
The corresponding plots are shown in Fig.~\ref{fig:ESPEs}(f-j).
The starting point of these calculations are ESPEs in $^{22}$O as obtained by different effective interactions
and shown in  Fig.~\ref{fig:ESPEs}(a-e).
The numerical values of the $N=14$ shell gaps from the monopole part of the interactions
are summarized in Table~\ref{tab:N14_shell_gaps}.
We notice that in $^{22}$O the $N=14$ subshell gaps given by DJ16$_4$ and DJ16$_6$, 
except for $d_{3/2}{-}d_{5/2}$, are somewhat smaller than the corresponding gaps from USDB.
However, let us concentrate on the evolution of the first subshell gap from $^{22}$O towards $^{28}$Si, 
produced by the USDB interaction which increases by 3.82~MeV.
DJ16$_4$ results in an increase of 2.41~MeV, while DJ16$_6$ produce an increase of 2.72~MeV.
This means that with DJ16$_6$ the difference between the corresponding centroids, 
$ V^{pn}_{d_{5/2} d_{5/2}}$ and $V^{pn}_{s_{1/2} d_{5/2}}$, trends closer to the USDB value.

From Table~\ref{tab:N14_shell_gaps} one observes that the spin-orbit splitting between neutron 
$0d_{3/2}$ and $0d_{5/2}$ states in $^{28}$Si stays about the same as in $^{22}$O for the both microscopic interactions,
as well as for USDB.

\begin{table*}[t!]
\vspace{-2ex}
\centering
\caption{Centroids (in MeV) of DJ16$_4$ and DJ16$_6$ and the changes (``Diff") which yield 
the centroids of DJ16$_4$A and DJ16$_6$A, respectively.}
\label{tab:centroids_mod}\vspace{1ex}
\begin{ruledtabular}
\begin{tabular}{l|r|r|r|r|r|r|} 
& \multicolumn{3}{c|}{$N_{\rm max}=4$} & \multicolumn{3}{c|}{$N_{\rm max}=6$}  \\
\hline
		& 		DJ16$_4$ & DJ16$_4$A & Diff 	& DJ16$_6$ &  DJ16$_6$A & Diff  \\[1mm]  
\hline
$V^{T=1}_{d_{5/2} d_{5/2}}$      & $-0.705$ & $-0.625$ & $\mathbf{ +0.080}$ &  $-0.647$ & $-0.627$ & $\mathbf{ +0.020}$ \\
$V^{T=1}_{d_{5/2} s_{1/2}}$      & $-0.335$ & $+0.015$ & $\mathbf{ +0.350}$ &  $-0.221$ & $+0.008$ & $\mathbf{ +0.230}$ \\
$V^{T=1}_{d_{5/2} d_{3/2}}$      & $-0.595$ & $-0.295$ & $\mathbf{ +0.300}$ &  $-0.479$ & $-0.269$ & $\mathbf{ +0.210}$  \\
$V^{T=1}_{d_{3/2} s_{1/2}}$      & $-0.282$ & $-0.082$ & $\mathbf{ +0.200}$ &  $-0.265$ & $-0.165$ & $\mathbf{ +0.100}$ \\
$V^{T=1}_{d_{3/2} d_{3/2}}$      & $-0.411$ & $-0.411$ &      -             &  $-0.300$ & $-0.400$ & $\mathbf{ -0.100}$  \\
$V^{T=1}_{s_{1/2} s_{1/2}}$      & $-2.017$ & $-2.017$ &      -             & $-1.910$ & $-1.910$  & - \\
\hline
$V^{T=0}_{d_{5/2} d_{5/2}}$      & $-2.745$ & $-2.825$ & $\mathbf{ -0.080}$ &  $-2.818$ & $-2.918$ & $\mathbf{ -0.100}$ \\
$V^{T=0}_{d_{5/2} s_{1/2}}$      & $-2.551$ & $-2.451$ & $\mathbf{ +0.100}$ &  $-2.577$ & $-2.527$ & $\mathbf{ +0.050}$ \\
$V^{T=0}_{d_{5/2} d_{3/2}}$      & $-3.213$ & $-3.213$ &       -            &  $-3.208$ & $-3.408$ & $\mathbf{ -0.200}$  \\
$V^{T=0}_{d_{3/2} s_{1/2}}$      & $-2.534$ & $-2.534$ &       -            &  $-2.760$ & $-2.760$ &  - \\
$V^{T=0}_{d_{3/2} d_{3/2}}$      & $-2.675$ & $-2.675$ &       -            &  $-2.655$ & $-2.655$ &  - \\
$V^{T=0}_{s_{1/2} s_{1/2}}$      & $-2.938$ & $-2.938$ &       -            &  $-2.851$ & $-2.851$ &  - \\
\end{tabular}
\end{ruledtabular}
\end{table*}

Although we have noted an encouraging trend in the monopole properties when moving from DJ16$_4$ to DJ16$_6$, 
we also note that the gaps and slopes governed by the neutron-neutron centroids
are still significantly different from the USDB benchmark.

\begin{figure*}[!t]
  \includegraphics[height=.22\textheight]{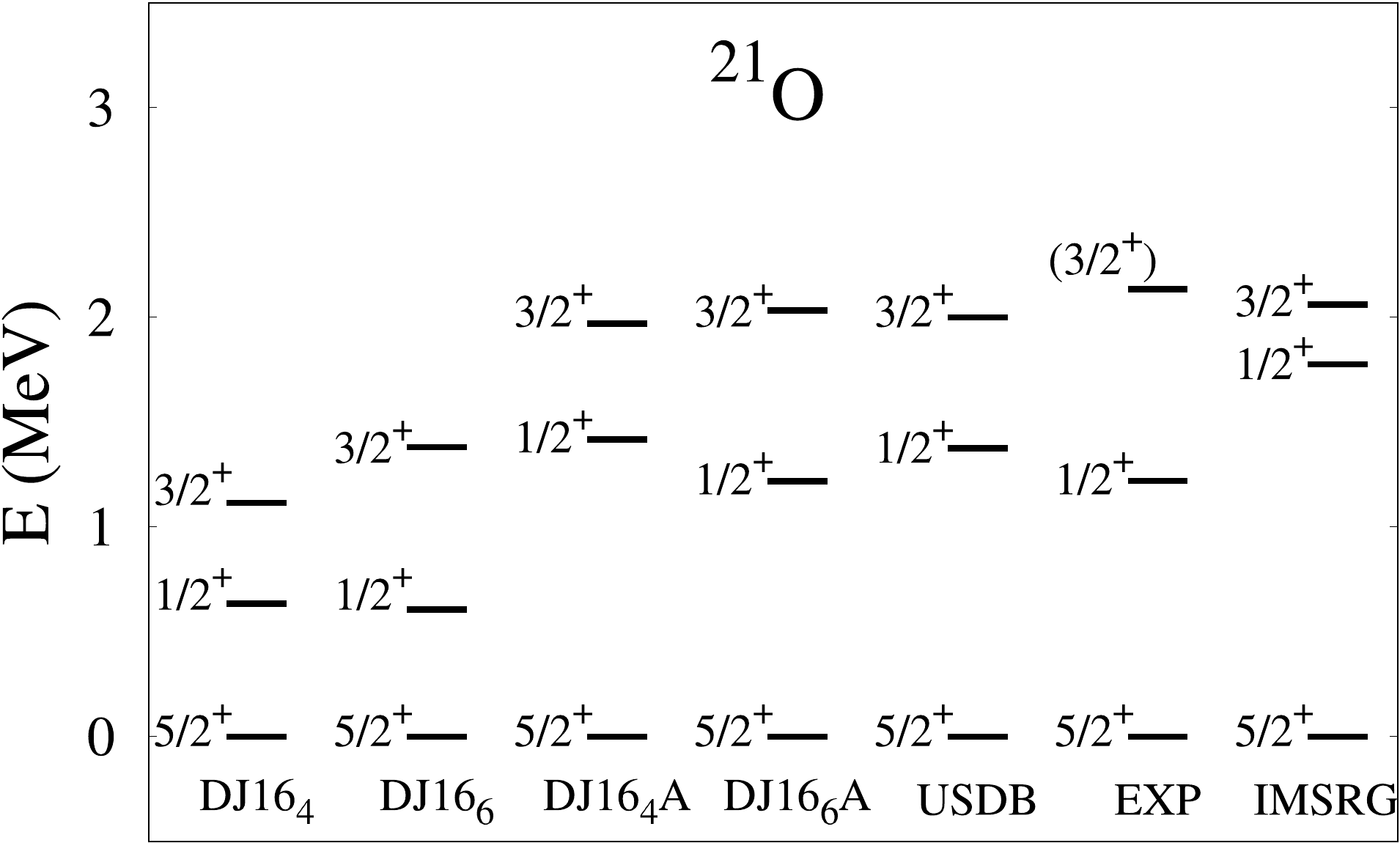} 
  \includegraphics[height=.22\textheight]{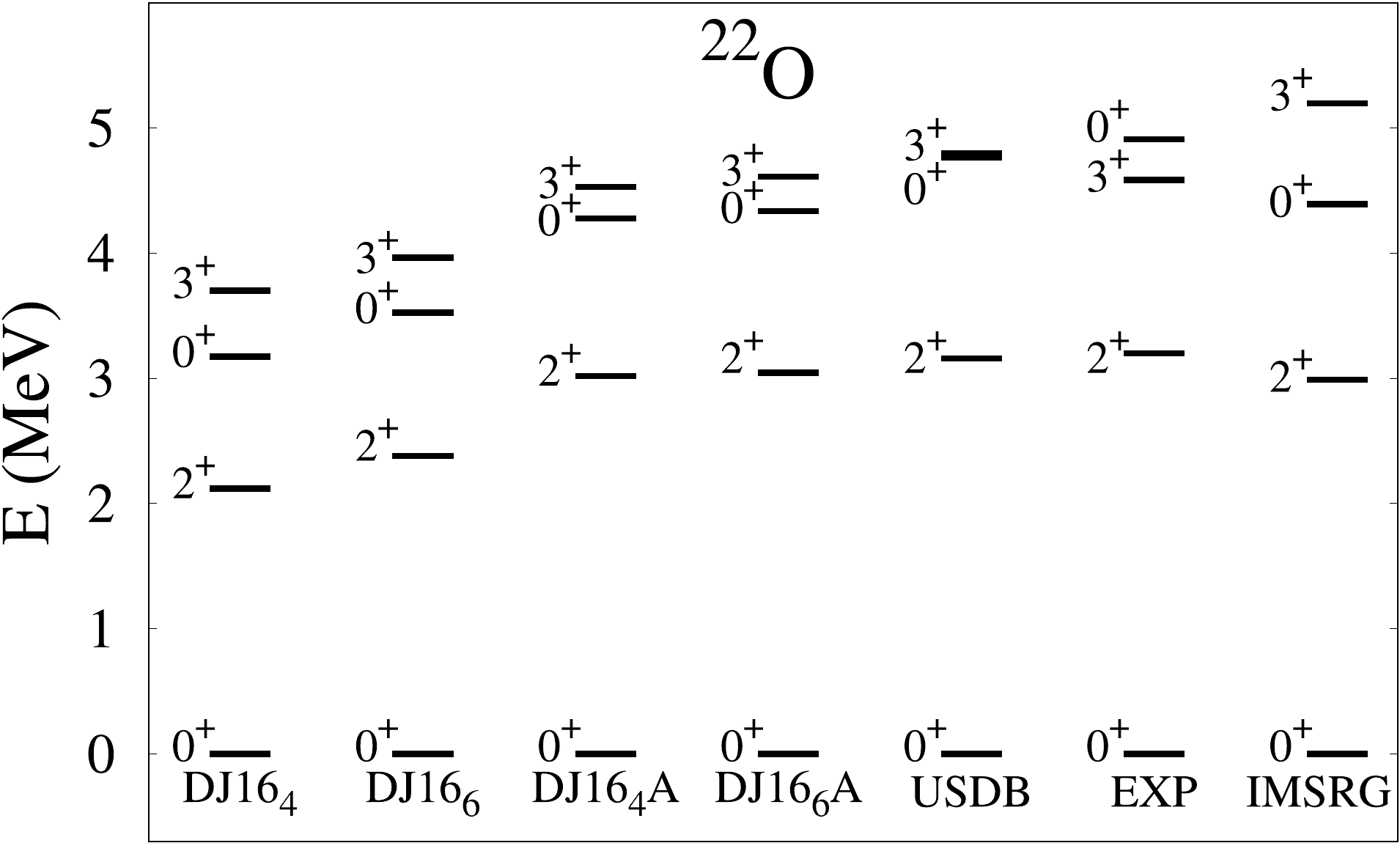} \\
  \includegraphics[height=.22\textheight]{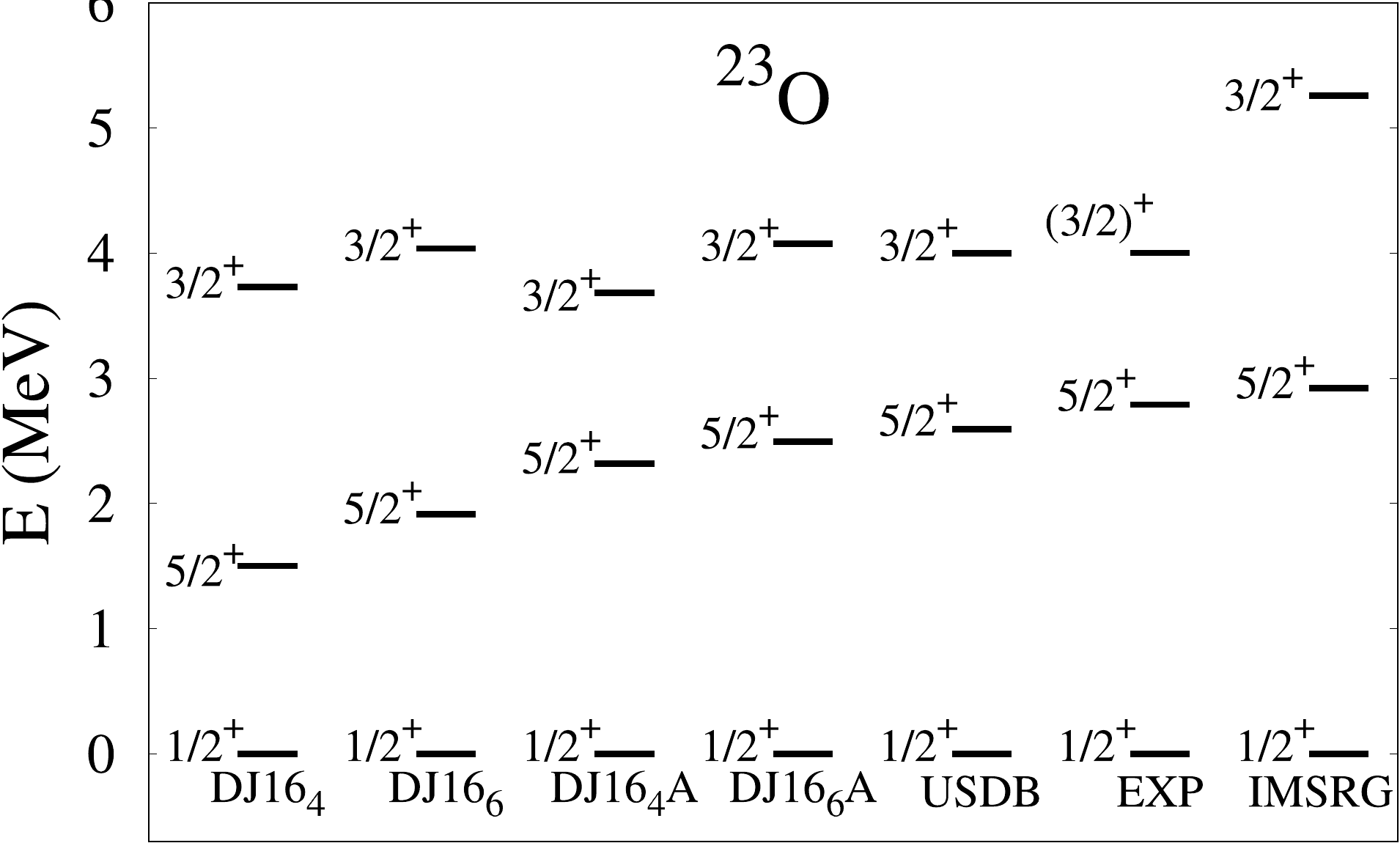} 
  \includegraphics[height=.22\textheight]{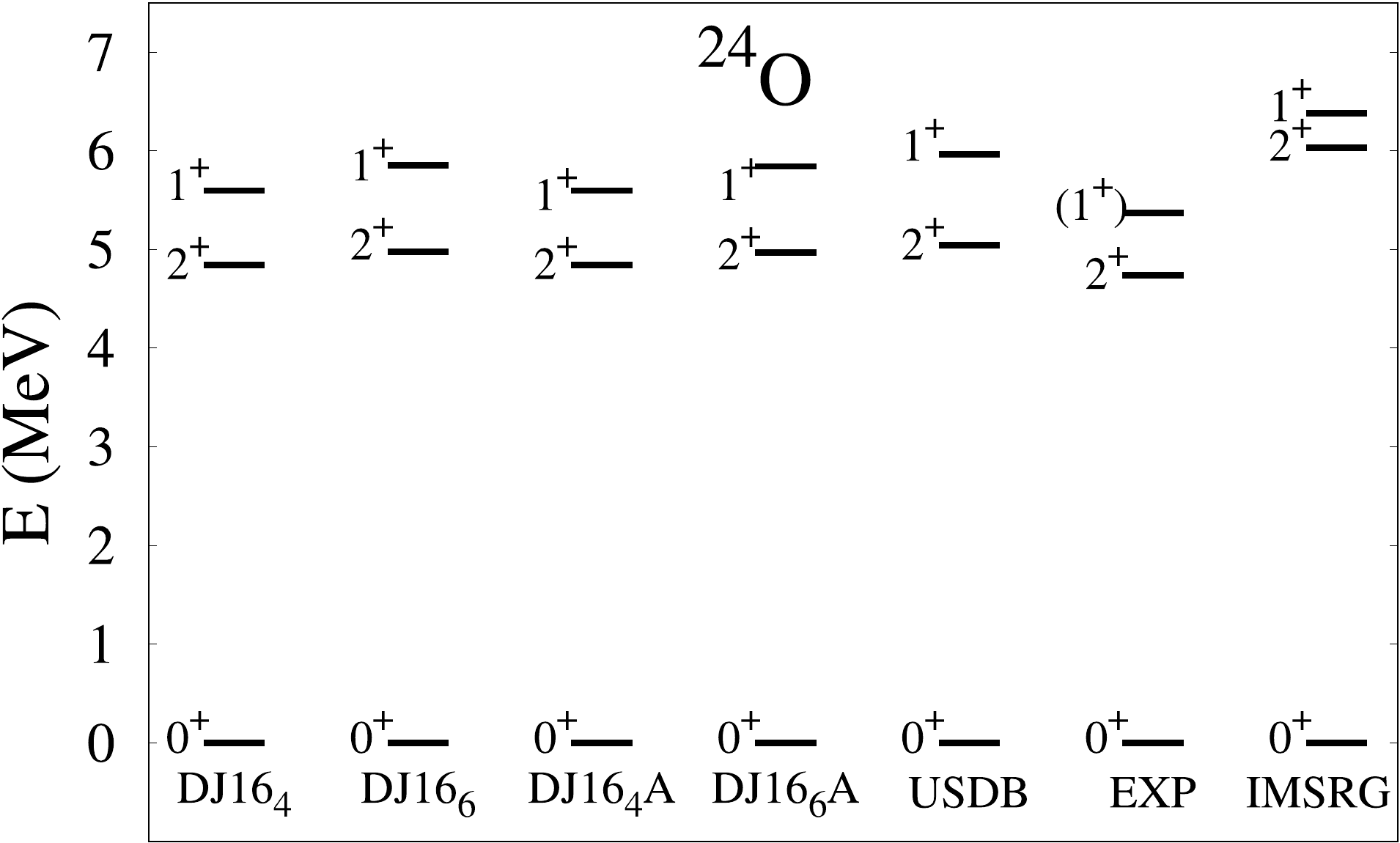} 
  \caption{\label{fig:O-spectra} Experimental low-energy spectrum of $^{21{-}24}$O in comparison with theoretical results, 
obtained from USDB and from the microscopic effective interactions.
The experimental data are from Ref.~\protect\cite{nndc}.
The results labeled ``IMSRG'' are from Ref.~\protect\cite{StrPRL118}.}
\vspace{-1ex}
\end{figure*}

\begin{figure}[!b]
  \includegraphics[width=\columnwidth]{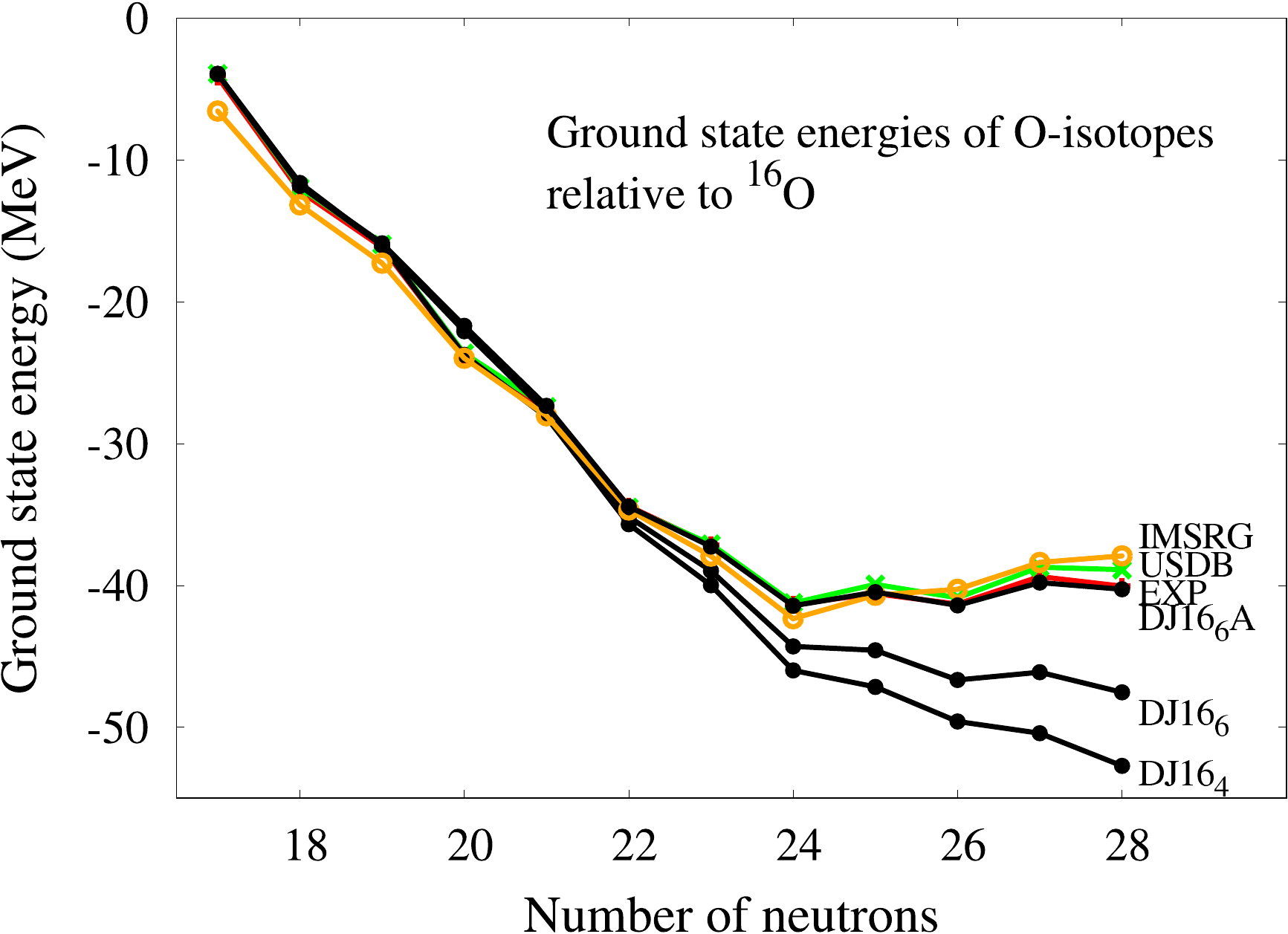} 
  \caption{\label{fig:BE_O} (Color online) Experimental ground state energies of O isotopes relative to the ground state energy of 
$^{16}$O, in comparison with theoretical results, obtained from USDB and from the microscopic effective interactions.
The experimental data (extrapolations included) are from AME2012~\protect\cite{AME2012}.
The results labeled ``IMSRG'' are from Ref.~\protect\cite{StrPRL118}.}
\vspace{-1ex}
\end{figure}

As in our previous study, with further guidance from USDB [Figs.~\ref{fig:ESPEs}(e) and (j)],
we now perform a few modifications of the DJ16$_6$ centroids to improve agreement with experiment.
Namely, we now modify the original DJ16$_6$ centroids
by introducing shifts that improve their alignment with USDB centroids as indicated in Table~\ref{tab:centroids_mod}. 
The monopole modified interaction is referred to as DJ16$_6$A.
In the same table, we show for comparison the modification performed
to the previously derived interaction, DJ16$_4$, to get DJ16$_4$A from Ref.~\cite{SmBa2019}.
First, we notice that the modifications are mainly of the same type: 
make the $T=1$ centroids of DJ16$_6$ more repulsive, which will tend to lessen 
the overbinding of the O isotopes. The most significant change is again required for
$V^{T=1}_{d_{5/2} s_{1/2}}$ to correct the $N=14$ subshell gap in $^{22}$O.
The $T=0$ centroids are little changed.
Second, we notice that modifications required for DJ16$_6$ are generally smaller than those applied to DJ16$_4$,
with a few exceptions. This is in line with the trend of ESPEs.
We distribute monopole modifications evenly among the TBMEs of different $J$. 
The TBMEs of the DJ16$_6$A interaction can be found in Table I of the Supplemental material.

The resulting neutron ESPEs from DJ16$_6$A are shown in Figs.~\ref{fig:ESPEs}(d) and (i).
As was intended with these modifications, the spherical mean-fields from DJ16$_4$A and DJ16$_6$A are close 
to the mean-field provided by USDB.
Therefore, the differences  between the spectroscopies emerging from modified Daejeon16 effective interactions 
(DJ16$_4$A, DJ16$_6$A) and the USDB interaction will be mainly related to differences in the other 
multipole terms of the effective interaction.


\section{Comparison with experiment}

\subsection{O isotopes}

We begin our discussion of spectroscopic properties with the spectra of the $^{21-24}$O 
isotopes as obtained from shell-model diagonalization with the microscopic interactions described above.
In addition, we include a comparison with results from the IMSRG approach from Ref.~\cite{StrPRL118}.
These theoretical results are shown in Fig.~\ref{fig:O-spectra} in comparison with the experimental spectrum and the spectrum from USDB.

Although the ${N=14}$ subshell gap from DJ16$_4$ and DJ16$_6$ is larger than that predicted by other microscopic interactions studied in Ref.~\cite{SmBa2019},
it is still not large enough to provide satisfactory spectra when comparing with experiment. 
This is manifested in low $1/2^+$ and $3/2^+$ states 
in  $^{21}$O and the low $2^+_1$ state in  $^{22}$O. 
Similarly, the lowest  $5/2^+$ state  in the spectrum of $^{23}$O is also lower than the experimental
counterpart (Fig.~\ref{fig:O-spectra}). 
These nuclei have one neutron hole or one particle beyond semi-magic $^{22}$O.
Monopole modifications included in DJ16$_4$A and DJ16$_6$A largely correct these shortcomings and provide closer agreement with experiment.

For comparison, we also show the IMSRG results obtained with the Hamiltonians from Refs.~\cite{Stroberg2016,StrPRL118}.
The corresponding spectra are in good agreement with experiment due, in large measure, 
to the satisfactory $T=1$ monopole component of the interaction as was discussed in Ref.~\cite{SmBa2019}. 
With IMSRG, a few low-lying states are positioned slightly higher in energy than their experimental counterparts.

The ground state energies of the O isotopes relative to the ground energy of $^{16}$O are shown in Fig.~\ref{fig:BE_O}.
Here, one observes that DJ16$_6$ reduces the overbinding of neutron-rich O isotopes, compared with DJ16$_4$.
Furthermore, DJ16$_6$A produces an excellent description of experimental binding energies.
The root-mean-square (rms) deviations in experiment versus theory binding energies  
are summarized in Table~\ref{tab:rms}.
Indeed, we notice that among the theoretical cases tabulated,
DJ16$_6$A provides the smallest rms deviation for the binding energy of the O isotopes (column 2 of Table~\ref{tab:rms}).

\begin{table*}[!t]
\centering
\caption{Root-mean-square deviations (in keV) between experimental and theoretical binding energies of O isotopes 
and between experimental and theoretical excitation energies of low-lying states of a few $sd$-shell nuclei
shown in Figs.~\ref{fig:O-spectra}--\ref{fig:Mg24} as obtained from different interactions.}
\label{tab:rms}\vspace{1ex}
\begin{ruledtabular}
\begin{tabular}{l|r|r|r|r|r|r|r} 
Interaction & BE(O) & $^{21-24}$O & $^{19,21,23,25,27}$F and $^{39}$K & $^{22}$Na & $^{28}$Si,$^{32}$S  & $^{24}$Mg  & $^{25}$Mg \\
\hline
DJ16$_4$        &  5960 &  931 &  700 &  429 & 1146 & 1096 & 1314 \\
DJ16$_6$        &  3671 &  741 &  649 &  336 & 1015 &  831 & 1149 \\
DJ16$_4$A       &   449 &  274 &  285 &  328 &  891 &  806 &  925 \\
DJ16$_6$A       &   235 &  248 &  308 &  300 &  795 &  781 &  790 \\
DJ16$_6$B       &   235 &  248 &  388 &  197 &  634 &  696 &  419 \\
USDB            &   467 &  251 &  437 &  155 &  234 &  313 &   75 \\
IMSRG           &  1177 &  728 &      &  445 & 1497 &  & \\
\end{tabular}
\end{ruledtabular}
\end{table*}


\subsection{\boldmath Odd-$A$ F isotopes and $^{39}$K}

\begin{figure*}[t!]
  \includegraphics[height=.22\textheight]{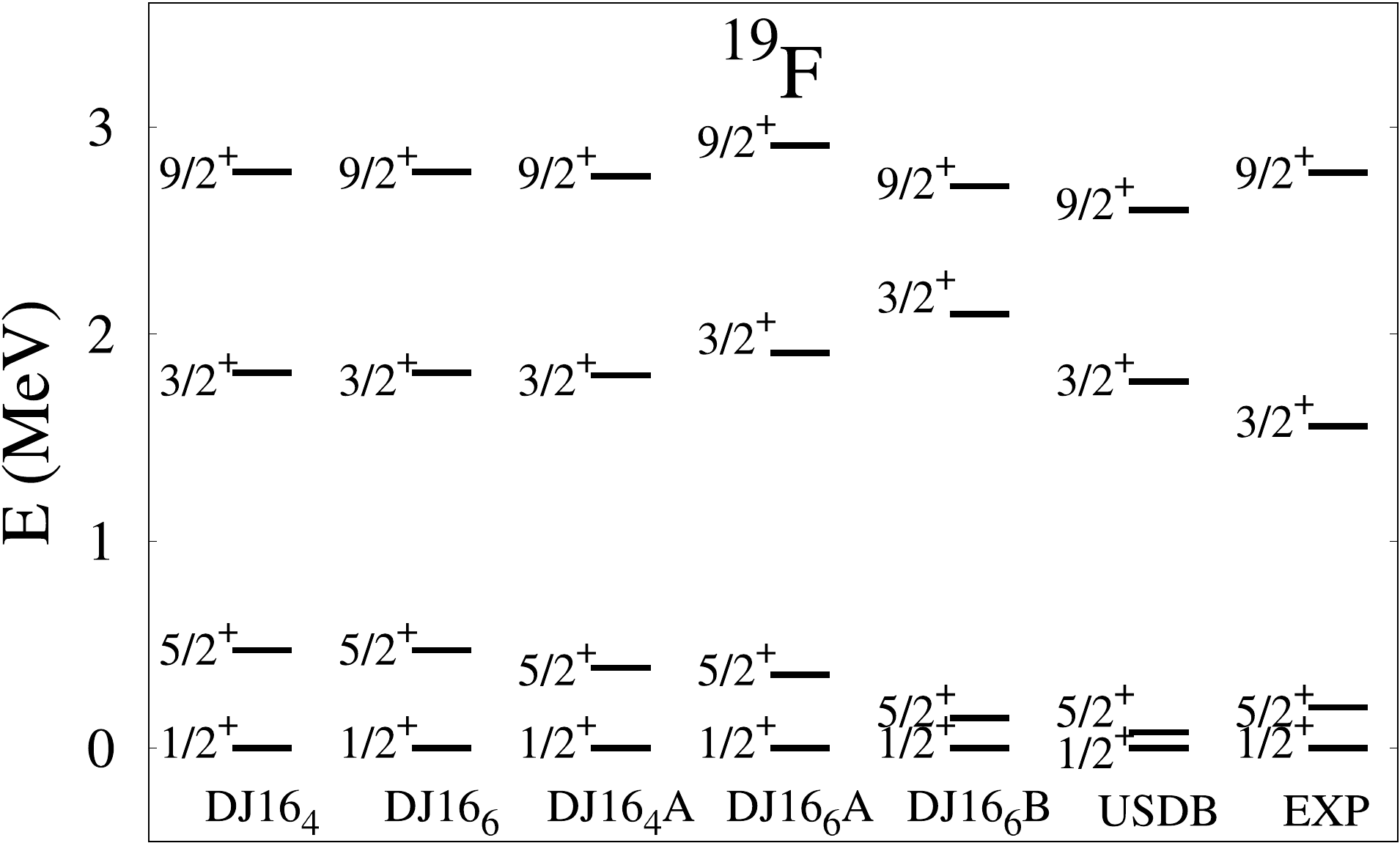}
  \includegraphics[height=.22\textheight]{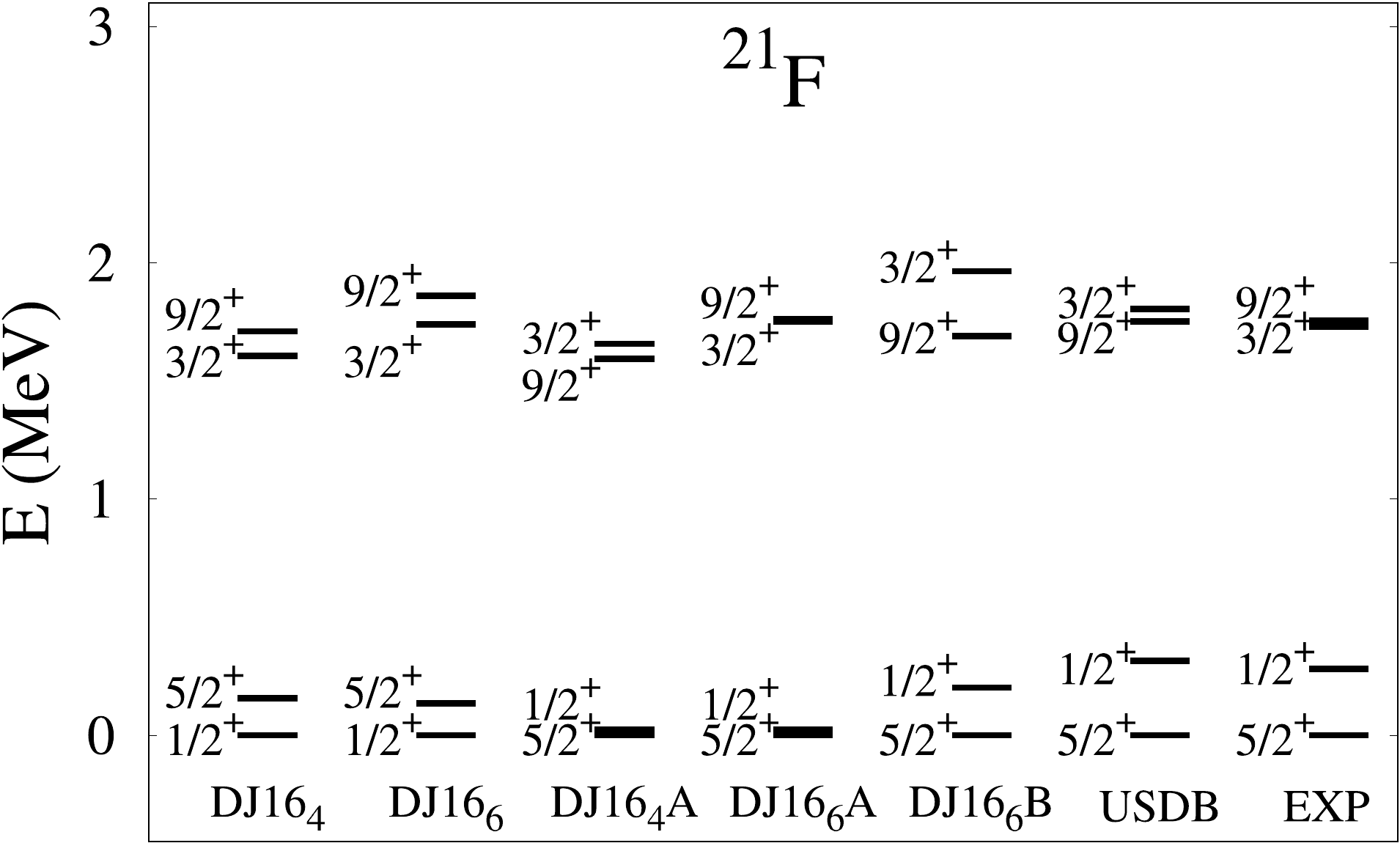} \\
  \includegraphics[height=.22\textheight]{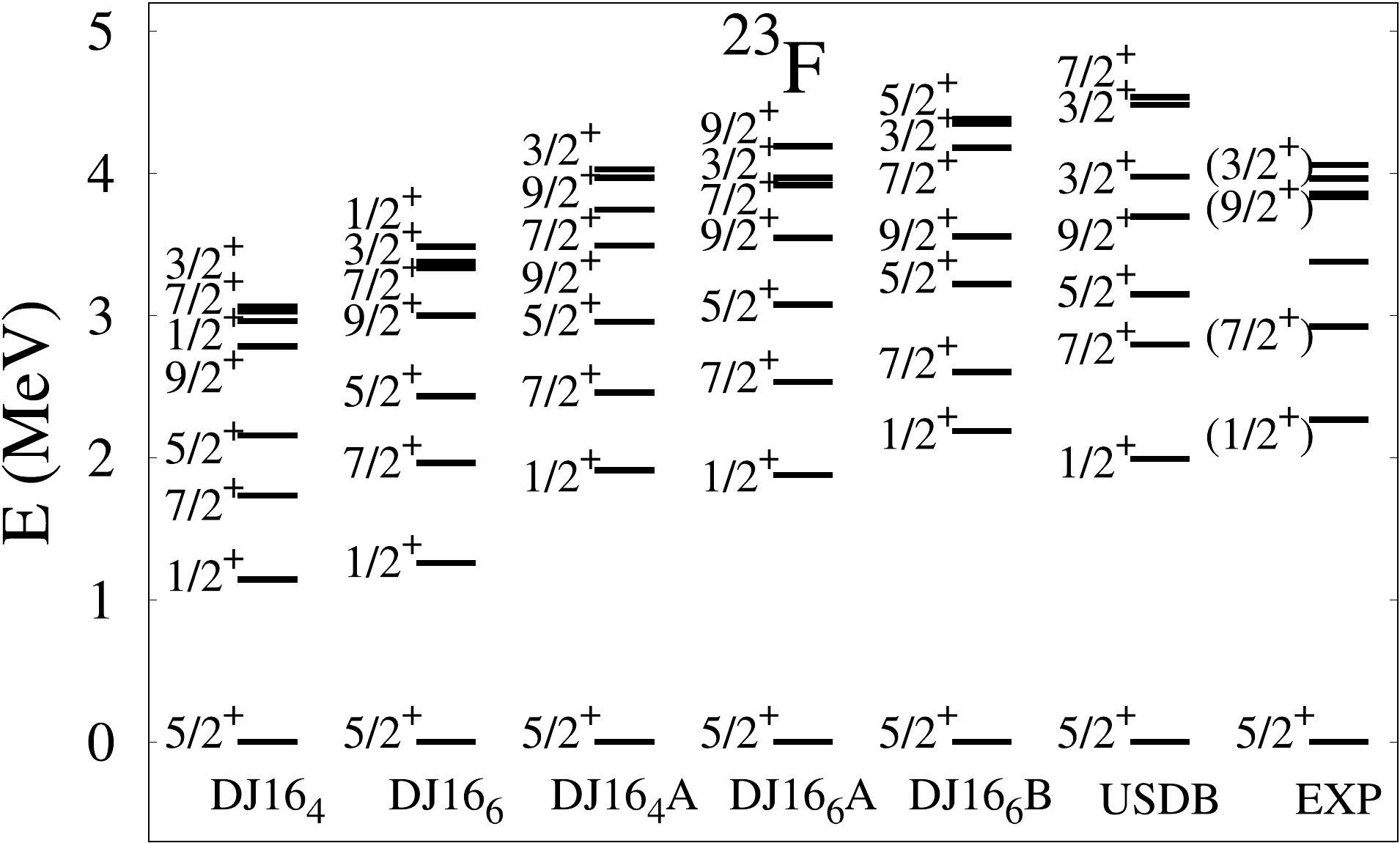}
  \includegraphics[height=.22\textheight]{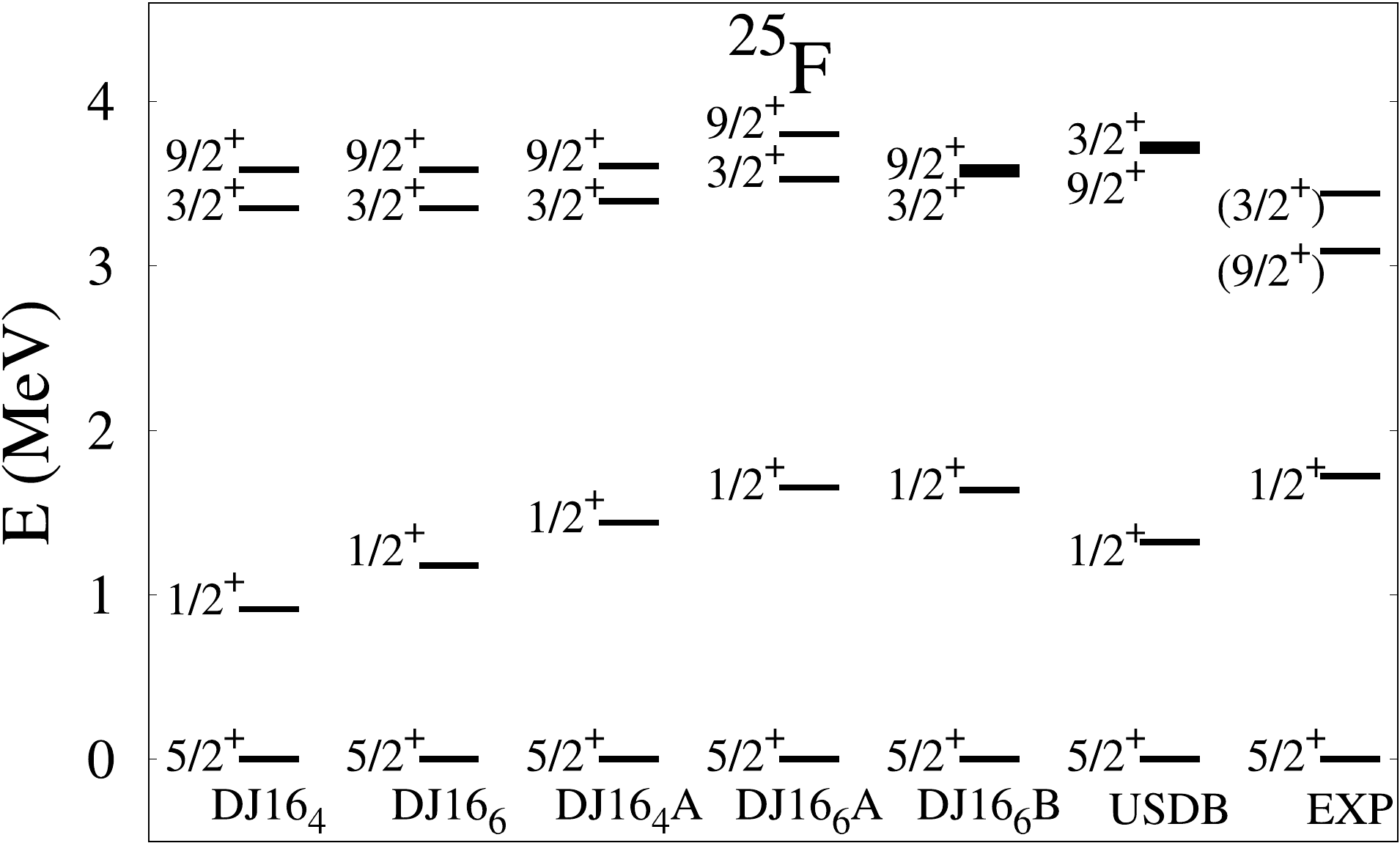} \\
  \includegraphics[height=.22\textheight]{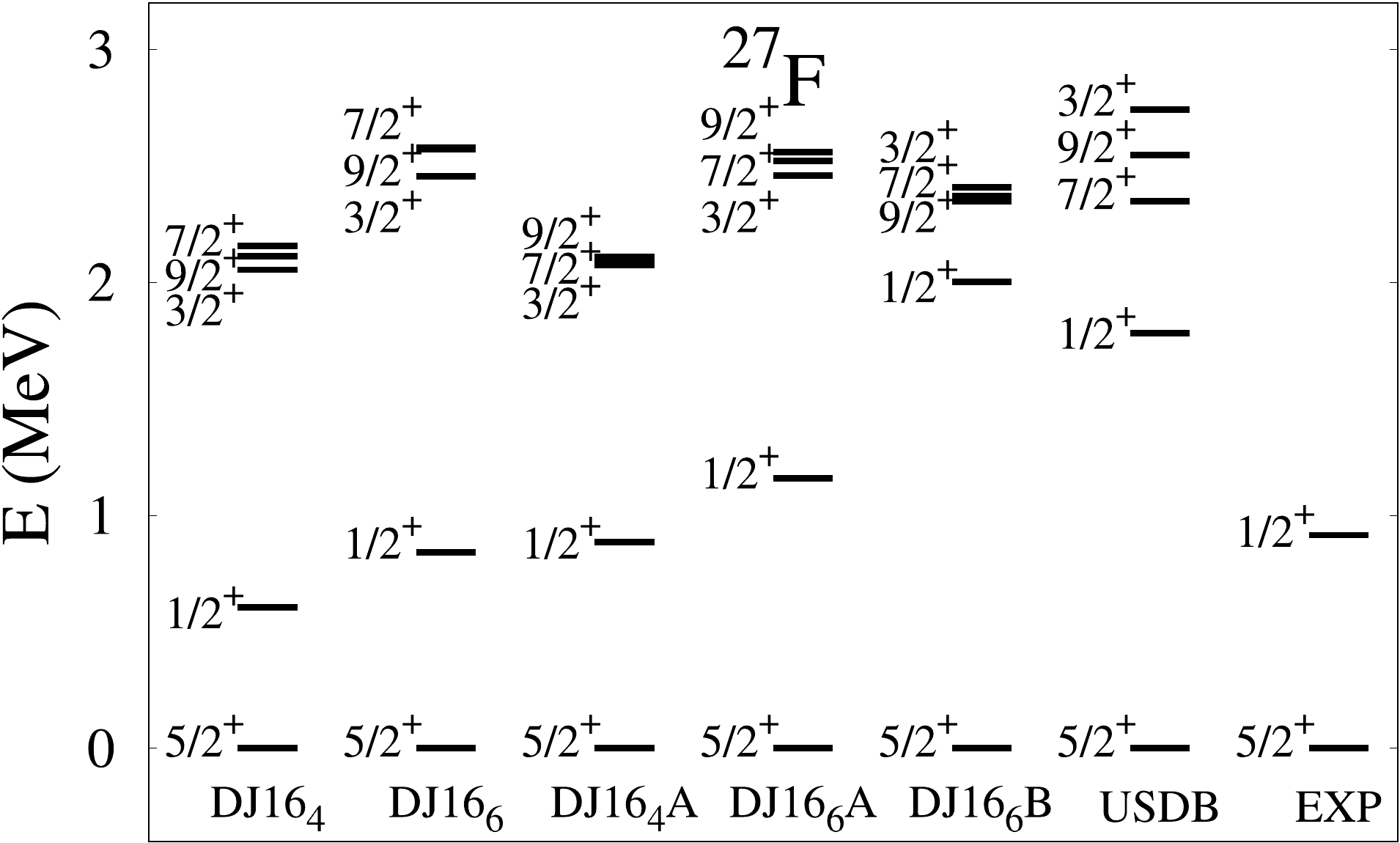}
  \includegraphics[height=.22\textheight]{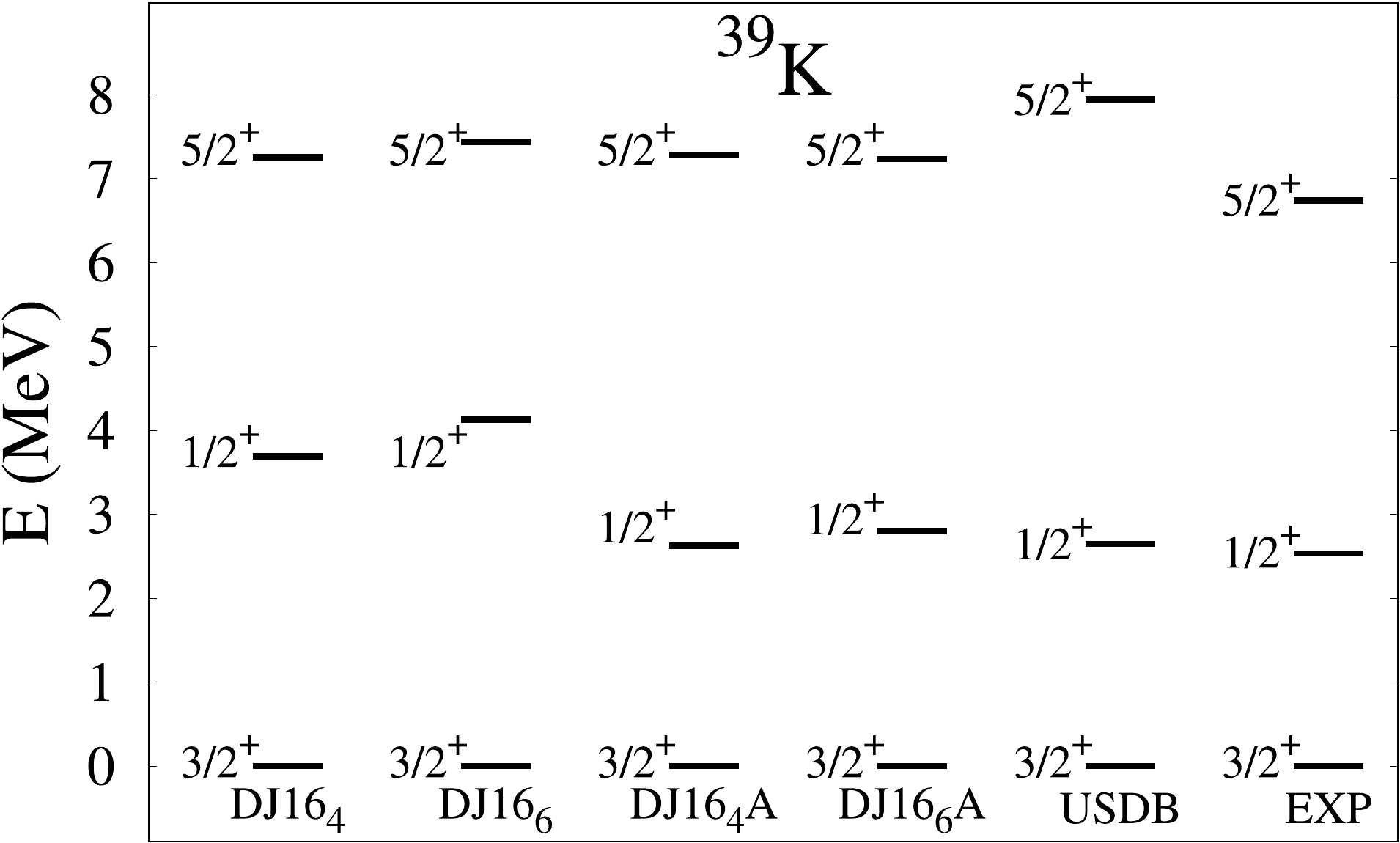}
  \caption{\label{fig:F-isotopes} Low-energy spectra of odd-$A$ $^{19-27}$F and $^{39}$K, 
obtained from USDB and microscopic effective interactions based on Daejeon16, in comparison with the experimental  data on positive-parity states from Refs.~\protect\cite{nndc,VanLe2017,Doorn2017}.
For $^{39}$K, we show experimentally deduced centroids from Ref.~\protect\cite{Doll76}.}
\end{figure*}

The odd-$A$ F isotopes are important because, while neutrons are affected by the pairing force, 
the proton single-particle centroids can provide direct information on the proton-neutron monopoles. 
In practice, it is difficult to obtain the experimental centroids 
due to the sparsity and imprecision of available data on the spectroscopic factors.
The low-energy theoretical spectra of odd-$A$ F isotopes are shown in Fig.~\ref{fig:F-isotopes} in comparison
with experiment.
In these calculations we introduce DJ16$_6$B which is DJ16$_6$A with modified quadrupole pairing TBMEs.
The idea is to improve a few characteristic spectra, such as $^{25}$Mg, see discussion in Ref.~\cite{DuZu96}.
To this end, we have modified three non-diagonal TBMEs, namely, $\langle d_{5/2} d_{3/2} | V|  d_{5/2} s_{1/2}\rangle_{JT} $,
$\langle d_{5/2} d_{3/2}| V|  d_{3/2} s_{1/2}\rangle_{JT}$ and 
$\langle d_{5/2} s_{1/2}| V|  d_{3/2} s_{1/2}\rangle_{JT}$ with $J=2$ and $T=0$ by making them
1.13~MeV, 1.83~MeV and 0.45~MeV more repulsive, respectively. 
The TBMEs of the DJ16$_6$B interaction are given in Table I of the Supplemental material.

When discussing odd-$A$ fluorine isotopes, we expect that only in $^{23,25}$F 
the low-lying $5/2^+$, $1/2^+$ and $3/2^+$ states may contain 
appreciable proton $d_{5/2}$, $s_{1/2}$ and $d_{3/2}$ single-particle components, respectively.
In other cases, neutron correlations have more heavily mixed configurations 
arising from coupling to proton degrees of freedom.

The low-lying states of $^{19}$F are relatively well-reproduced by all interactions,
which is a typical trend for nuclei  with a small number of valence particles.
In the case of $^{21}$F, there is an inversion of the lowest $1/2^+$ and $5/2^+$ states
in the spectra obtained by DJ16$_{4}$ and DJ16$_{6}$ and other microscopic interactions
obtained from various realistic $NN$ potentials when compared with experiment or with USDB. 
One of the possible reasons is the insufficient $N=14$ shell gap seen in Fig.~\ref{fig:ESPEs}.
Modifications made to the monopoles to produce
the DJ16$_4$A interaction, succeed in yielding the $5/2^+$ ground state, but produce inverted
higher lying $3/2^+$ and $9/2^+$ states. 
DJ16$_6$A provides correct ordering of $3/2^+$ and $9/2^+$ contrary to any other interaction, including USDB, 
while DJ16$_6$B spoils this ordering of high-lying states although clearly improves splitting of low-lying $5/2^+$ and $1/2^+$.

The small $N=14$ subshell gap in $^{22}$O, as obtained from the valence-space effective interactions DJ16$_4$ and DJ16$_6$, 
manifests itself in a shift down excitation spectrum of $^{23}$F relative to experiment shown in Fig.~\ref{fig:F-isotopes}. 
In particular, the first $1/2^+$ and $3/2^+$ states, which contain large components of proton $s_{1/2}$ and $d_{3/2}$ single-particle states,
are too low with respect to the experimental data and to the USDB calculation. 
Similarly, the small gap $N=14$ gap will lead to the low $1/2^+$ state in $^{25}$F.
Again, we see that these features are improved from DJ16$_4$ to DJ16$_6$, and that there
is a rather good agreement between DJ16$_4$A, DJ16$_6$A, USDB and experiment.
It is interesting that the position of the $1/2^+$ first excited state in $^{27}$F, 
observed in Ref.~\cite{Doorn2017}, is better described by DJ16$_6$, DJ16$_4$A and DJ16$_6$A 
than by USDB. 
We can make the same remark for DJ16$_6$B: while being slightly better than DJ16$_4$A and DJ16$_6$A for light fluorine
isotopes, it fails to produce a sufficiently low $1/2^+$ state in $^{27}$F. 
Since an elevated position of this state is common to both DJ16$_6$B and USDB, one infers a connection with 
quadrupole pairing.


\begin{figure*}[!ht]
\centerline{\includegraphics[width=.9\textwidth]{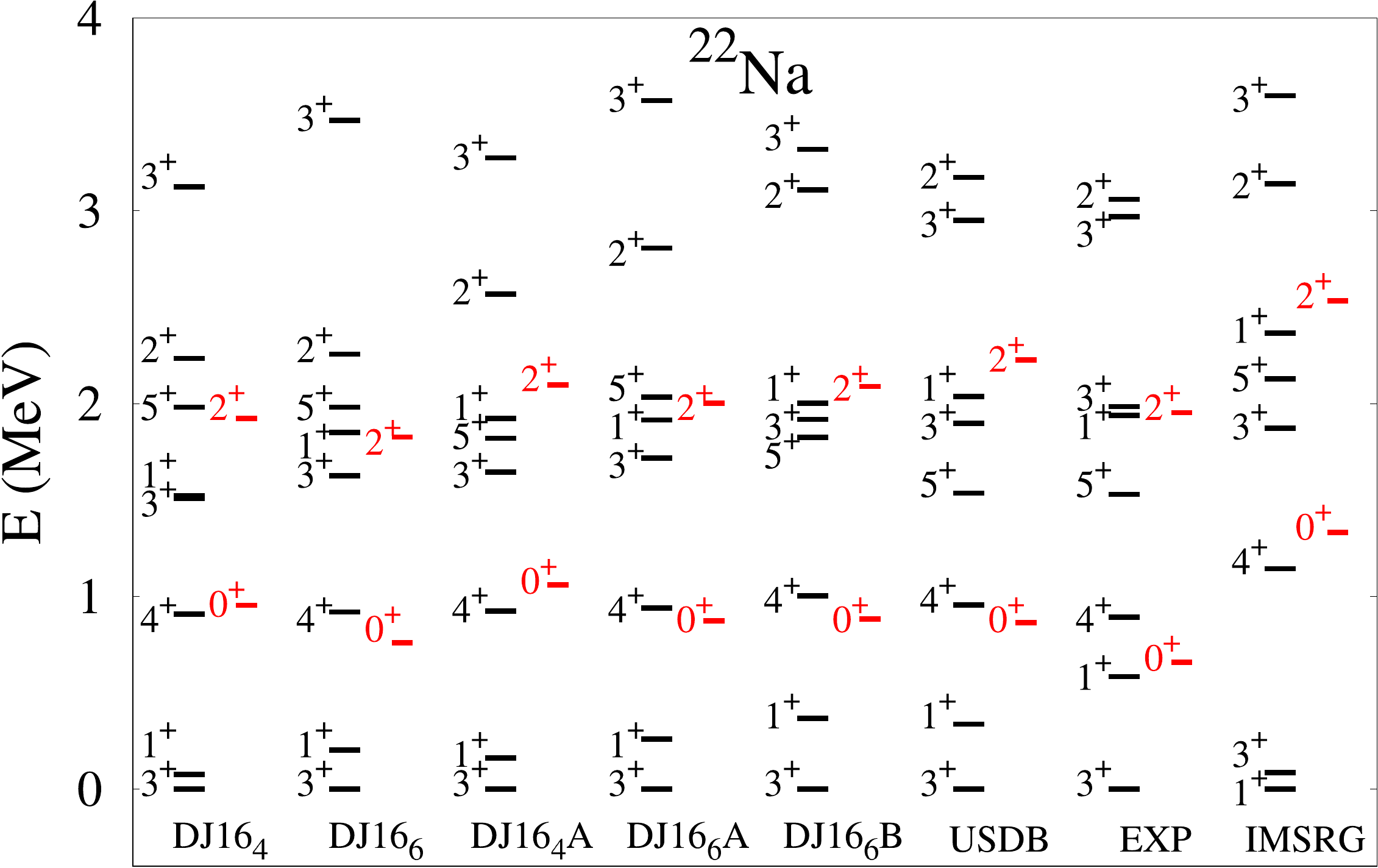} }
  \caption{\label{fig:Na22-spectrum}(Color online) 
Low-energy spectrum of $^{22}$Na  obtained from USDB and microscopic effective interactions, 
in comparison with the experimental  data on positive-parity states from Ref.~\protect\cite{nndc}.
$T=0$ states are shown in black, while $T=1$ states are plotted in red.}
\end{figure*}

Finally, to see what happens as the $sd$ shell becomes almost filled, we present the spectrum of $^{39}$K  
which may  shed additional light on the evolution of the nuclear mean field.
The experimental spectrum of $^{39}$K (Fig.~\ref{fig:F-isotopes}) 
shows the centroids of the single-particle states, as extracted from Ref.~\cite{Doll76}. 
They can be directly compared to the theory.
Although the DJ16$_4$ and DJ16$_6$ interactions show the $1/2^+$ state is slightly elevated, there is
in general a robust agreement with the experiment.
Since the spectrum of $^{39}$K is sensitive only to the monopole part of the $sd$-shell Hamiltonian, 
DJ16$_6$B produces exactly the same spectrum as DJ16$_6$A and, hence, we do not show it in the figure.

The rms deviations of the excitation energies relative to experiment for the  odd-$A$ fluorine isotopes 
and $^{39}$K can be found in the 4th column of Table~\ref{tab:rms}.
Note that among other microscopic interactions, DJ16$_4$A produces the smallest rms deviation 
for this particular group of six nuclei.

\subsection{\boldmath $^{22}\rm Na$}

The case of $^{22}$Na with 3 protons and 3 neutrons in the valence space is considered to be 
an important benchmark of the $3N$ forces~\cite{Zuker3N}.
As seen from Fig.~\ref{fig:Na22-spectrum}, the $T=0$ spectra from DJ16-family of
valence-space effective interactions
provide robust overall agreement with experiment, although the details are not fully reproduced.
The crucial thing is that the ground state is correctly found to be $3^+$ in all theoretical cases except
IMSRG, with the splitting between $1^+$ and $3^+$ state
being in the best agreement for DJ16$_6$B. 
Indeed, besides the improvement in monopole properties, the latter
suggests that further improvements will accrue with 
modified quadrupole pairing TBMEs.

The rms deviations between theory and experiment for excitation energies of $^{22}$Na are given in Table~\ref{tab:rms} 
(the fifth column). 
A continuous reduction of the rms deviations is evident from DJ16$_4$ to DJ16$_6$
and from DJ16$_4$A to DJ16$_6$A.  
Thus, expanding our NCSM basis space and including the monopole modifications which, combined, results in DJ16$_6$A, 
leads to an essential decrease of the rms deviation (from 439 keV to 300 keV). It is also noteworthy that
DJ16$_6$B is characterized by an even smaller rms deviation (197~keV), approaching the USDB value  (155~keV) 
where the main difference in rms values arises from the locations of their $5^+$ states.

%
\section{Quadrupole properties and $\mbox{Mg}$ isotopes}

\begin{figure*}[!t]
\includegraphics[width=\textwidth]{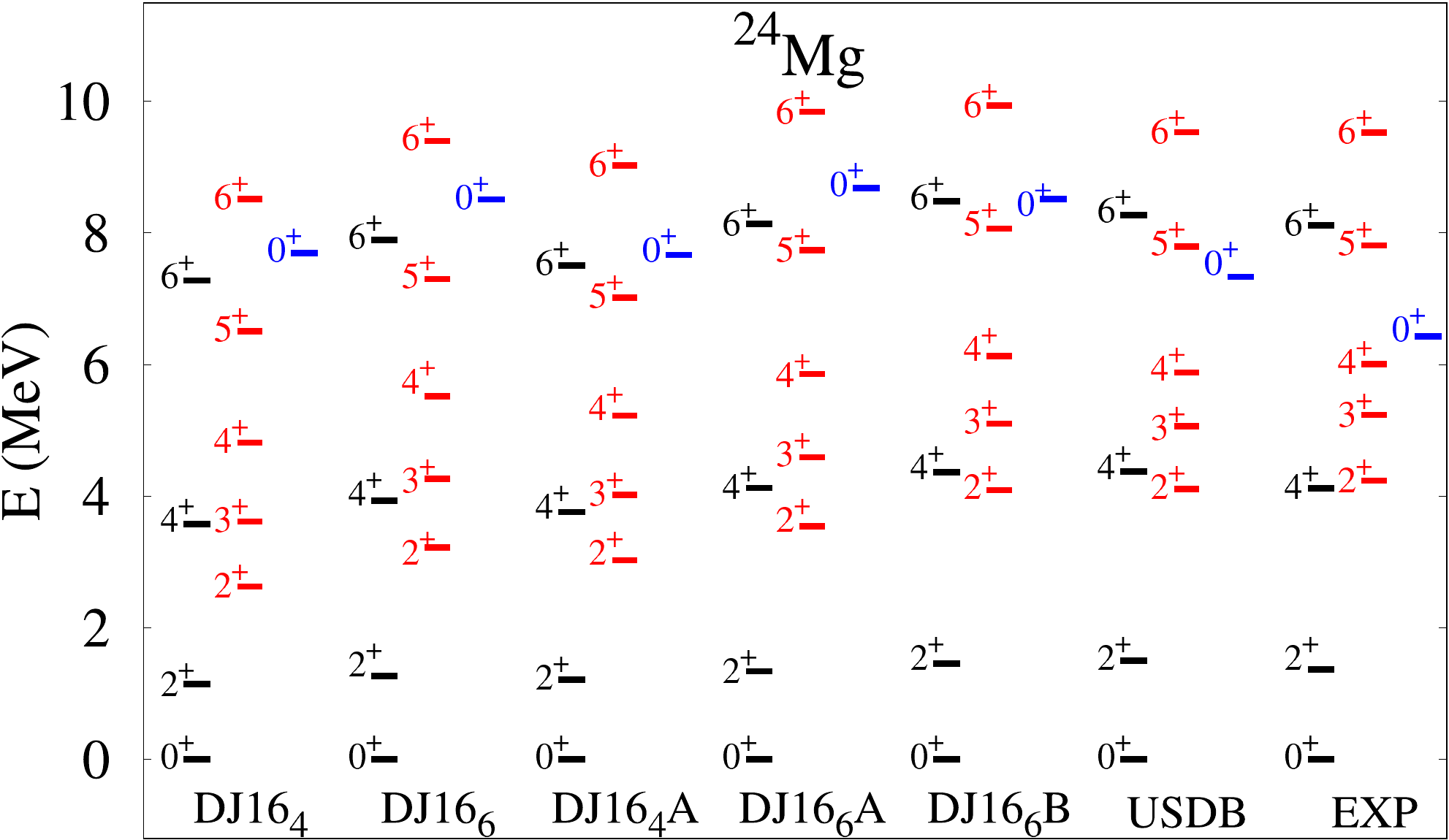}
\caption{\label{fig:Mg24}(Color online) 
Low-energy spectrum of $^{24}$Mg  obtained from USDB and microscopic effective interactions, 
in comparison with the experimental  data on positive-parity states from Ref.~\protect\cite{nndc}.
Different rotational bands are distinguished by color.
}
\end{figure*}

\begin{figure}[!b]
\includegraphics[width=0.45\textwidth]{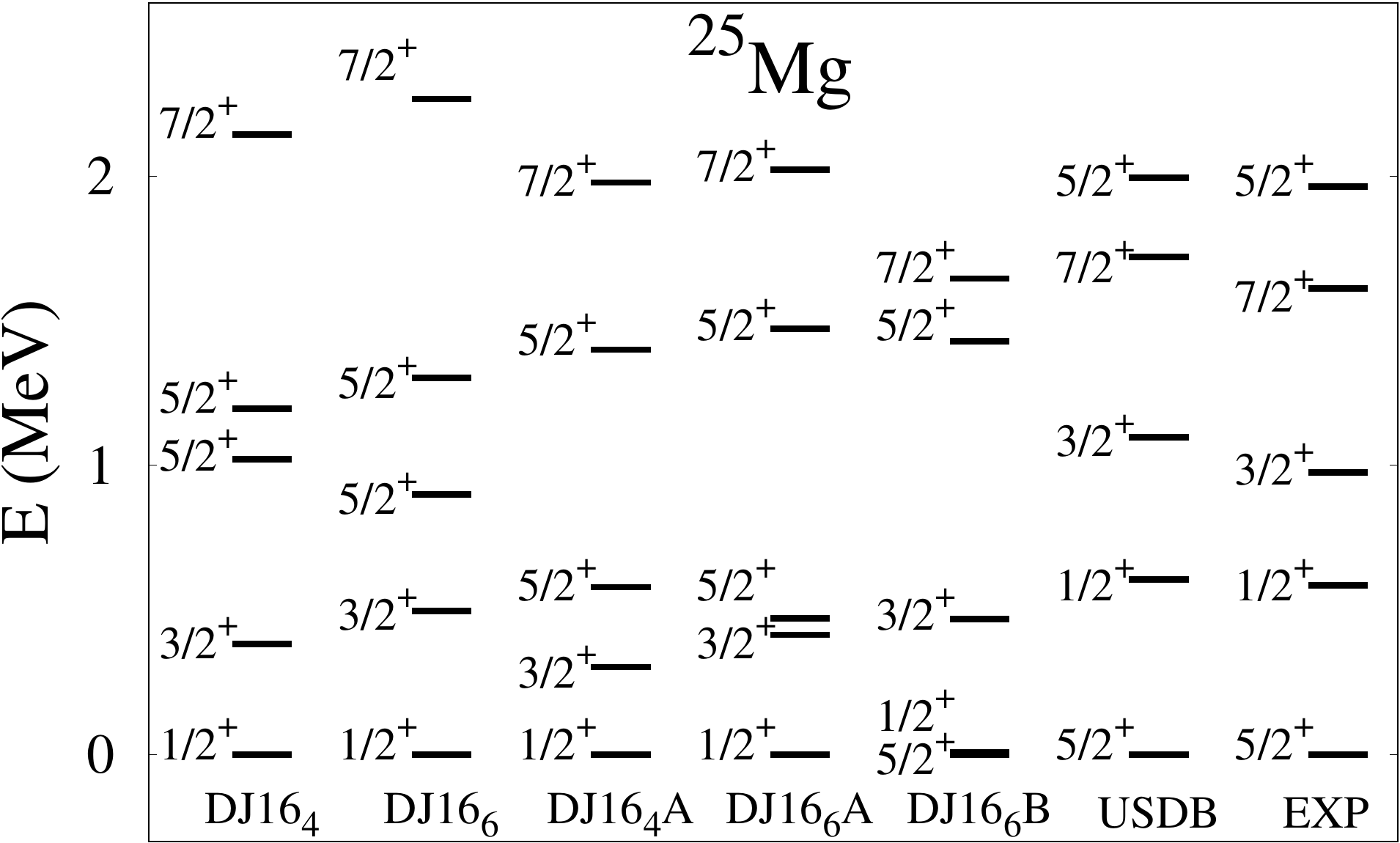}
\caption{\label{fig:Mg25} Low-energy spectrum of $^{25}$Mg  obtained from USDB and microscopic effective interactions, 
in comparison with the experimental  data on positive-parity states from Ref.~\protect\cite{nndc}.
}
\vspace{-1ex}
\end{figure}

To characterize the proton-neutron quadrupole component of the microscopic interactions, 
we consider the spectrum of a well-known  $sd$-shell rotor, $^{24}$Mg, see Fig.~\ref{fig:Mg24}. 
We remark that the ground-state band is relatively well described by all interactions presented. 
The DJ16$_4$ and DJ16$_6$ interactions produce a somewhat lower lying and slightly more stretched \mbox{$\gamma $-band}, 
as compared to experiment and to the USDB result. This feature is corrected partially by the monopole
modifications.  We also find that there is a further improvement in the location of the \mbox{${\gamma }$-band} with DJ16$_6$B.


In addition, we have calculated a neighboring nucleus, $^{25}$Mg,  
with results presented in Fig.~\ref{fig:Mg25} and compared with experiment. This nucleus is known to be a difficult case
for microscopic interactions. Indeed, a microscopic theory prefers to describe this nucleus as a neutron coupled
to the well-deformed core of $^{24}$Mg and, therefore, we see at low energies a typical rotational band
from $1/2^+$, $3/2^+$ and $5/2^+$. This is not at all what is observed 
experimentally where, instead, a $5/2^+$ ground state is determined. 
The USDB interaction is known to cope with this feature by a strong quadrupole pairing.
Monopole corrections proposed by DJ16$_4$A and DJ16$_6$A are seen to be inline with this changes, since we notice
a lowering of the $5/2^+$ state.
Furthermore, DJ16$_6$B demonstrates that increasing the amount of quadrupole pairing finally produces a 
$5/2^+$ ground state, although not well-separated from the first excited state $1/2^+$.




\section{Conclusions and summary}

In the present work we have compared the general properties of the new microscopic effective $sd$ shell 
interactions obtained from the NCSM wave functions via the OLS transformation.
The NCSM calculations have been performed using the Daejeon16 $NN$ potential in the model space truncated at $N_{\rm max}=6$
with a fixed value of $\hbar \Omega =14$~MeV.

Since the theoretical single-particle energies show major deficiencies when compared with empirical values, 
we have adopted the empirical single-particle energies for these investigations.
In addition, to accommodate the expected dependence of the mean-field with increasing $A$, 
we used the USDB scaling of the TBMEs.

The monopole components of the derived microscopic effective interaction are compared to those of the 
phenomenological USDB interaction and to an earlier effective interaction constructed in a similar manner from
NCSM calculations with Daejeon16 at $N_{\rm max}=4$~\cite{SmBa2019}. 
We obtain a general improvement in the monopole evolution
and in the binding energy of O isotopes using a larger model space.
The improving trend suggests the need for even larger model spaces.

To improve the description of the data,
we propose minimal modifications to the centroids of the derived effective interaction,
with the aim to  reproduce the $N=14$ shell gap, important for the O isotopes.
The monopole-modified interaction (DJ16$_6$A) provides excellent agreement for binding energies of the O isotopes and
greatly improves the excitation spectra of the Oxygen chain as well as the odd-$A$ F isotopes.
Furthermore, we show that modification of the quadrupole pairing helps to reproduce the spectrum of $^{25}$Mg, for example.
Extended calculations through the $sd$ shell are summarized in Supplemental Material.

\acknowledgments{
N.~A.~Smirnova and Zhen Li acknowledge the financial support of CNRS/IN2P3, France, via ENFIA Master project. 
The work was supported by 
the Rare Isotope Science Project of Institute for Basic Science funded by Ministry of Science
and ICT and National Research Foundation of Korea (2013M7A1A1075764).
This work was in part supported by the Institute for Basic Science, Korea (IBS-R031-D1).
This work was supported in part by the US Department of Energy (DOE) under Grant Nos. DE-FG02-87ER40371,
DE-SC0018223 (SciDAC-4/NUCLEI) and DE-SC0023495 (SciDAC-5/NUCLEI).
Computational resources were provided by the National Energy Research Scientific Computing Center (NERSC), 
which is supported by the US DOE Office of Science under Contract No. DE-AC02-05CH11231, and
also by the National Supercomputing Center with supercomputing resources including technical support (KSC-2020-CRE-0027). 
The work was supported by the JSPS Grant-in-Aid for Scientific Research (S) under Grant No. 20H05648.
N.~A.~Smirnova and A.~M.~Shirokov
thank the Institute for Basic Science, Daejeon, for hospitality and financial support of their visits.
The work of A.~M.~Shirokov is supported by the Russian Foundation for Basic Research under Grant No.~20-02-00357.
A.~M.~Shirokov also thanks the University of Bordeaux for hospitality and financial support of his visit 
to CENBG (LP2IB) where a part of this work was done.
}

\bibliography{Nm6}

\newpage
\widetext
\clearpage
\begin{center}
\textbf{\large Supplemental Material for\\[5mm] 
\textit{Improved \boldmath$sd$ shell effective interactions from Daejeon16}}
\end{center}
\setcounter{equation}{0}
\setcounter{figure}{0}
\setcounter{table}{0}
\setcounter{page}{1}
\setcounter{section}{0}
\makeatletter
\renewcommand{\theequation}{S\arabic{equation}}
\renewcommand{\thetable}{S\arabic{table}}
\renewcommand{\thefigure}{S\arabic{figure}}
\renewcommand{\thesection}{S\arabic{section}}
\renewcommand{\bibnumfmt}[1]{[S#1]}
\renewcommand{\citenumfont}[1]{S#1}


\section{Tabulation of derived two-body matrix elements}
\label{app-TBMEs}

\begin{longtable}{rrrrrrrrrr}
\mbox{}\\[-4ex]
\caption{\label{tab:TBMEs_DJ16} 
The TBMEs (in MeV) of the secondary $sd$-shell effective Hamiltonian $H_{18}^{P'}$
obtained from the NCSM calculation with $N_{\rm max}=6$, $\hbar \Omega=14$ MeV, 
and Daejeon16 potential for $^{18}$F are shown as well as the TBMEs of its residual 
valence effective interaction, $V_2^{P'}$. 
DJ16$_6$A and DJ16$_6$B are modified versions (see text of the paper for details).
The TBMEs are presented for $A=18$. Calculating spectra of nuclei with a different $A$ value, we applied
a scaling factor $(18/A)^{0.3}$, as was established for USDB.
}\\ 
\hline \hline \\
\hspace{2mm} $2j_a$ & \hspace{2mm} $2j_b$ & \hspace{2mm} $2j_c$ & \hspace{2mm} $2j_d$ & \hspace{2mm} $J$ & \hspace{2mm} $T$ & \hspace{2mm} 
$\quad H_{18}^{P'}$ & \hspace{2mm} \quad $V_2^{P'}$ & \hspace{2mm} $\quad V_2^{P'}$ & \hspace{2mm} $\quad V_2^{P'}$ \\
\hline
       &        &        &        &     &     &               &  \hspace{2mm} DJ16$_6$  & \hspace{2mm}    DJ16$_6$A & \hspace{2mm} DJ16$_6$B \\
\hline
   1 &   1 &   1 &   1 &   0 &   1 & $ -130.345$ & $   -1.910$ & $   -1.910$ & $   -1.910$ \\
   1 &   1 &   3 &   3 &   0 &   1 & $   -0.274$ & $   -0.274$ & $   -0.274$ & $   -0.274$ \\
   1 &   1 &   5 &   5 &   0 &   1 & $   -1.564$ & $   -1.564$ & $   -1.564$ & $   -1.564$ \\
   3 &   3 &   3 &   3 &   0 &   1 & $ -111.153$ & $   -1.554$ & $   -1.654$ & $   -1.654$ \\
   3 &   3 &   5 &   5 &   0 &   1 & $   -3.279$ & $   -3.279$ & $   -3.279$ & $   -3.279$ \\
   5 &   5 &   5 &   5 &   0 &   1 & $ -129.953$ & $   -2.459$ & $   -2.439$ & $   -2.439$ \\
   1 &   1 &   1 &   1 &   1 &   0 & $ -131.287$ & $   -2.851$ & $   -2.851$ & $   -2.851$ \\
   1 &   1 &   1 &   3 &   1 &   0 & $    0.388$ & $    0.388$ & $    0.388$ & $    0.388$ \\
   1 &   1 &   3 &   3 &   1 &   0 & $    1.147$ & $    1.147$ & $    1.147$ & $    1.147$ \\
   1 &   1 &   3 &   5 &   1 &   0 & $   -2.165$ & $   -2.165$ & $   -2.165$ & $   -2.165$ \\
   1 &   1 &   5 &   5 &   1 &   0 & $   -1.210$ & $   -1.210$ & $   -1.210$ & $   -1.210$ \\
   1 &   3 &   1 &   3 &   1 &   0 & $ -123.026$ & $   -4.009$ & $   -4.009$ & $   -4.009$ \\
   1 &   3 &   3 &   3 &   1 &   0 & $    1.143$ & $    1.143$ & $    1.143$ & $    1.143$ \\
   1 &   3 &   3 &   5 &   1 &   0 & $    1.487$ & $    1.487$ & $    1.487$ & $    1.487$ \\
   1 &   3 &   5 &   5 &   1 &   0 & $   -0.937$ & $   -0.937$ & $   -0.937$ & $   -0.937$ \\
   3 &   3 &   3 &   3 &   1 &   0 & $ -110.164$ & $   -0.565$ & $   -0.565$ & $   -0.565$ \\
   3 &   3 &   3 &   5 &   1 &   0 & $    0.738$ & $    0.738$ & $    0.738$ & $    0.738$ \\
   3 &   3 &   5 &   5 &   1 &   0 & $    1.961$ & $    1.961$ & $    1.961$ & $    1.961$ \\
   3 &   5 &   3 &   5 &   1 &   0 & $ -124.252$ & $   -5.705$ & $   -5.905$ & $   -5.905$ \\
   3 &   5 &   5 &   5 &   1 &   0 & $   -3.229$ & $   -3.229$ & $   -3.229$ & $   -3.229$ \\
   5 &   5 &   5 &   5 &   1 &   0 & $ -128.812$ & $   -1.318$ & $   -1.418$ & $   -1.418$ \\
   1 &   3 &   1 &   3 &   1 &   1 & $ -118.750$ & $    0.268$ & $    0.368$ & $    0.368$ \\
   1 &   3 &   3 &   5 &   1 &   1 & $   -0.081$ & $   -0.081$ & $   -0.081$ & $   -0.081$ \\
   3 &   5 &   3 &   5 &   1 &   1 & $ -118.403$ & $    0.144$ & $    0.354$ & $    0.354$ \\
   1 &   3 &   1 &   3 &   2 &   0 & $ -121.028$ & $   -2.011$ & $   -2.011$ & $   -2.011$ \\
   1 &   3 &   1 &   5 &   2 &   0 & $    2.951$ & $    2.951$ & $    2.951$ & $    2.501$ \\
   1 &   3 &   3 &   5 &   2 &   0 & $   -2.259$ & $   -2.259$ & $   -2.259$ & $   -0.430$ \\
   1 &   5 &   1 &   5 &   2 &   0 & $ -128.544$ & $   -0.579$ & $   -0.529$ & $   -0.529$ \\
   1 &   5 &   3 &   5 &   2 &   0 & $    1.765$ & $    1.765$ & $    1.765$ & $    0.635$ \\
   3 &   5 &   3 &   5 &   2 &   0 & $ -121.960$ & $   -3.413$ & $   -3.613$ & $   -3.613$ \\
   1 &   3 &   1 &   3 &   2 &   1 & $ -119.602$ & $   -0.584$ & $   -0.484$ & $   -0.484$ \\
   1 &   3 &   1 &   5 &   2 &   1 & $   -1.587$ & $   -1.587$ & $   -1.587$ & $   -1.587$ \\
   1 &   3 &   3 &   3 &   2 &   1 & $   -0.013$ & $   -0.013$ & $   -0.013$ & $   -0.013$ \\
   1 &   3 &   3 &   5 &   2 &   1 & $    0.477$ & $    0.477$ & $    0.477$ & $    0.477$ \\
   1 &   3 &   5 &   5 &   2 &   1 & $   -1.211$ & $   -1.211$ & $   -1.211$ & $   -1.211$ \\
   1 &   5 &   1 &   5 &   2 &   1 & $ -129.227$ & $   -1.262$ & $   -1.032$ & $   -1.032$ \\
   1 &   5 &   3 &   3 &   2 &   1 & $   -1.114$ & $   -1.114$ & $   -1.114$ & $   -1.114$ \\
   1 &   5 &   3 &   5 &   2 &   1 & $    0.383$ & $    0.383$ & $    0.383$ & $    0.383$ \\
   1 &   5 &   5 &   5 &   2 &   1 & $   -0.544$ & $   -0.544$ & $   -0.544$ & $   -0.544$ \\
   3 &   3 &   3 &   3 &   2 &   1 & $ -109.649$ & $   -0.049$ & $   -0.149$ & $   -0.149$ \\
   3 &   3 &   3 &   5 &   2 &   1 & $    0.849$ & $    0.849$ & $    0.849$ & $    0.849$ \\
   3 &   3 &   5 &   5 &   2 &   1 & $   -0.551$ & $   -0.551$ & $   -0.551$ & $   -0.551$ \\
   3 &   5 &   3 &   5 &   2 &   1 & $ -118.644$ & $   -0.097$ & $    0.113$ & $    0.113$ \\
   3 &   5 &   5 &   5 &   2 &   1 & $    0.269$ & $    0.269$ & $    0.269$ & $    0.269$ \\
   5 &   5 &   5 &   5 &   2 &   1 & $ -128.633$ & $   -1.139$ & $   -1.119$ & $   -1.119$ \\
   1 &   5 &   1 &   5 &   3 &   0 & $ -131.969$ & $   -4.004$ & $   -3.954$ & $   -3.954$ \\
   1 &   5 &   3 &   3 &   3 &   0 & $   -0.438$ & $   -0.438$ & $   -0.438$ & $   -0.438$ \\
   1 &   5 &   3 &   5 &   3 &   0 & $   -1.412$ & $   -1.412$ & $   -1.412$ & $   -1.412$ \\
   1 &   5 &   5 &   5 &   3 &   0 & $   -2.018$ & $   -2.018$ & $   -2.018$ & $   -2.018$ \\
   3 &   3 &   3 &   3 &   3 &   0 & $ -113.150$ & $   -3.550$ & $   -3.550$ & $   -3.550$ \\
   3 &   3 &   3 &   5 &   3 &   0 & $   -1.526$ & $   -1.526$ & $   -1.526$ & $   -1.526$ \\
   3 &   3 &   5 &   5 &   3 &   0 & $    0.710$ & $    0.710$ & $    0.710$ & $    0.710$ \\
   3 &   5 &   3 &   5 &   3 &   0 & $ -119.340$ & $   -0.793$ & $   -0.993$ & $   -0.993$ \\
   3 &   5 &   5 &   5 &   3 &   0 & $   -1.969$ & $   -1.969$ & $   -1.969$ & $   -1.969$ \\
   5 &   5 &   5 &   5 &   3 &   0 & $ -128.317$ & $   -0.823$ & $   -0.923$ & $   -0.923$ \\
   1 &   5 &   1 &   5 &   3 &   1 & $ -127.443$ & $    0.521$ & $    0.751$ & $    0.751$ \\
   1 &   5 &   3 &   5 &   3 &   1 & $    0.298$ & $    0.298$ & $    0.298$ & $    0.298$ \\
   3 &   5 &   3 &   5 &   3 &   1 & $ -118.230$ & $    0.317$ & $    0.527$ & $    0.527$ \\
   3 &   5 &   3 &   5 &   4 &   0 & $ -122.688$ & $   -4.141$ & $   -4.341$ & $   -4.341$ \\
   3 &   5 &   3 &   5 &   4 &   1 & $ -120.066$ & $   -1.519$ & $   -1.309$ & $   -1.309$ \\
   3 &   5 &   5 &   5 &   4 &   1 & $    1.380$ & $    1.380$ & $    1.380$ & $    1.380$ \\
   5 &   5 &   5 &   5 &   4 &   1 & $ -127.666$ & $   -0.172$ & $   -0.152$ & $   -0.152$ \\
   5 &   5 &   5 &   5 &   5 &   0 & $ -131.991$ & $   -4.497$ & $   -4.597$ & $   -4.597$ \\
\hline\hline \\ \\ 
\end{longtable}

\section{Additional spectra of $sd$-shell nuclei.}

This section contains additional spectra of selected $sd$ shell nuclei which have not been presented in the main text.
Please, consult the main paper for other plots and for the definitions of the interactions employed.
\begin{figure}[htb]
    \includegraphics[width=.55\textwidth]{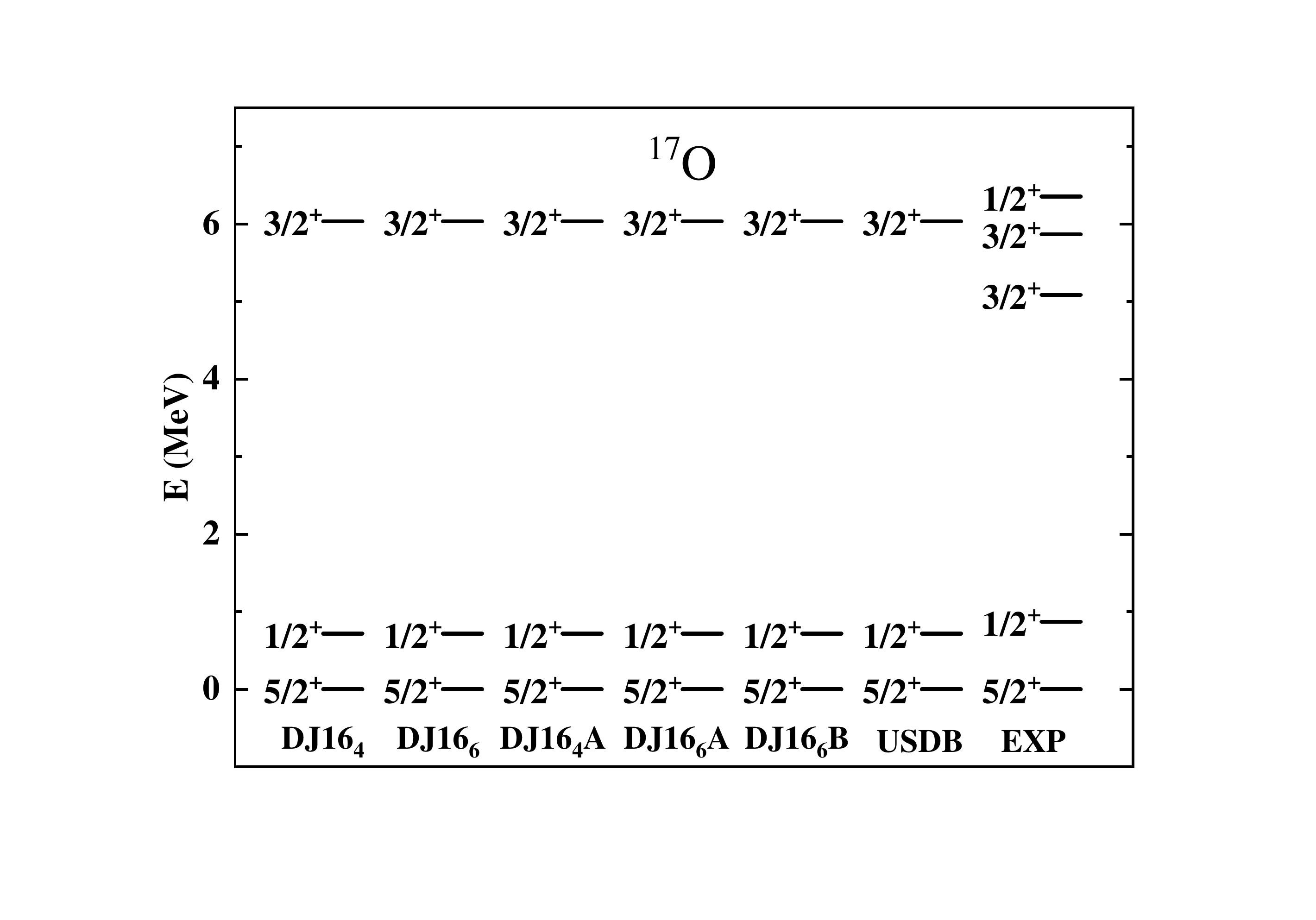}  \hspace{-2cm}
  \includegraphics[width=.55\textwidth]{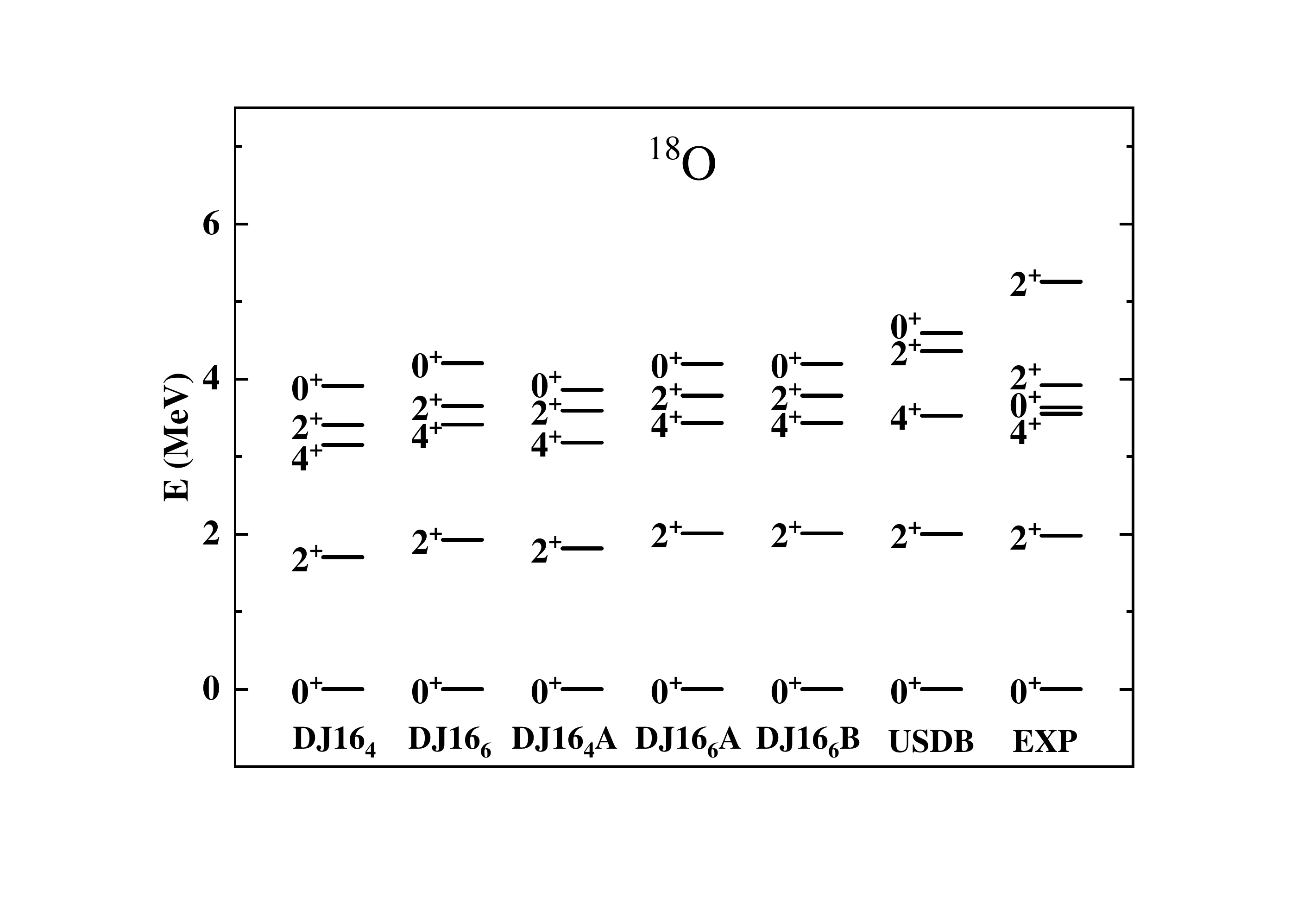}\\[-1.2cm]
   
  \includegraphics[width=.55\textwidth]{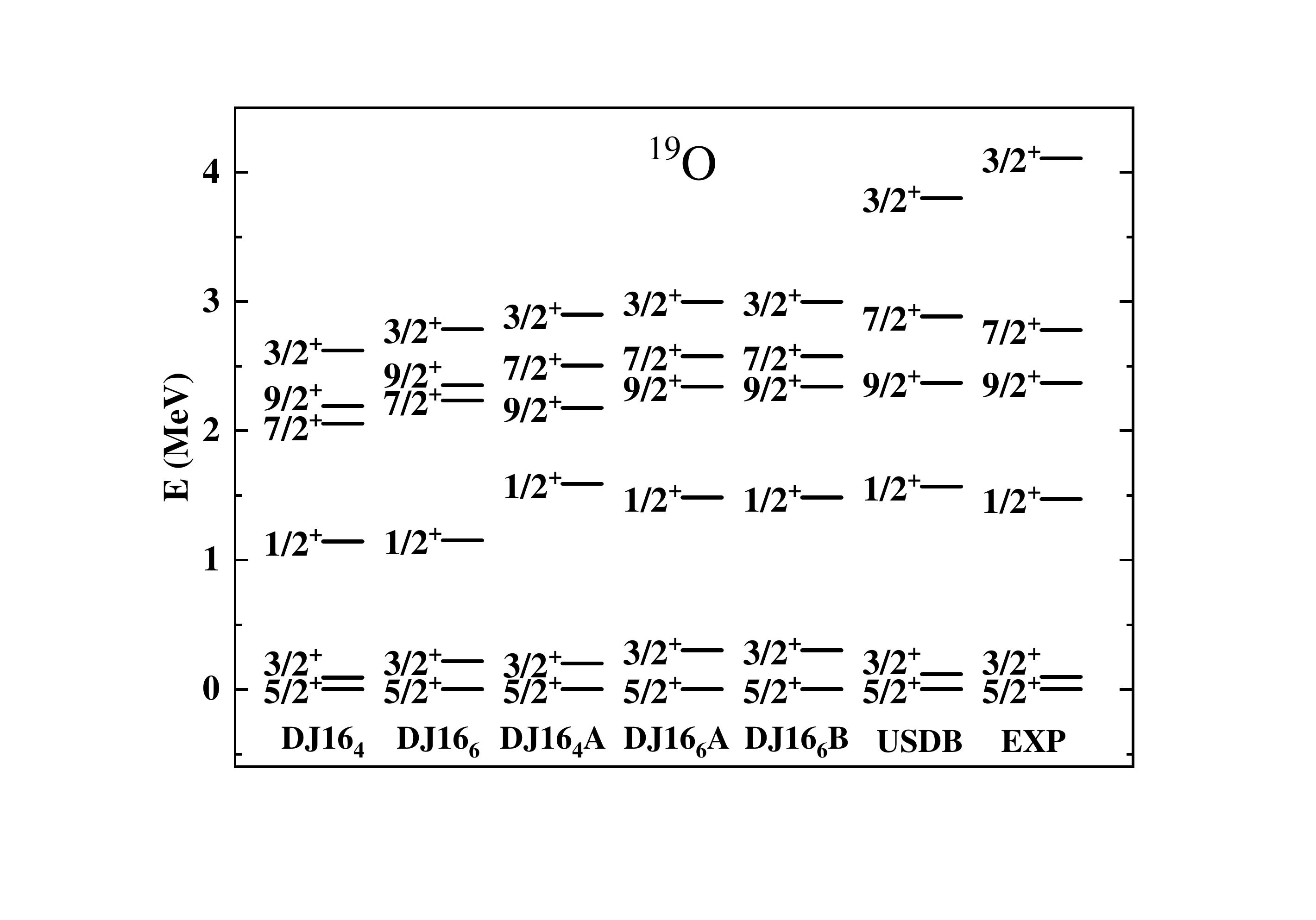} \hspace{-2cm}
  \includegraphics[width=.55\textwidth]{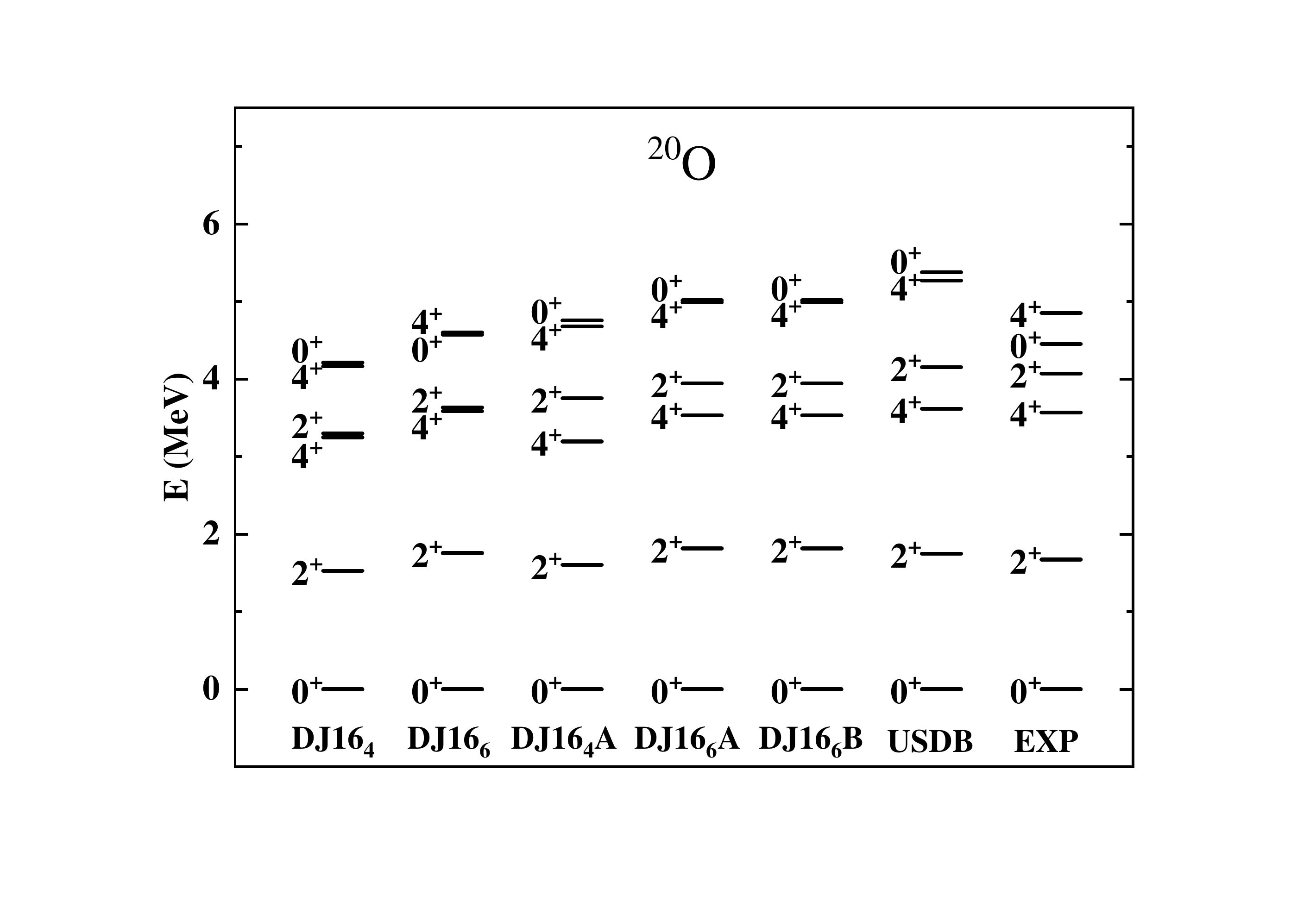}\\[-1.2cm]

  \caption{ Theoretical and experimental low-energy spectra of $^{17-20}$O 
(up to six lowest positive-parity states are presented, when known).}
\end{figure}


\begin{figure}
    
  \includegraphics[width=.55\textwidth]{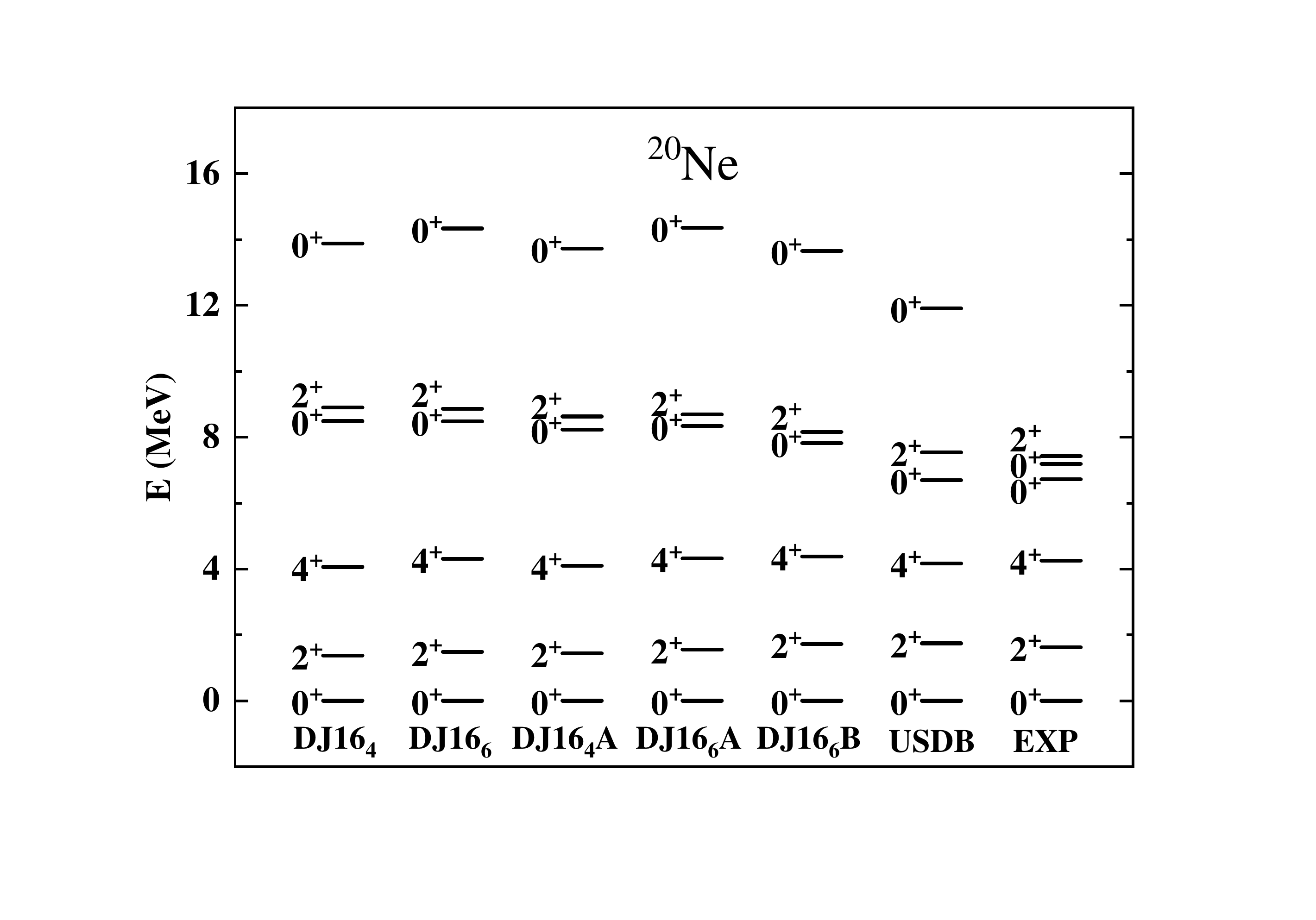} \hspace{-2cm}
  \includegraphics[width=.55\textwidth]{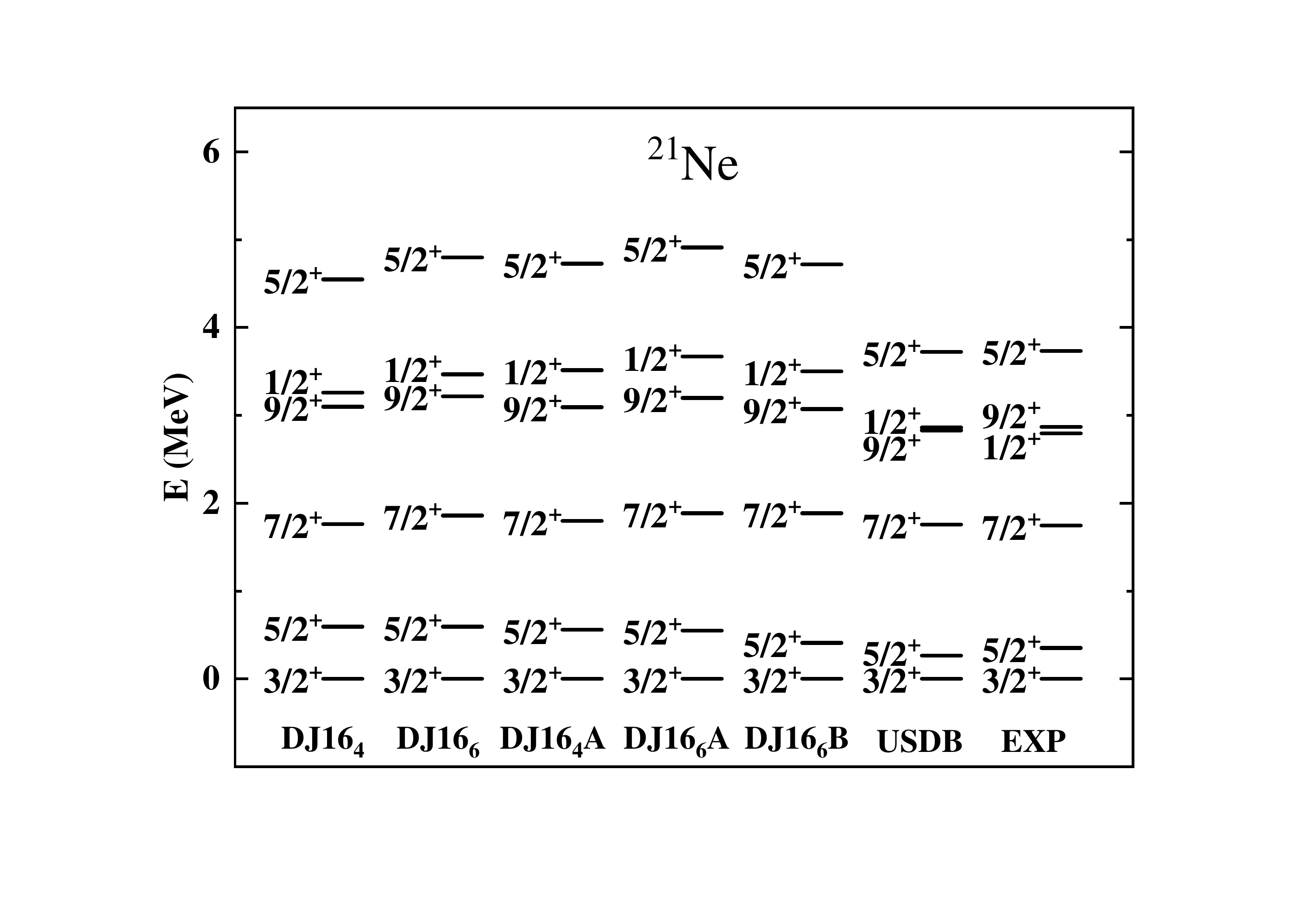}\\[-1.2cm]
  
  \includegraphics[width=.55\textwidth]{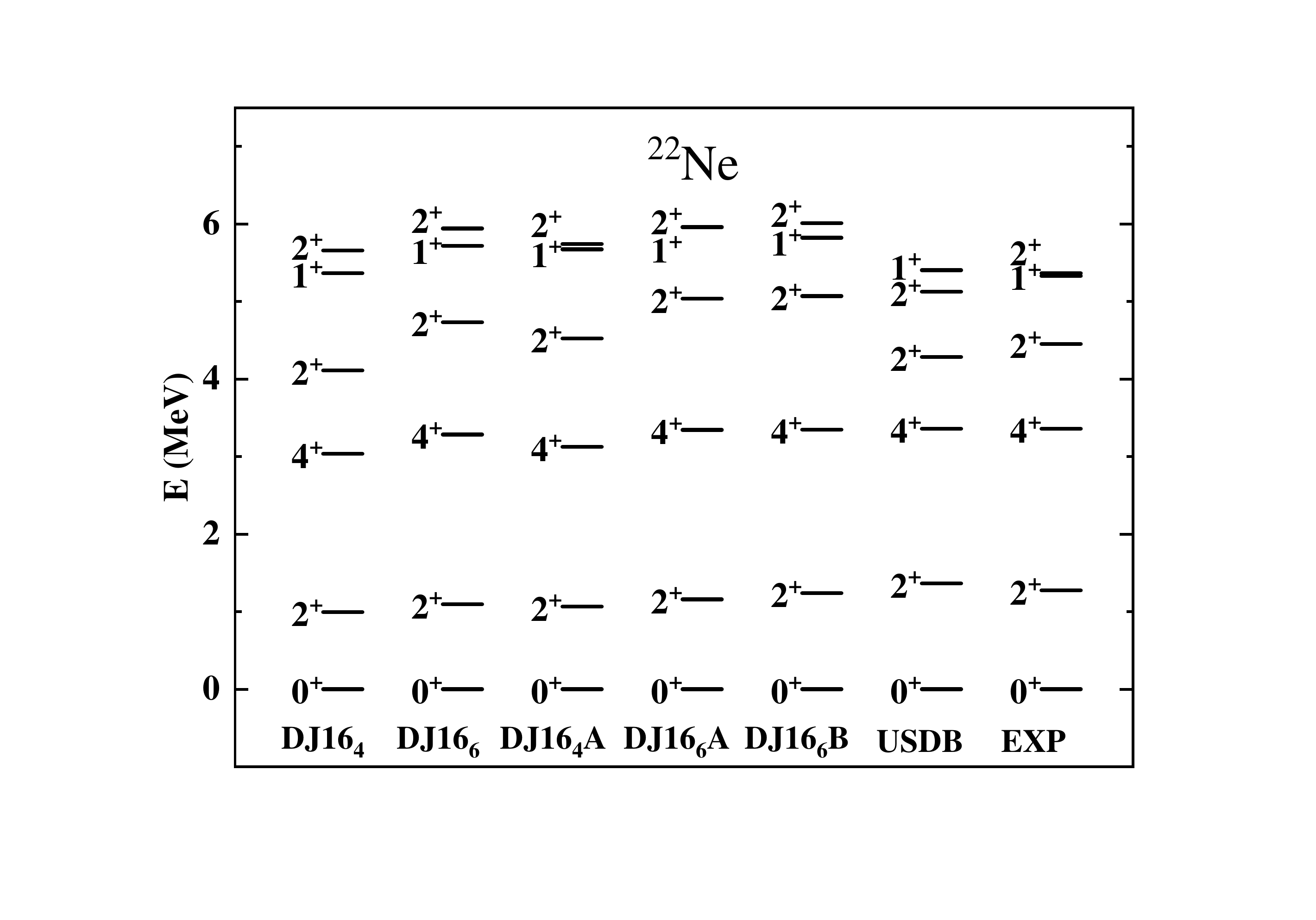} \hspace{-2cm}
  \includegraphics[width=.55\textwidth]{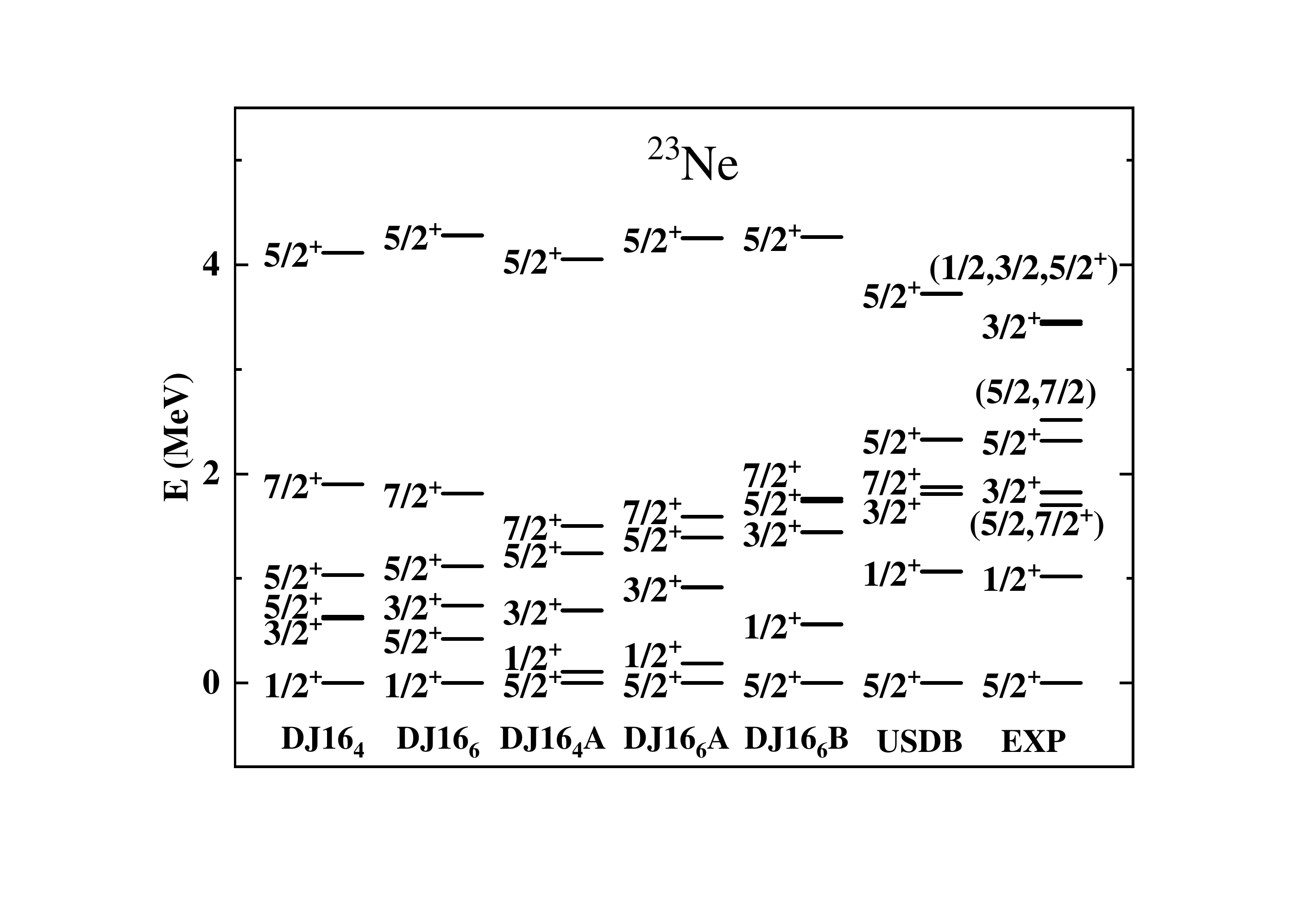}\\[-5mm]
  \caption{Theoretical and experimental low-energy spectra of $^{20-23}$Ne 
(eight lowest positive-parity $T=0$ states are shown for $^{20}$Ne, 
six lowest positive-parity states are presented for $^{21-22}$Ne and eight in the experimental spectrum of $^{23}$Ne because of uncertain spin and parity assignments).}
\end{figure}

\begin{figure}
   
  \includegraphics[width=.55\textwidth]{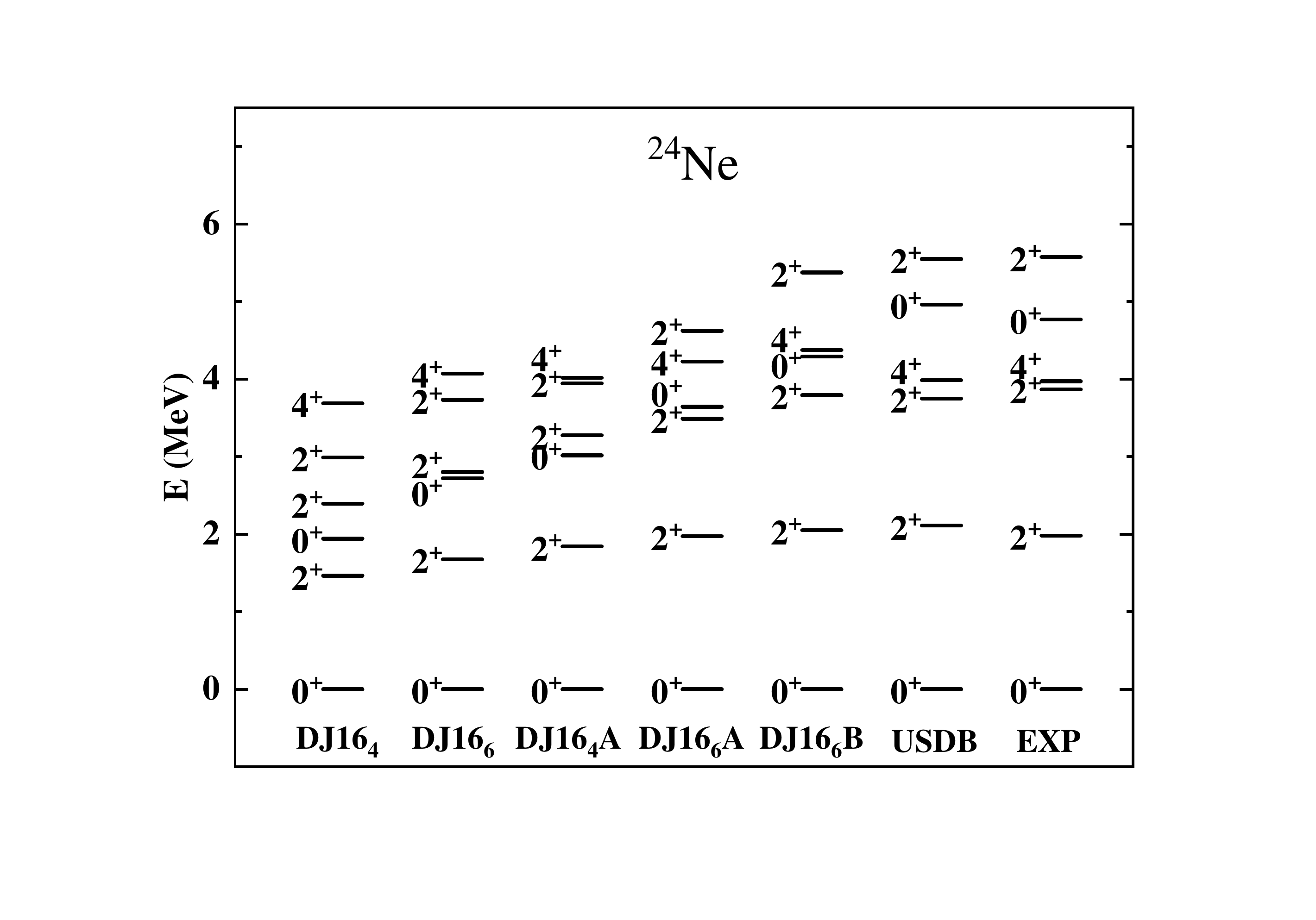}  \hspace{-2cm}
  \includegraphics[width=.55\textwidth]{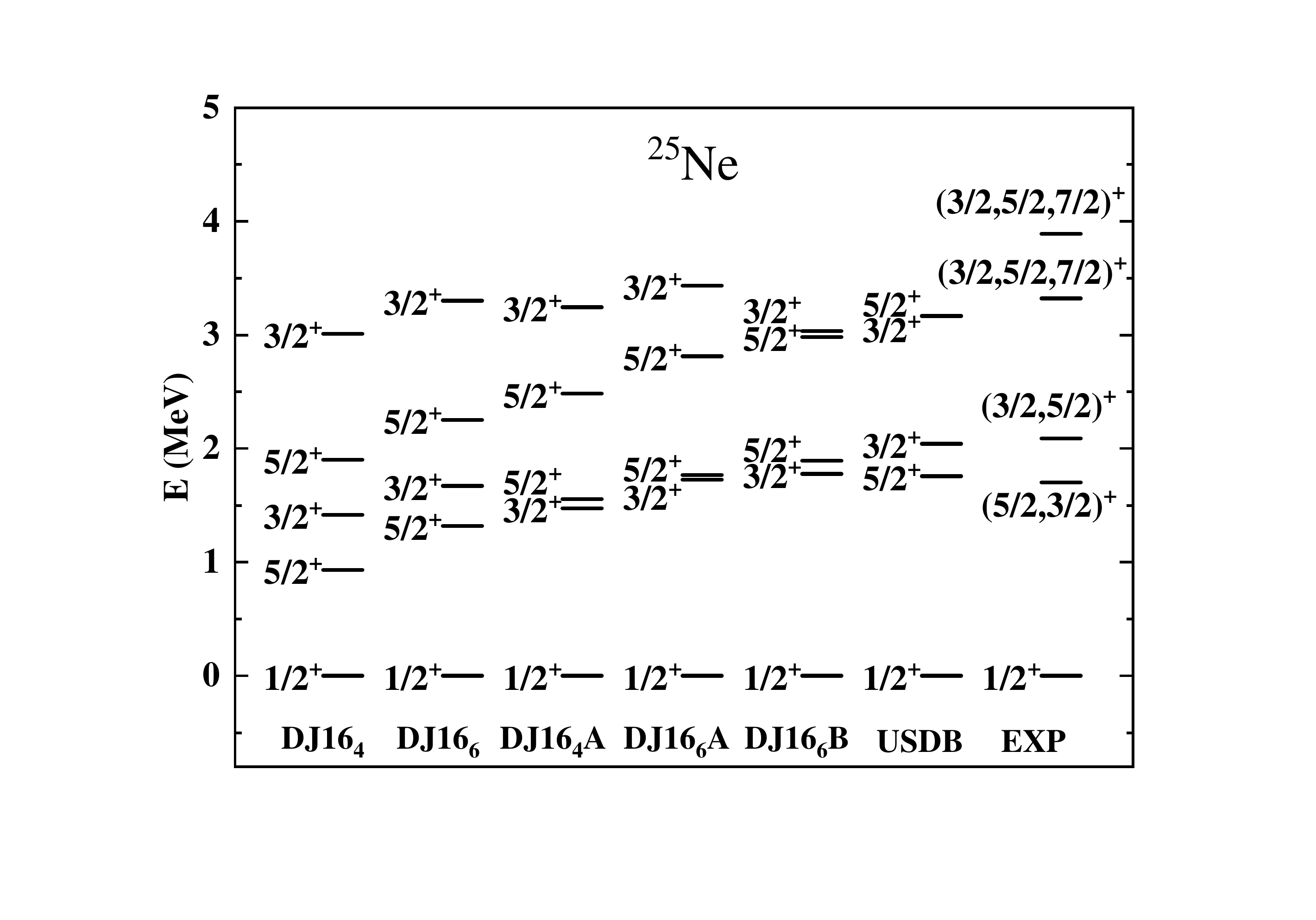}\\[-1.2cm]
  
  \includegraphics[width=.55\textwidth]{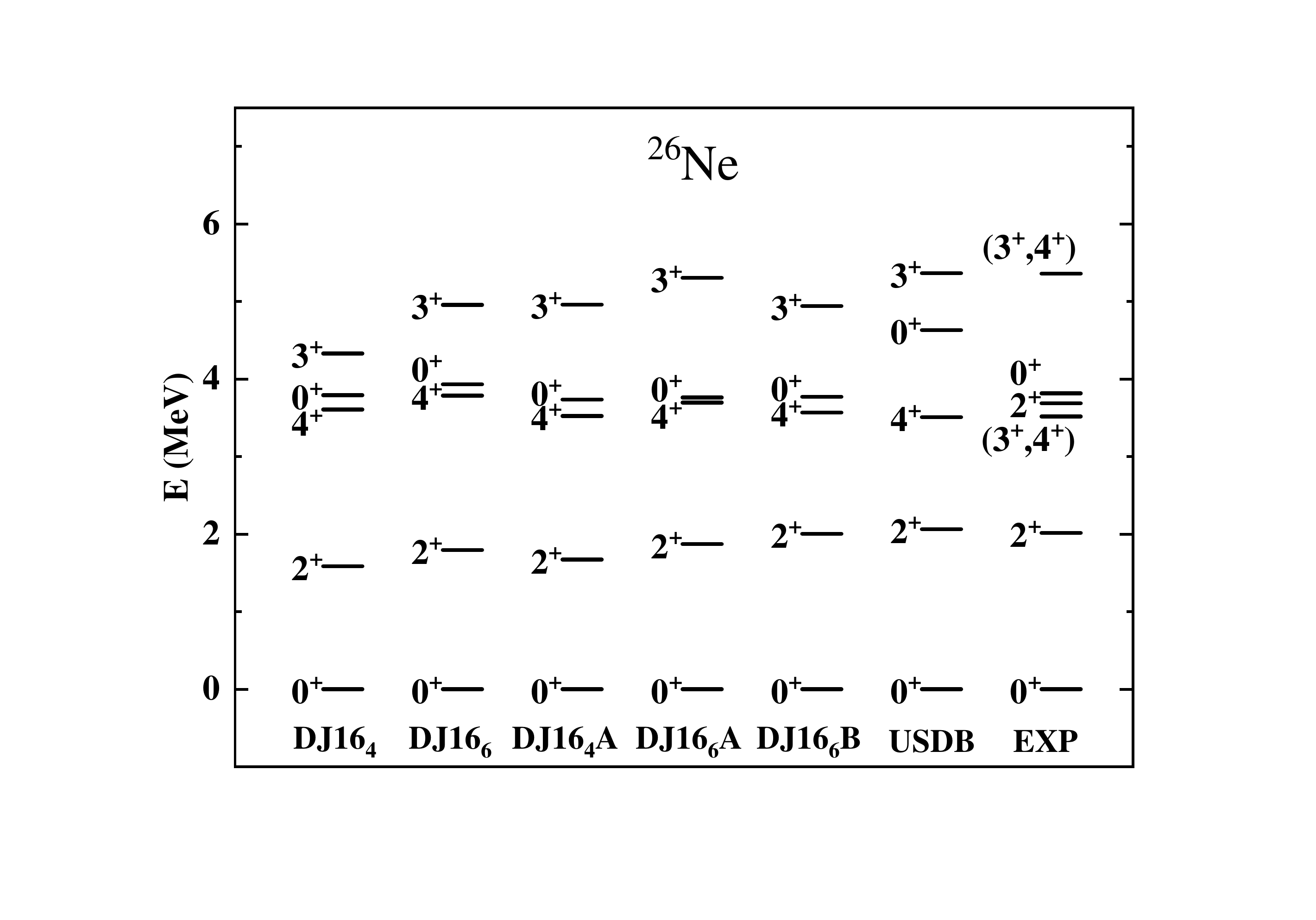} \hspace{-2cm}
  \includegraphics[width=.55\textwidth]{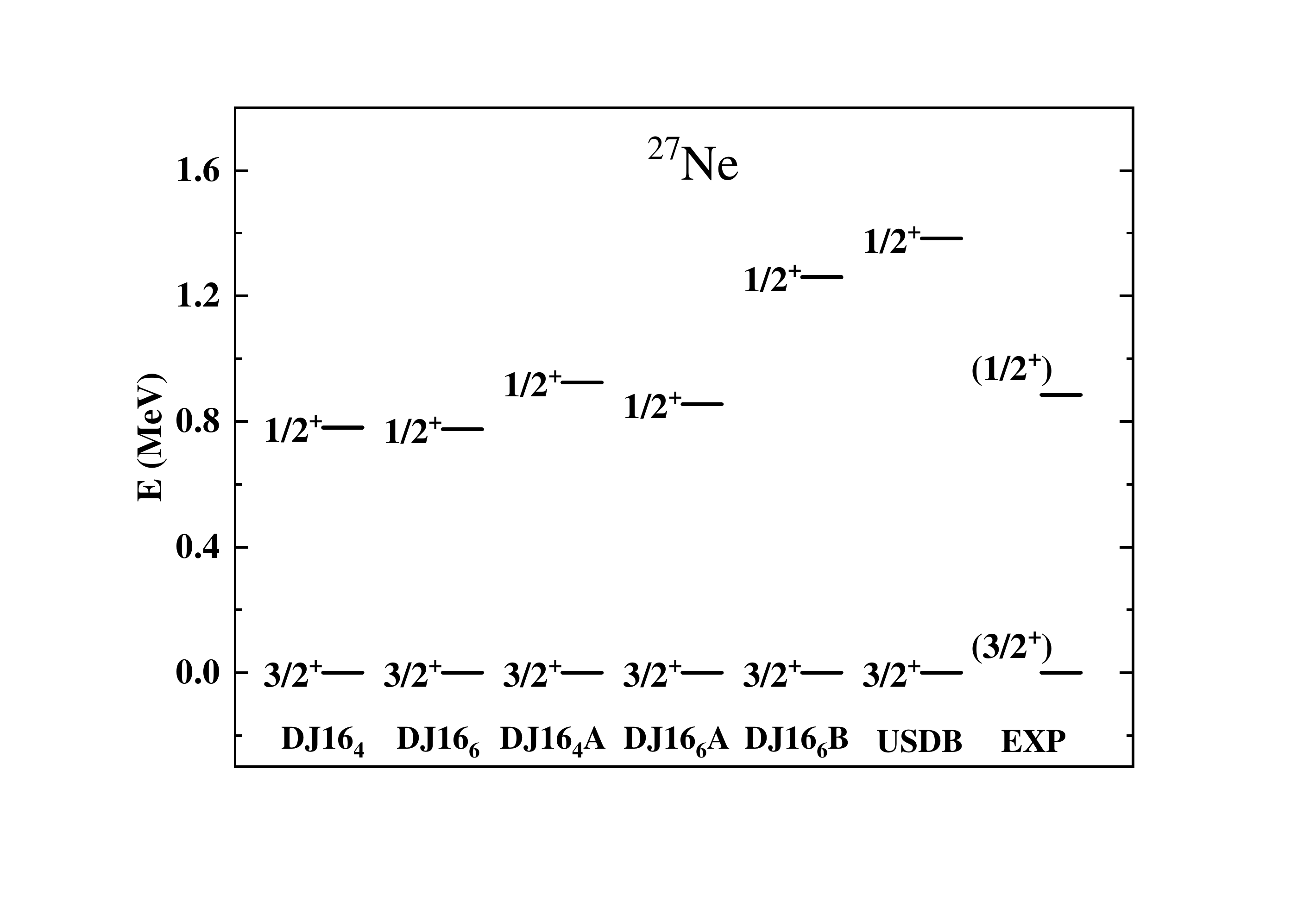}\\[-1.2cm]
  \includegraphics[width=.55\textwidth]{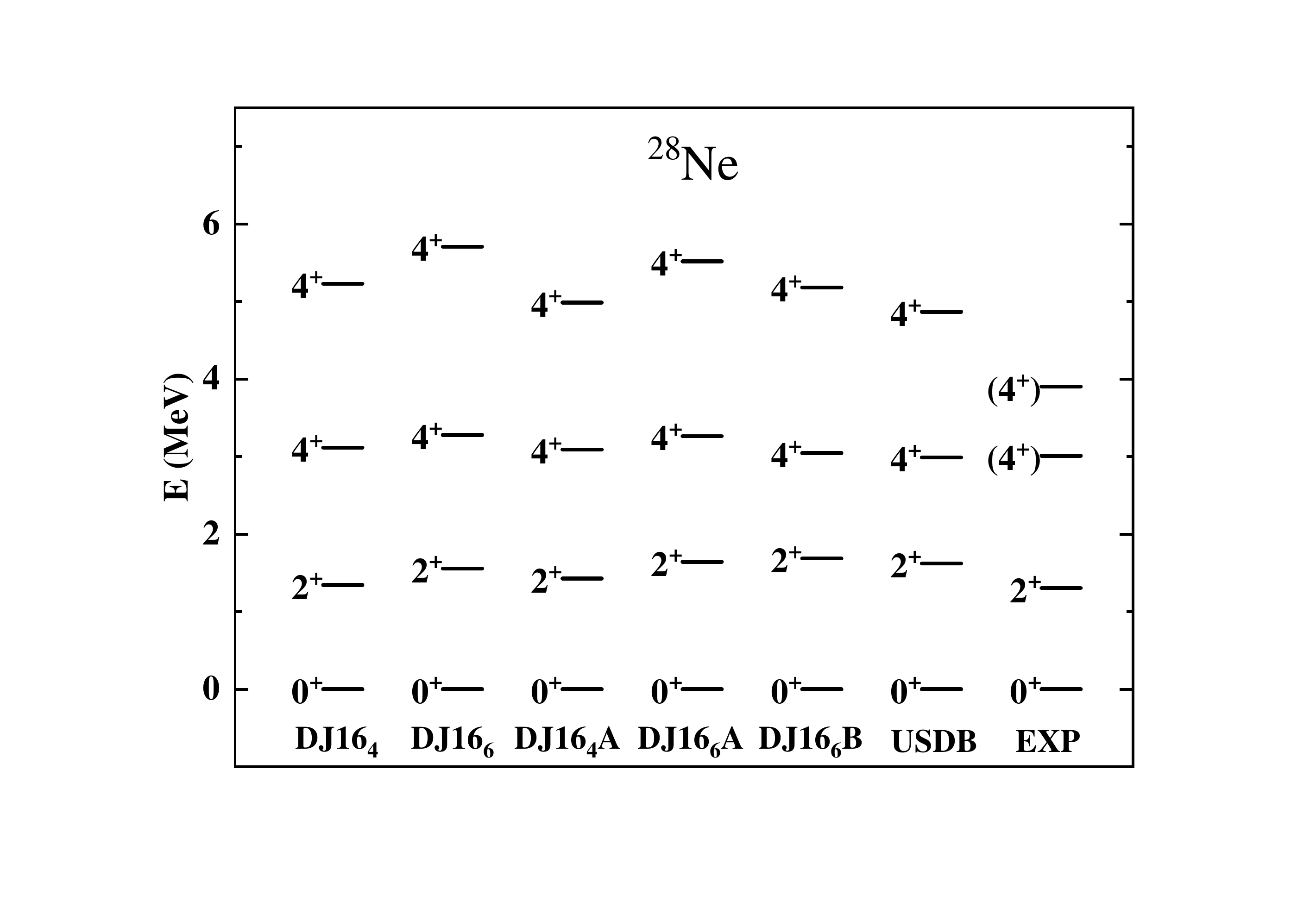}\\[-5mm]
  \caption{Theoretical and experimental low-energy spectra of $^{24-28}$Ne
(up to six lowest positive-parity states are presented, when known).}
\end{figure}

\begin{figure}
  
  \includegraphics[width=.55\textwidth]{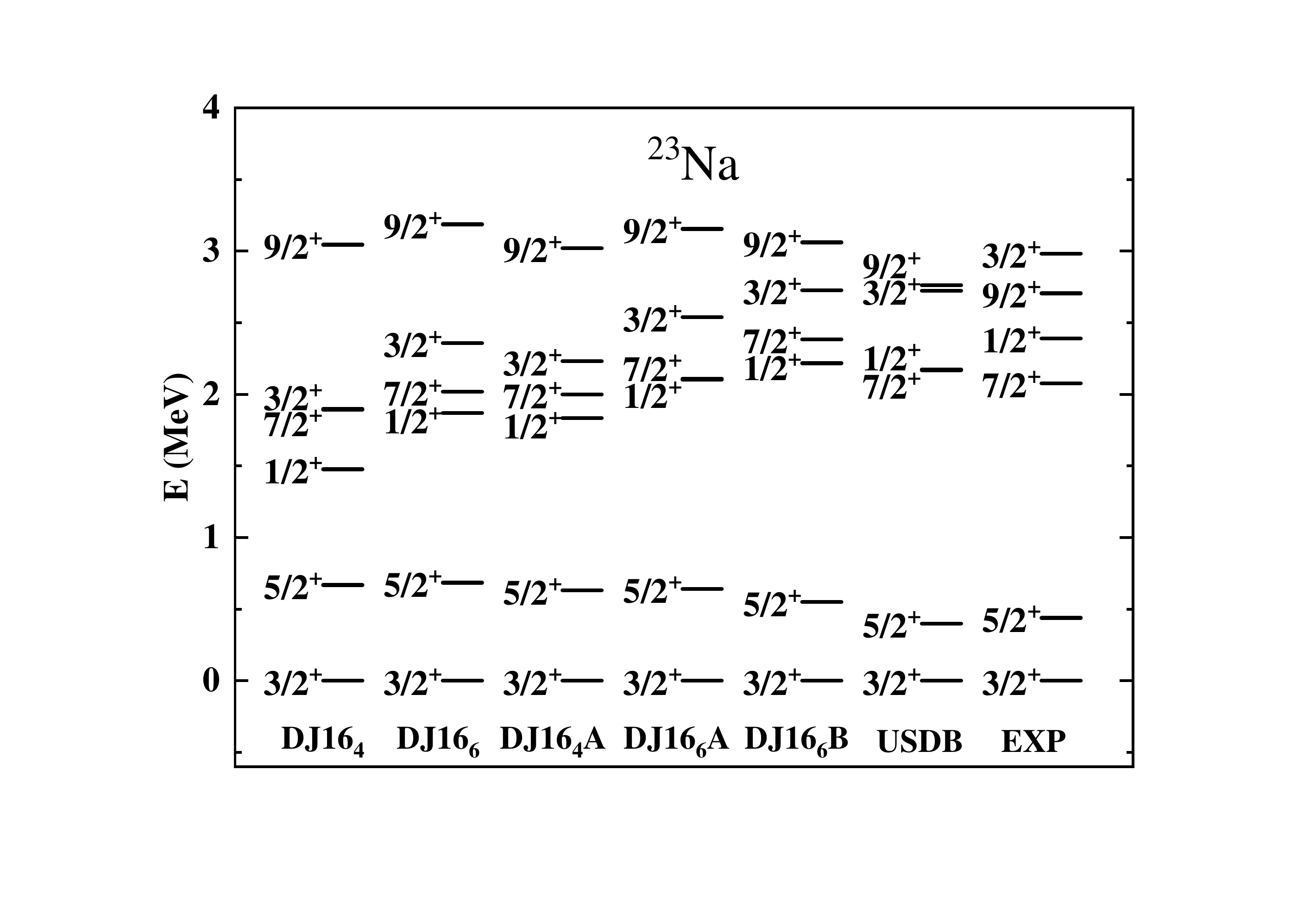} \hspace{-2cm}
  \includegraphics[width=.55\textwidth]{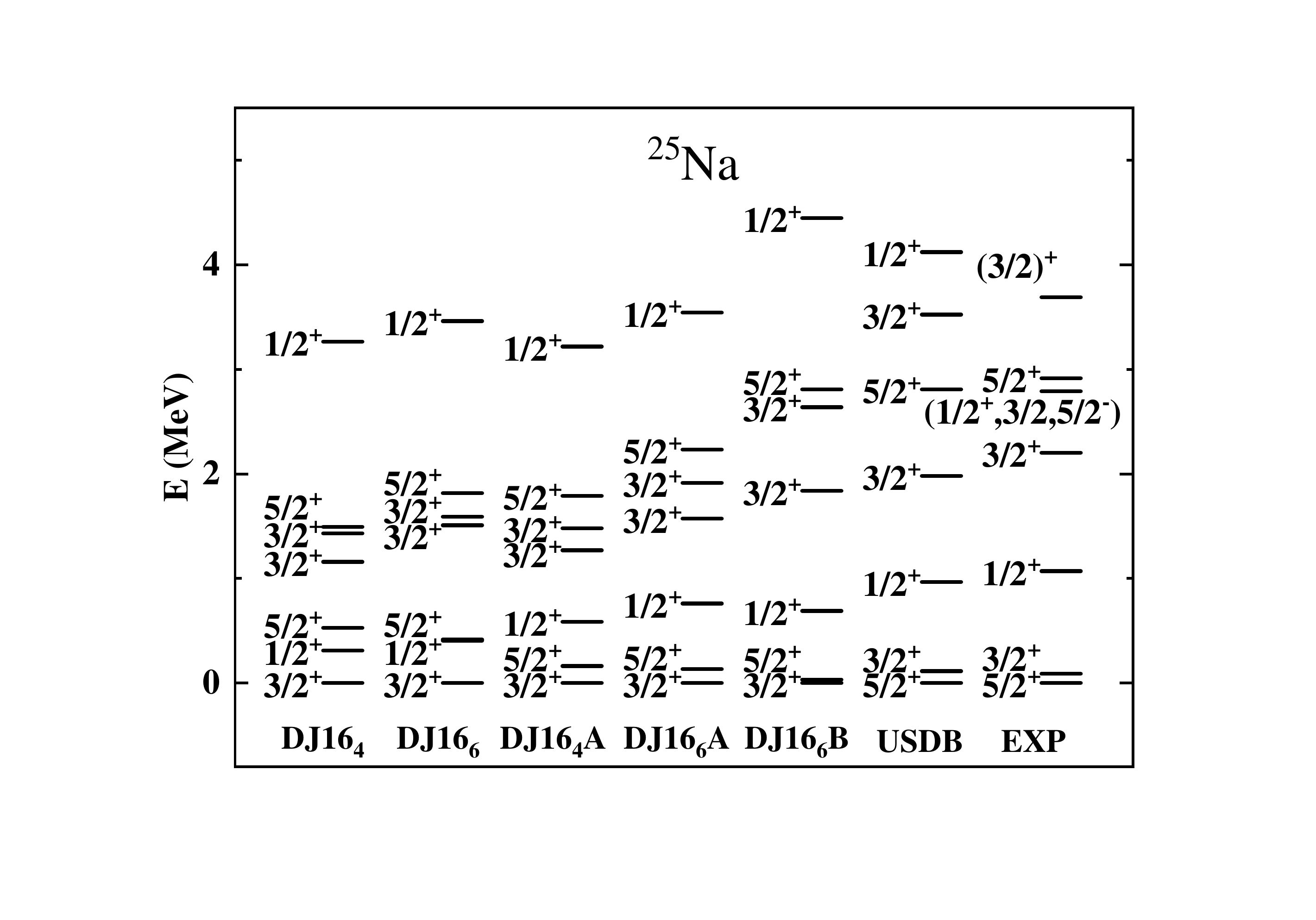}\\[-1.2cm]
  
  \includegraphics[width=.55\textwidth]{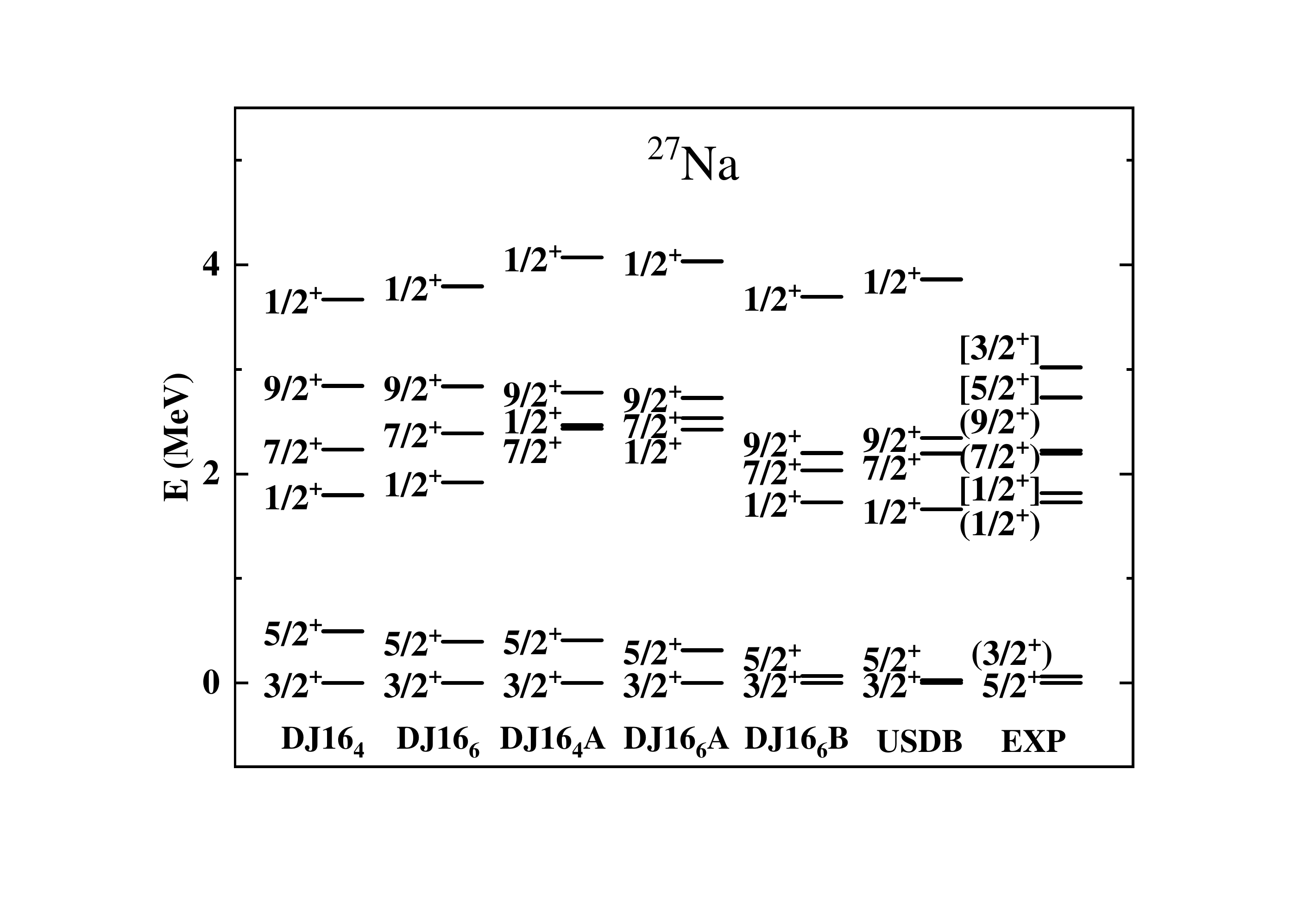} \hspace{-2cm}
  \includegraphics[width=.55\textwidth]{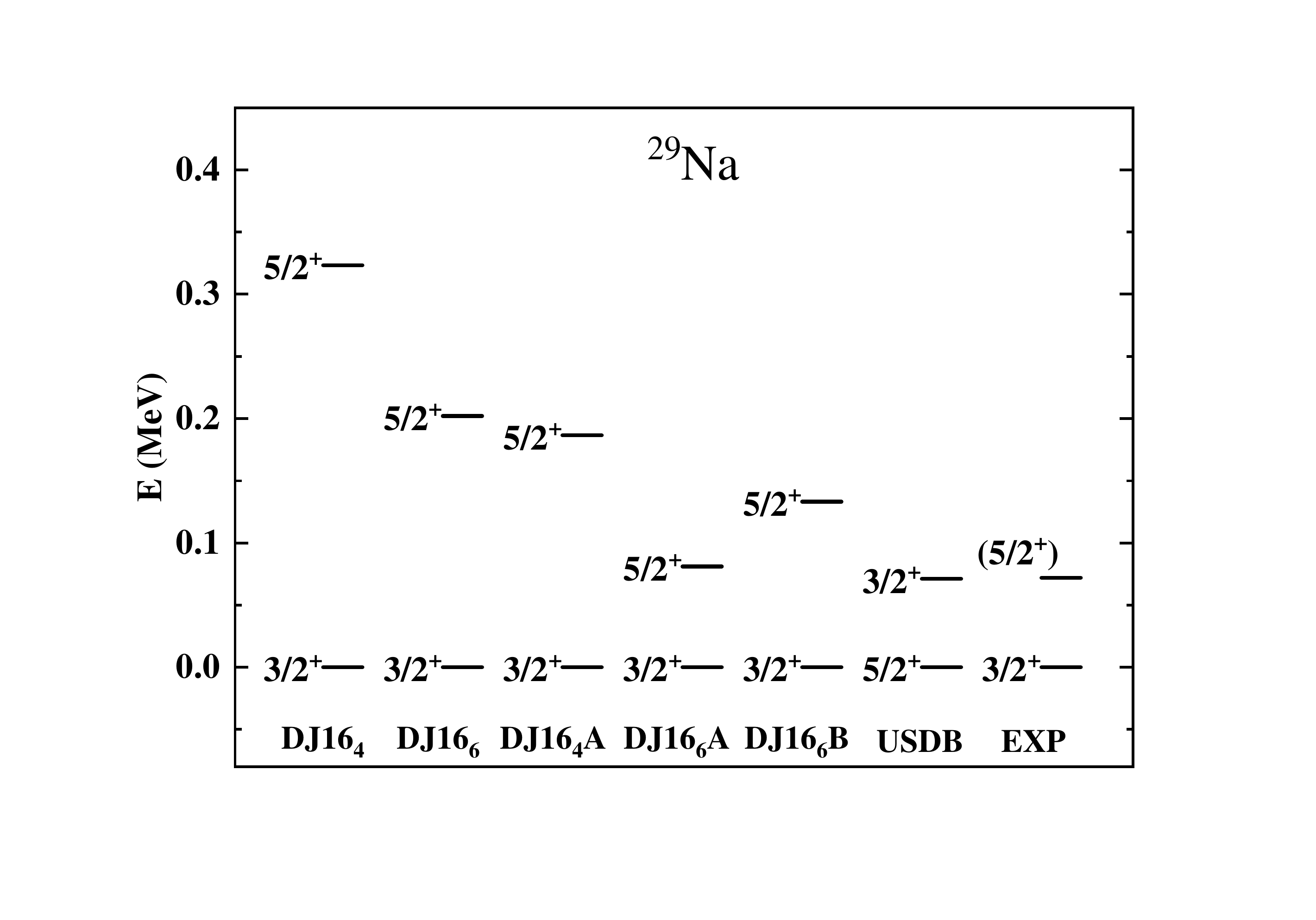}\\[-5mm]
  \caption{Theoretical and experimental low-energy spectra of odd-$A$ $^{23-29}$Na
(six for $^{23}$Na, seven for $^{25}$Na, eight for $^{27}$Na and two for $^{29}$Na lowest positive-parity states are presented).}
\end{figure}

\begin{figure}

  \includegraphics[width=.55\textwidth]{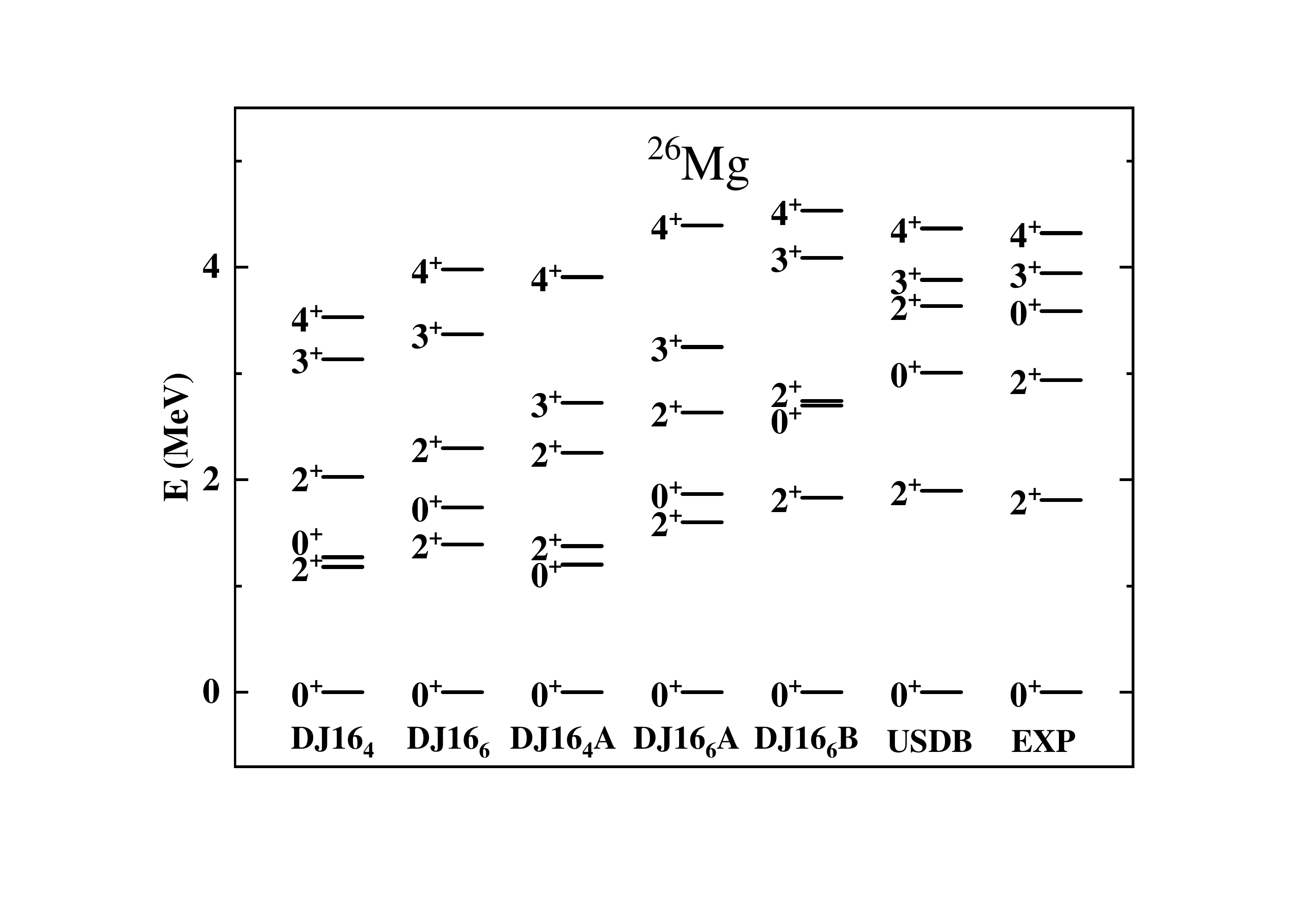} \hspace{-2cm}
  \includegraphics[width=.55\textwidth]{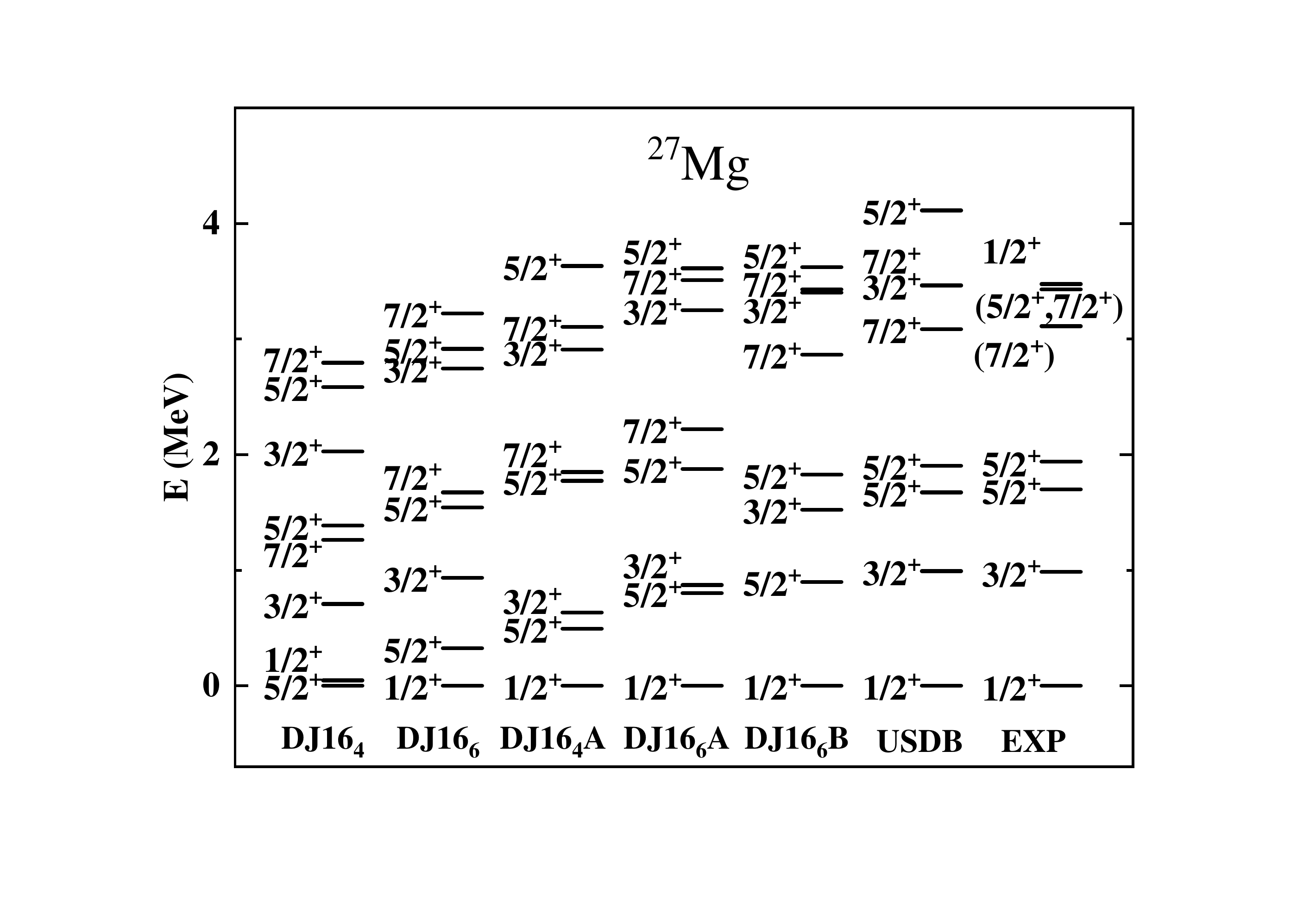}\\[-1.2cm]
  \includegraphics[width=.55\textwidth]{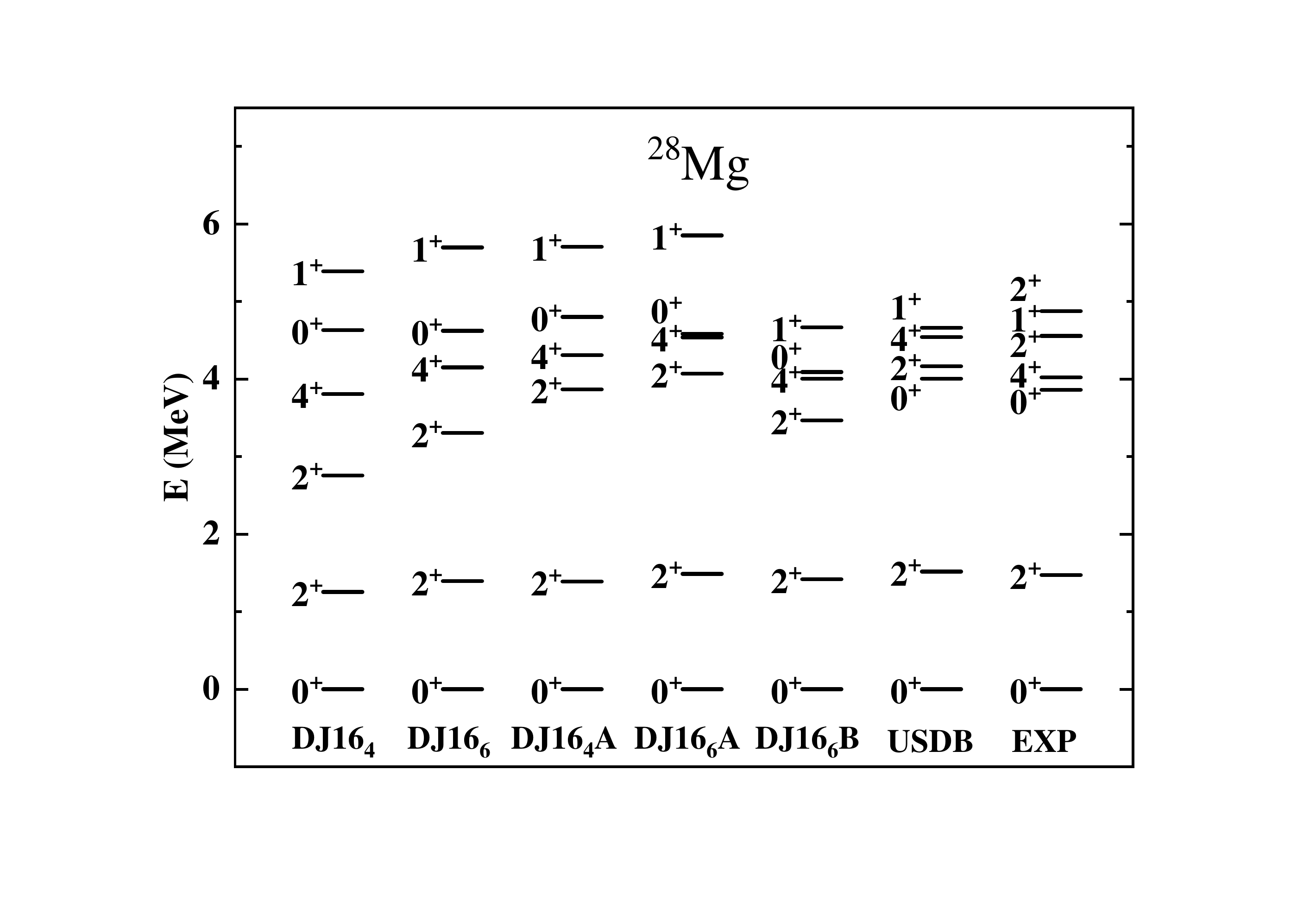} \hspace{-2cm}
  \includegraphics[width=.55\textwidth]{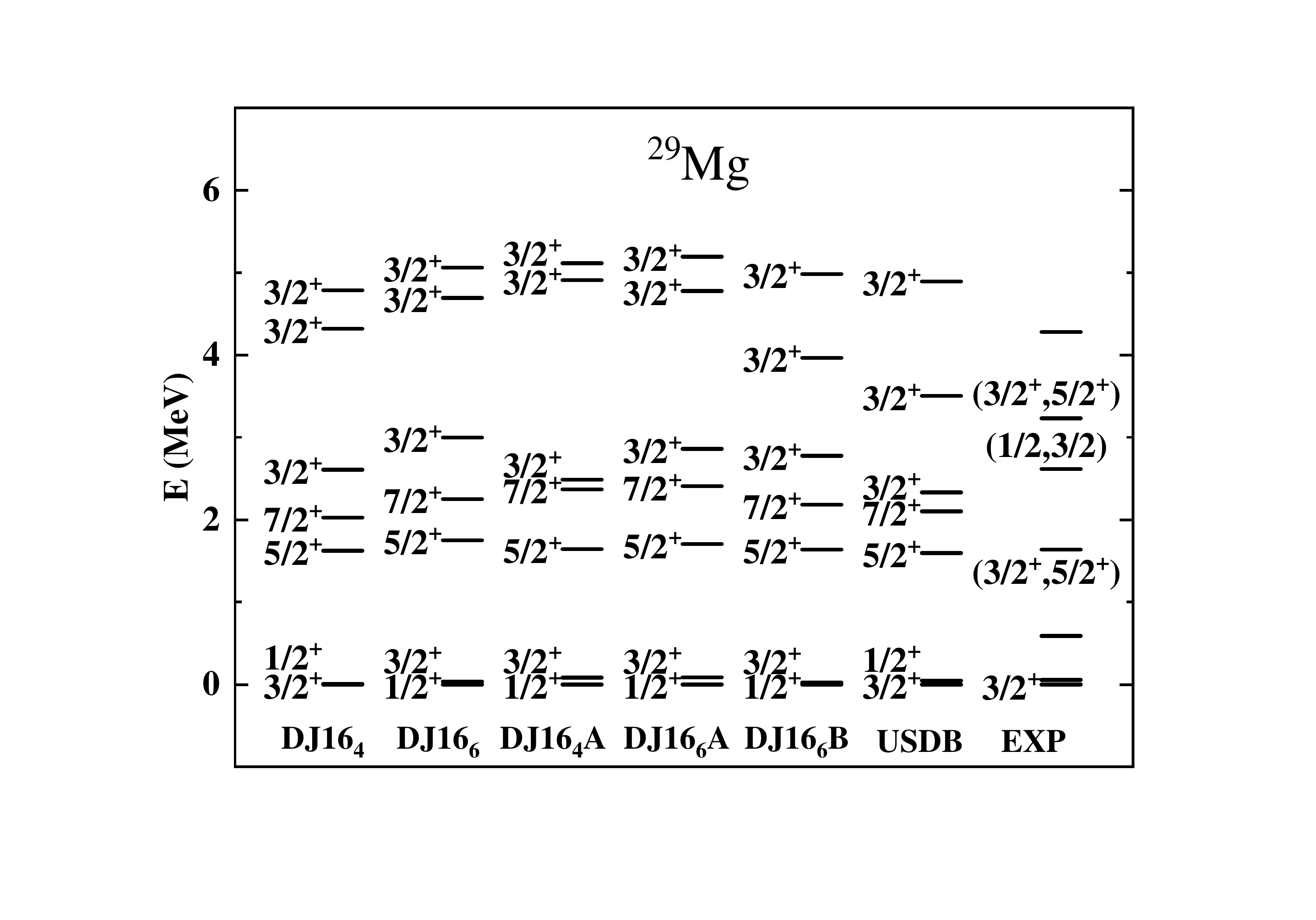}\\[-1.2cm]
  \includegraphics[width=.55\textwidth]{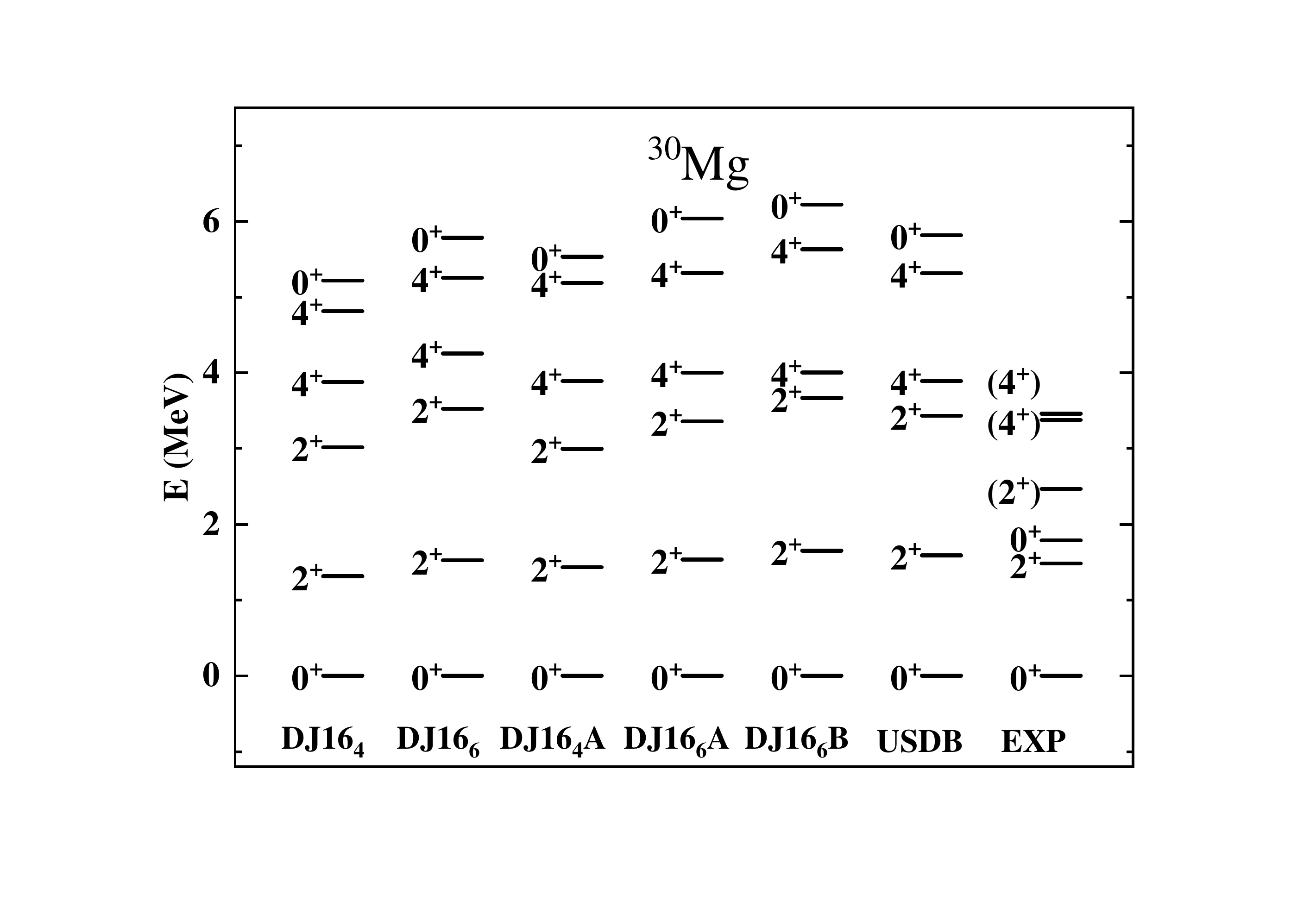}\\[-5mm]
  \caption{Theoretical and experimental low-energy spectra of $^{25-30}$Mg 
(six lowest positive-parity states are presented for all Mg isotopes, except for the experimental spectra of $^{27,29}$Mg 
where seven states are shown because of unknown spin and parity assignments).}
\end{figure}

\begin{figure}[tbh]
 
  \includegraphics[width=.55\textwidth]{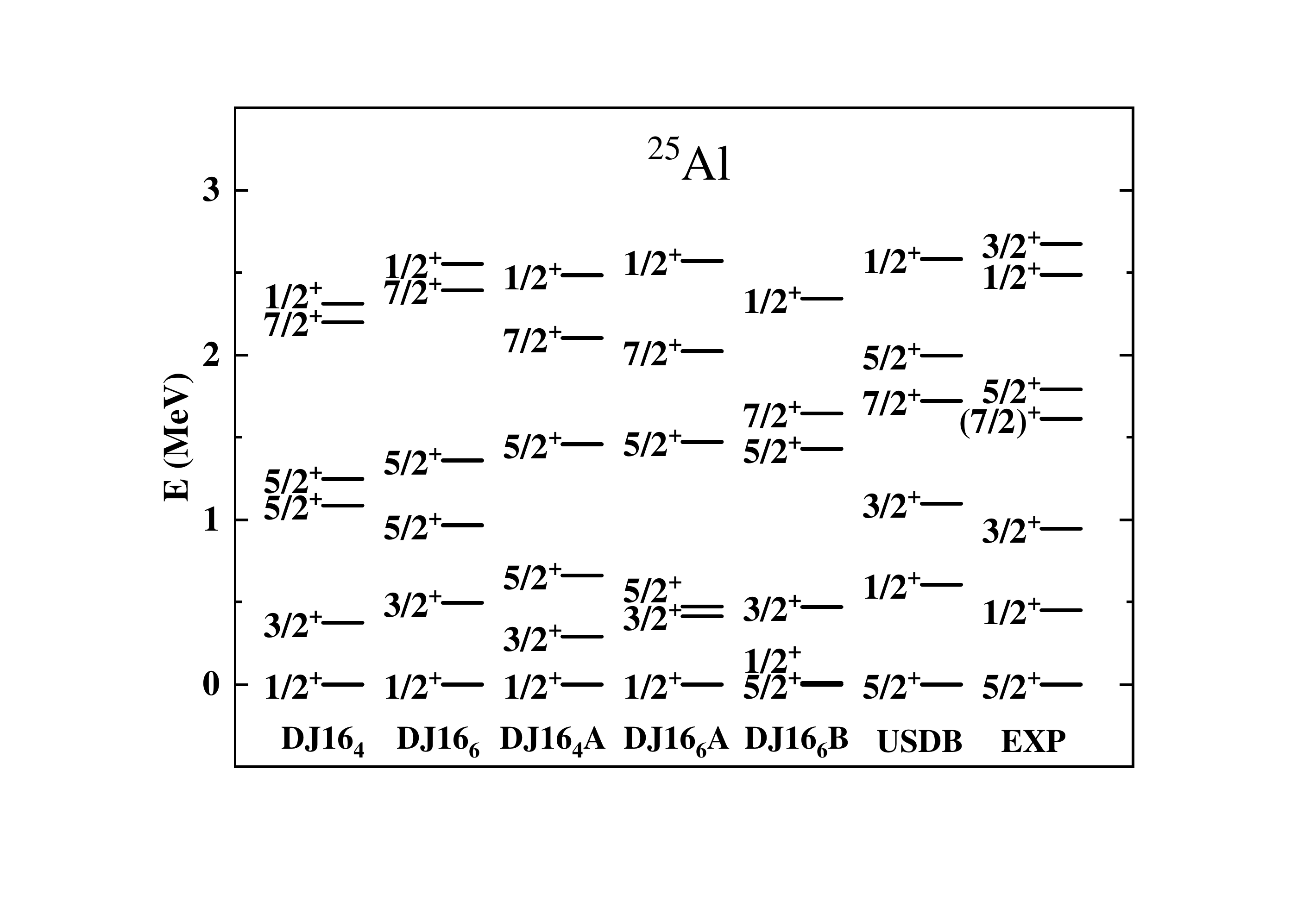}\hspace{-2cm}
  \includegraphics[width=.55\textwidth]{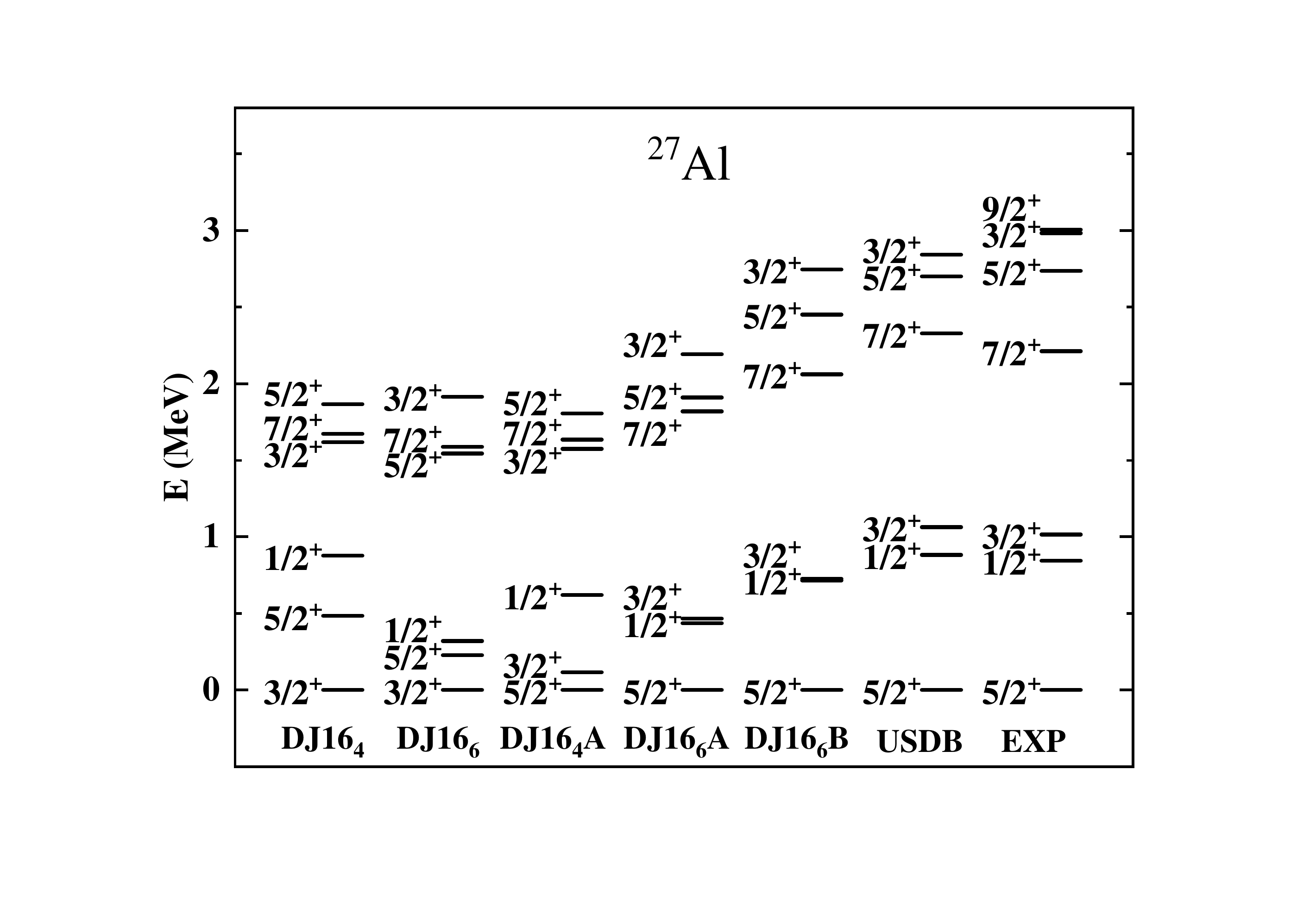}\\[-1.2cm]
  \includegraphics[width=.55\textwidth]{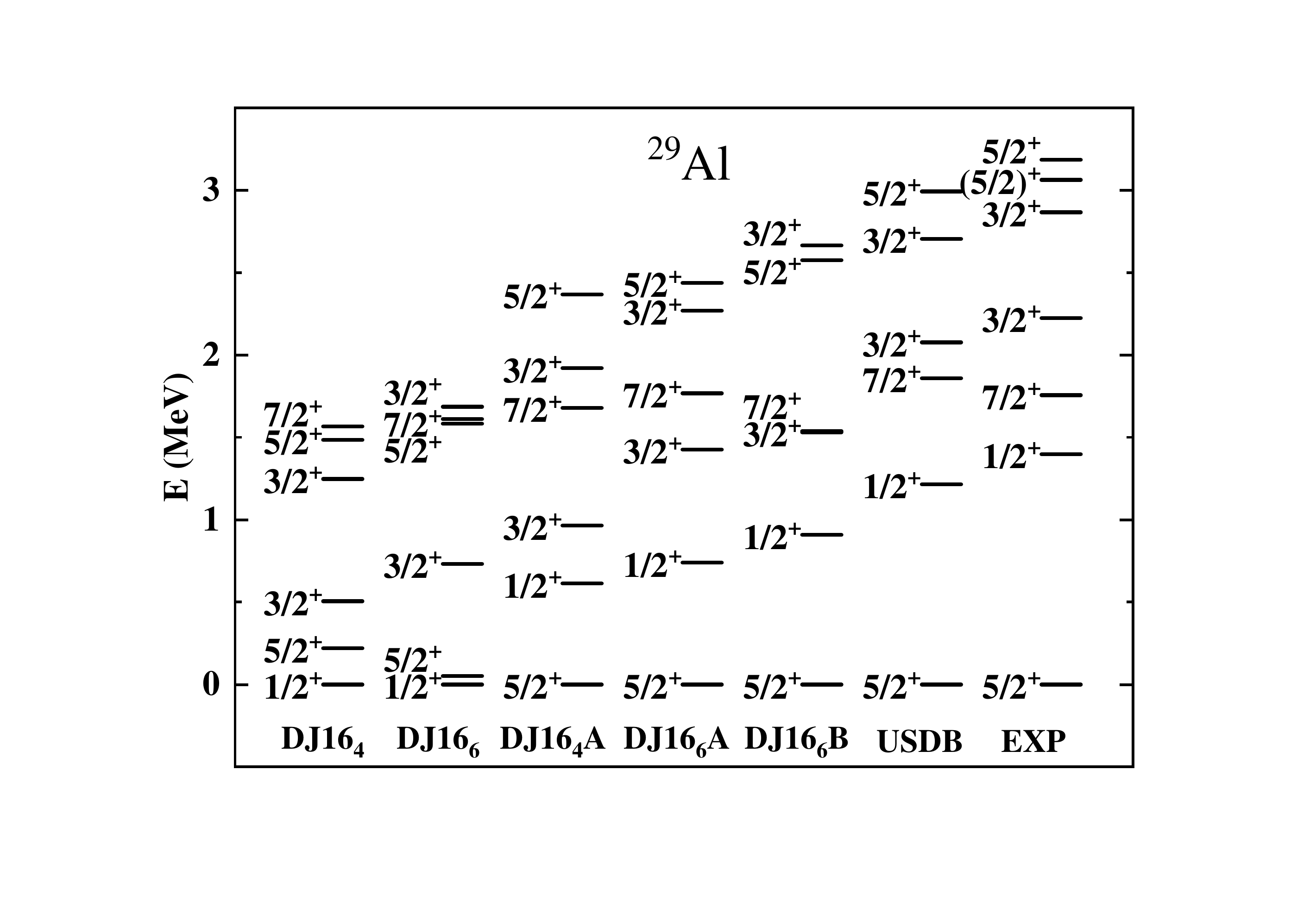} \\[-5mm]
  \caption{Theoretical and experimental low-energy spectra of $^{25,27,29}$Al 
(seven lowest positive-parity states are presented).}
\end{figure}

\begin{figure}[tbh]
 
  \includegraphics[width=.55\textwidth]{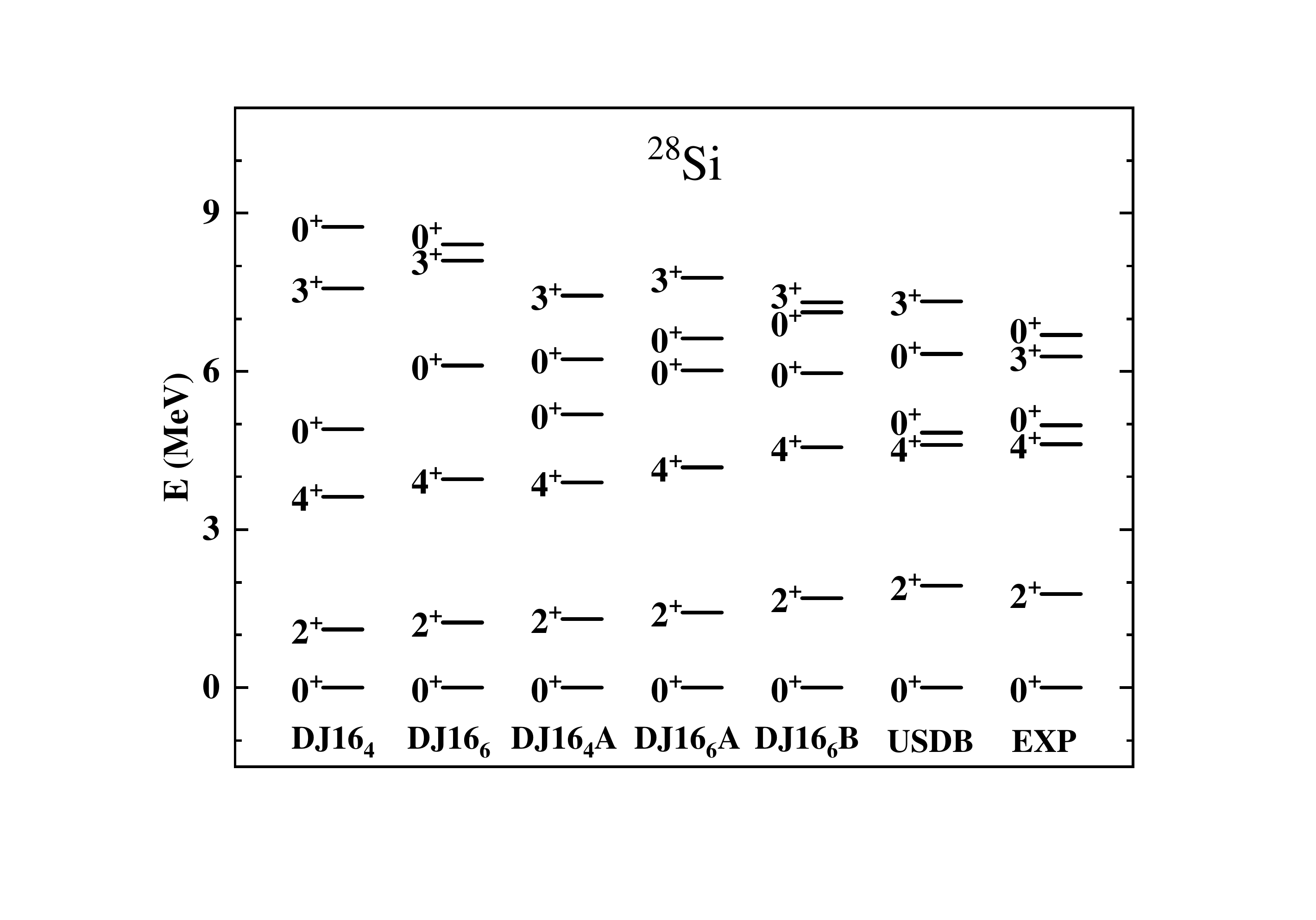}\hspace{-2cm}
  \includegraphics[width=.55\textwidth]{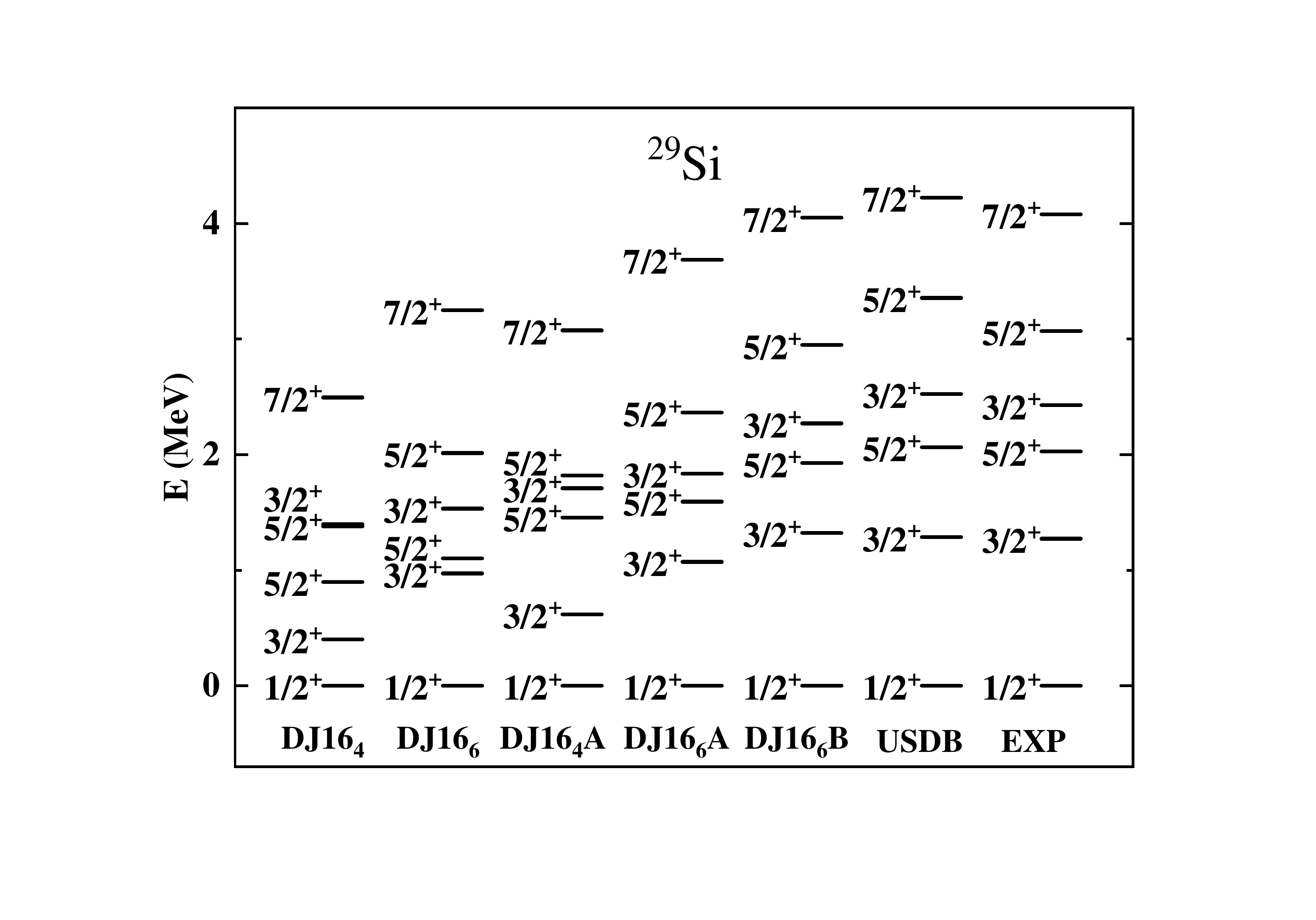}\\[-1.2cm]
 
  \includegraphics[width=.55\textwidth]{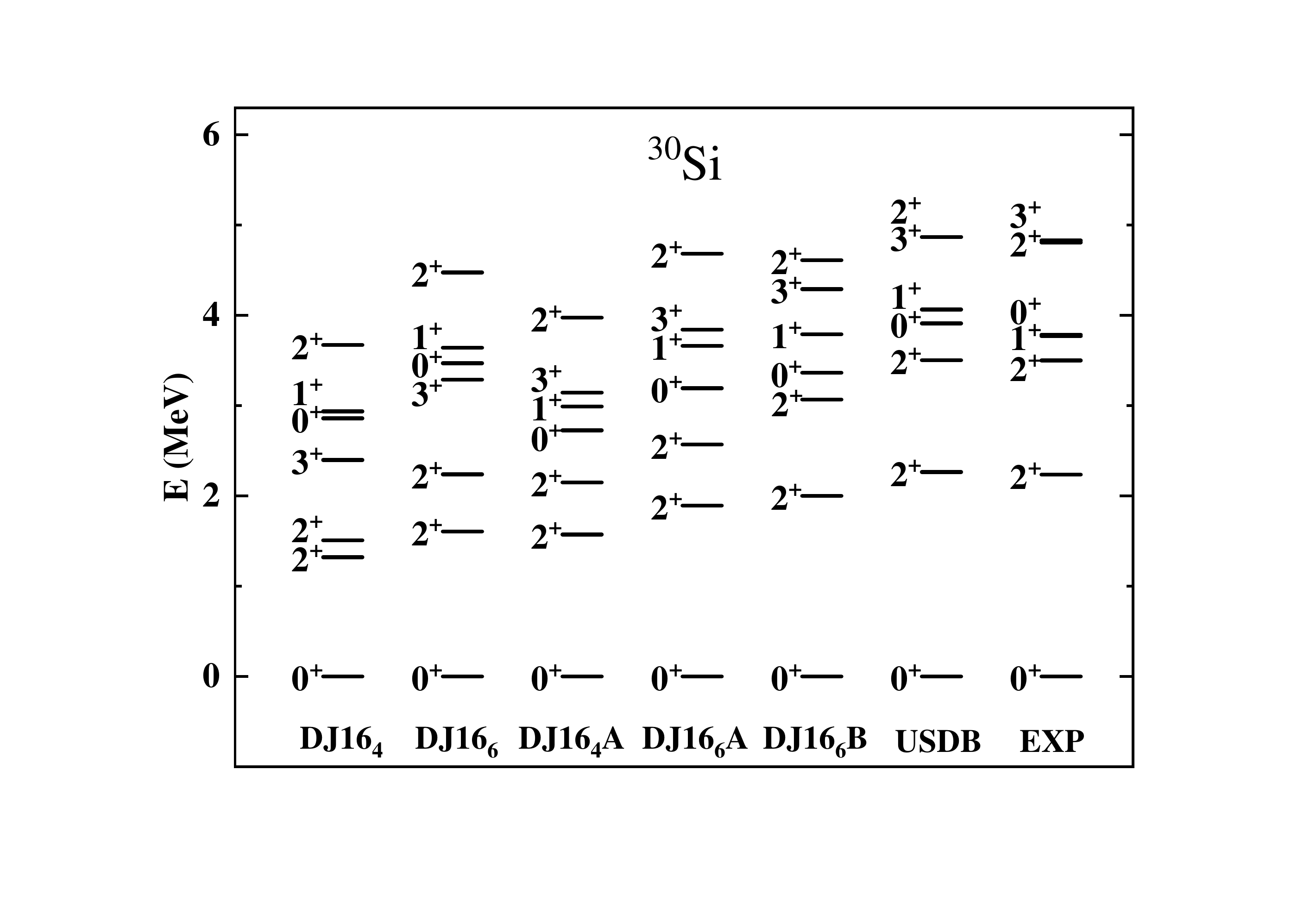}\hspace{-2cm}
  \includegraphics[width=.55\textwidth]{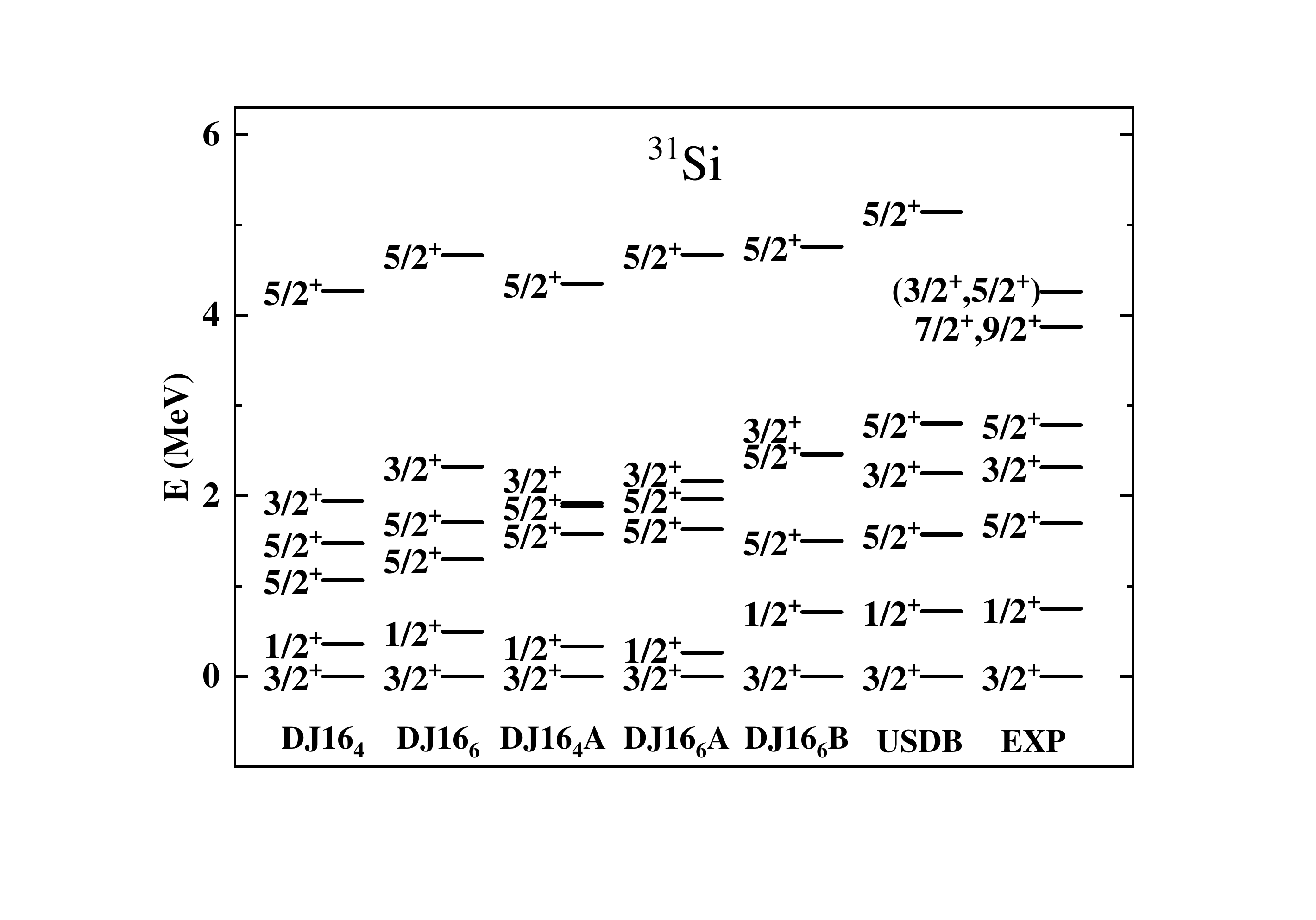}\\[-1.2cm]
  \includegraphics[width=.55\textwidth]{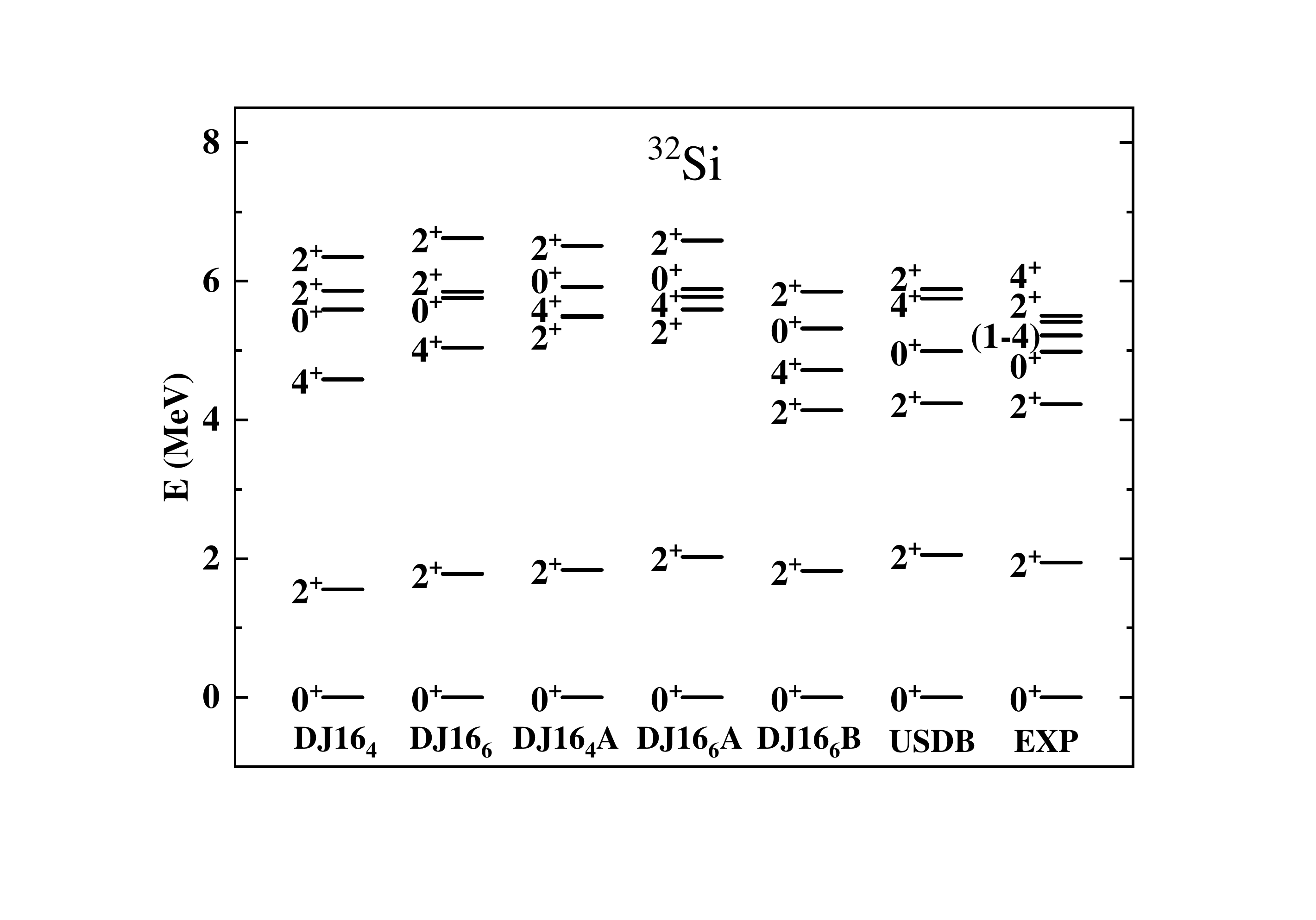}\\[-5mm]
  \caption{Theoretical and experimental low-energy spectra of $^{28-32}$Si 
(six lowest positive-parity states are presented, except for the experimental spectra of $^{31,32}$Si where
seven lowest positive parity states are shown, and the spectra of $^{28}$Si where eight lowest positive-parity states are shown.}
\end{figure}

\begin{figure}[tbh]
 
  \includegraphics[width=.55\textwidth]{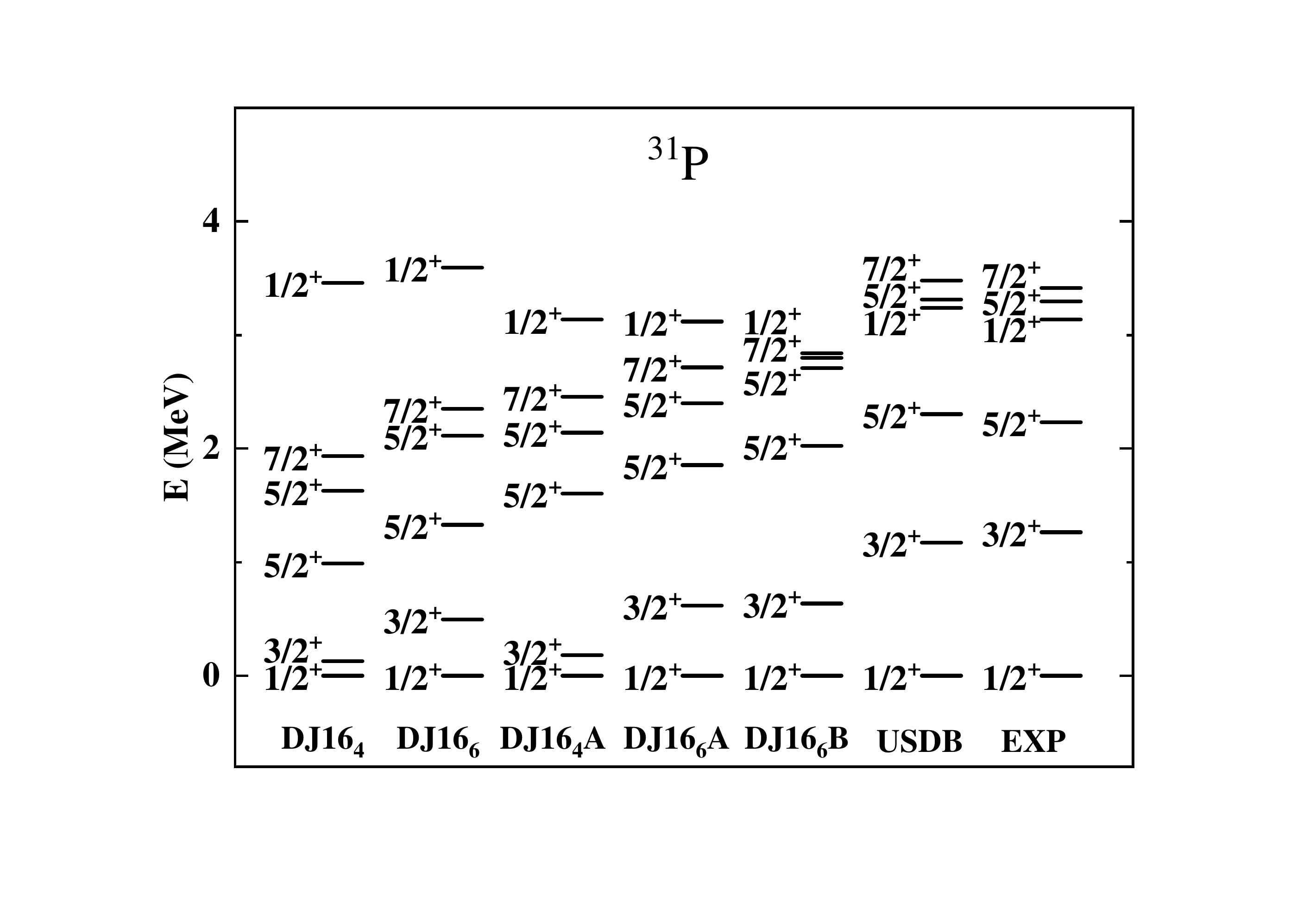}\hspace{-2cm}
  \includegraphics[width=.55\textwidth]{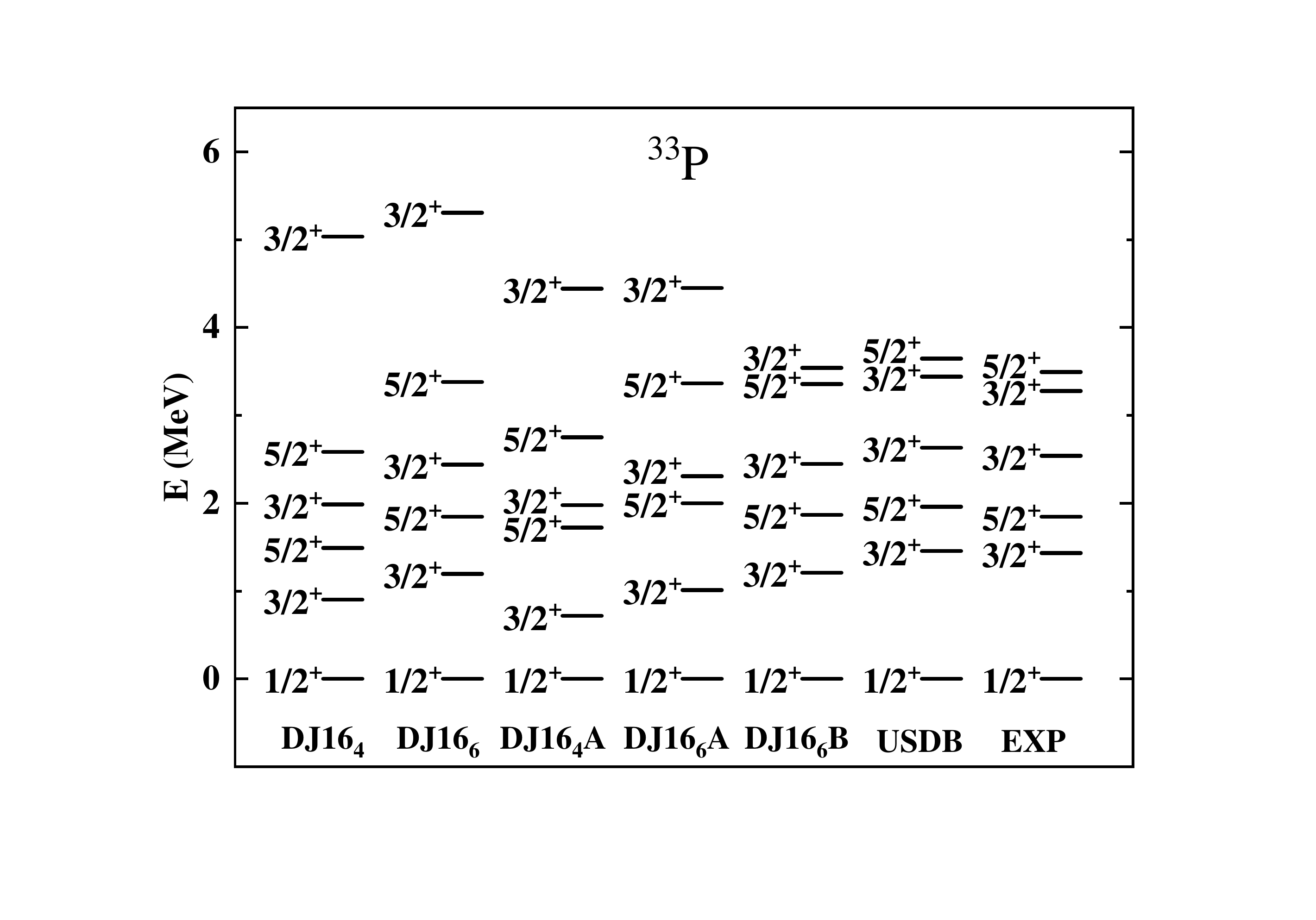}\\[-1.2cm]
  \includegraphics[width=.55\textwidth]{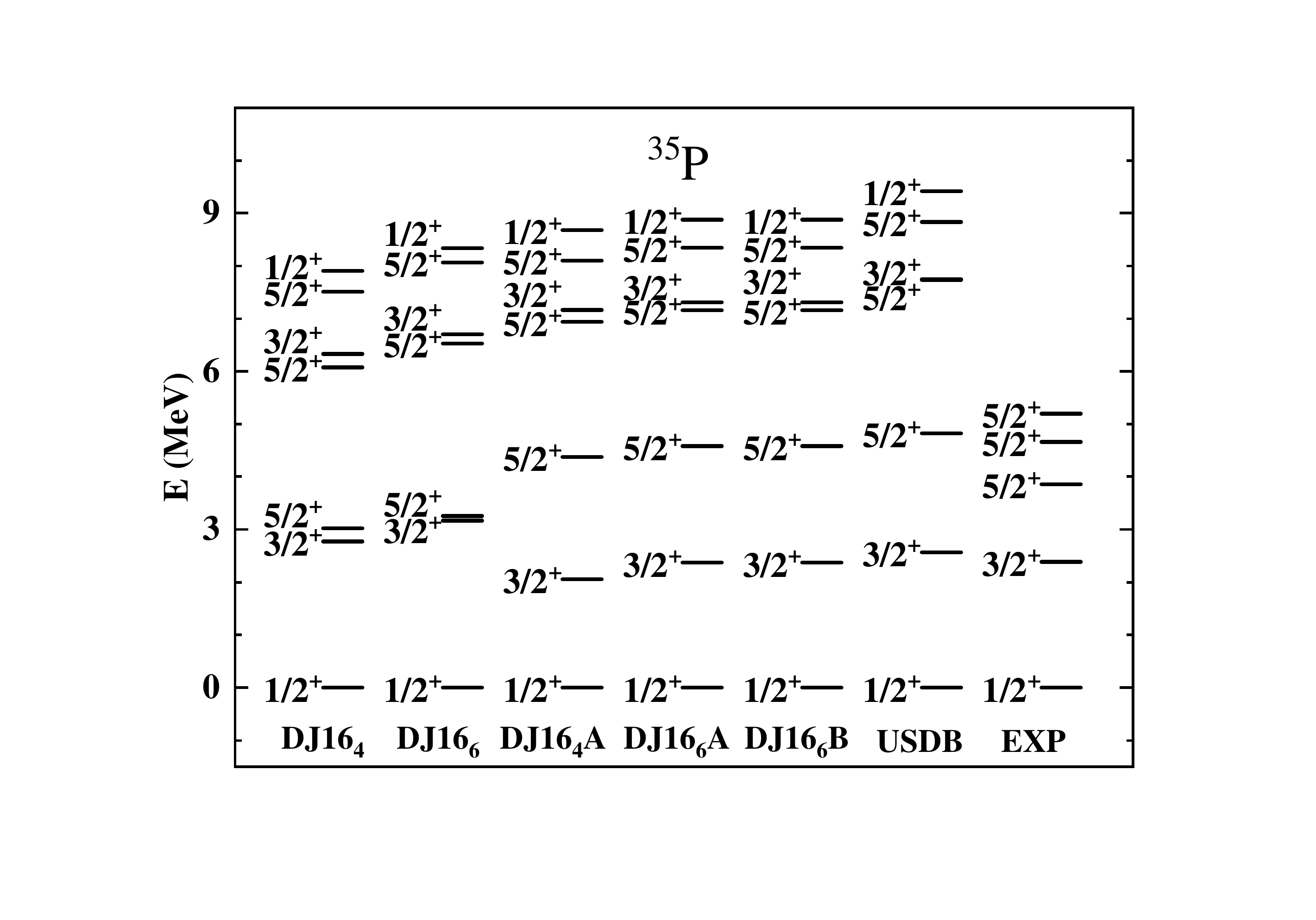}\\[-5mm]
  \caption{Theoretical and experimental low-energy spectra of odd-$A$ $^{31,33,35}$P 
(six lowest positive-parity states are presented for $^{31,33}$P and five states for $^{35}$P).}
\end{figure}

\begin{figure}[tbh]
 
  \includegraphics[width=.55\textwidth]{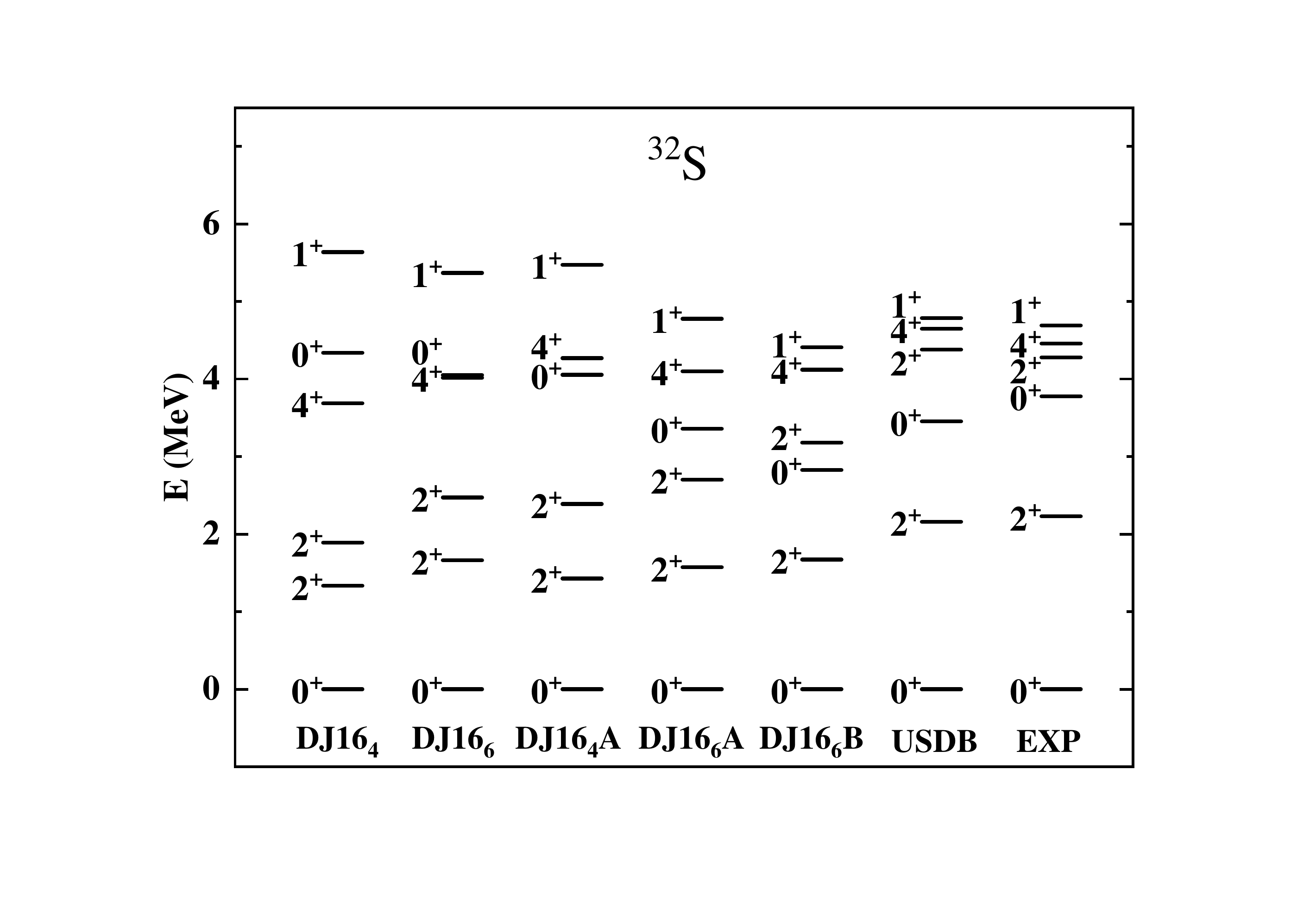}\hspace{-2cm}
  \includegraphics[width=.55\textwidth]{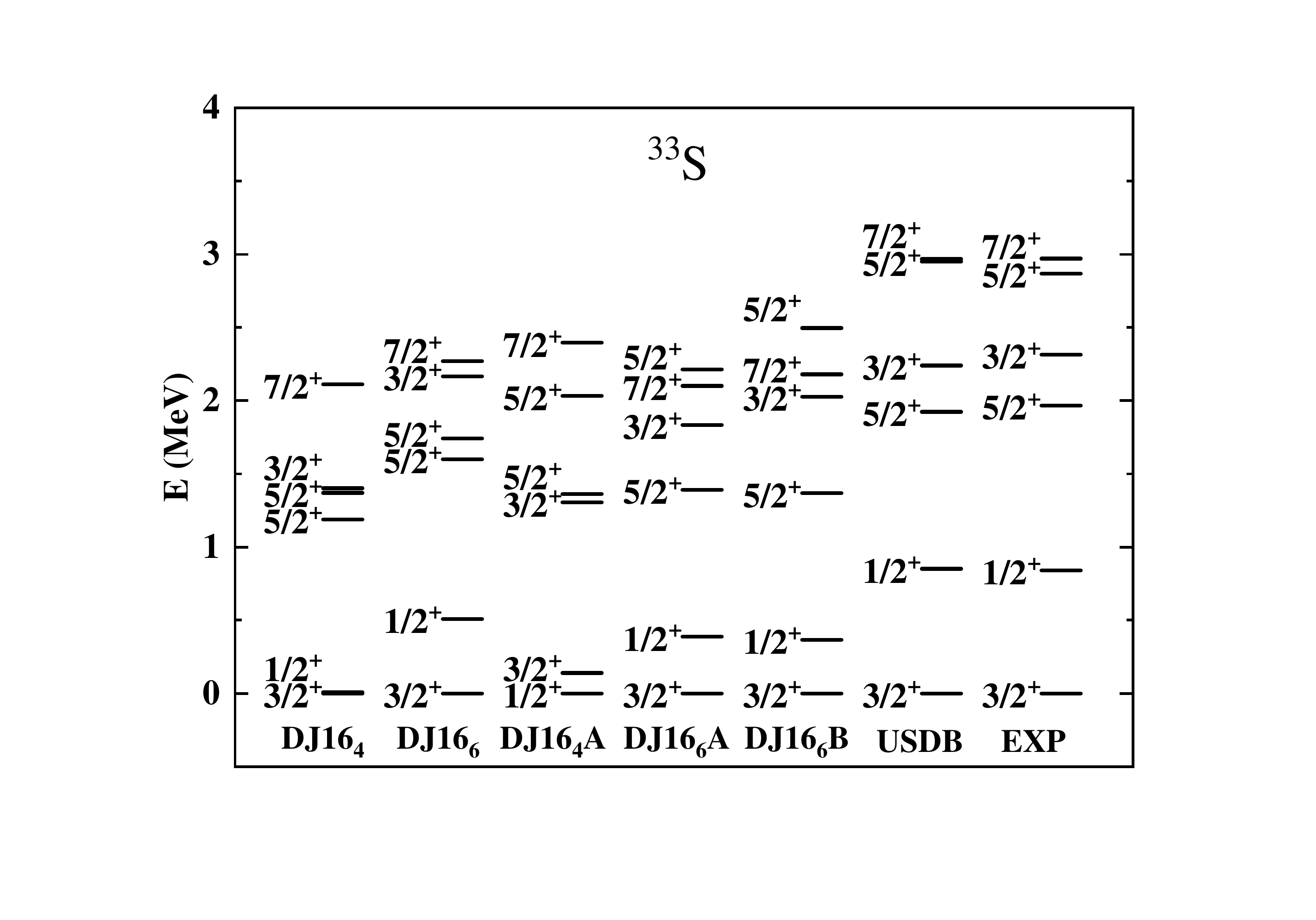}\\[-1.2cm]
 
  \includegraphics[width=.55\textwidth]{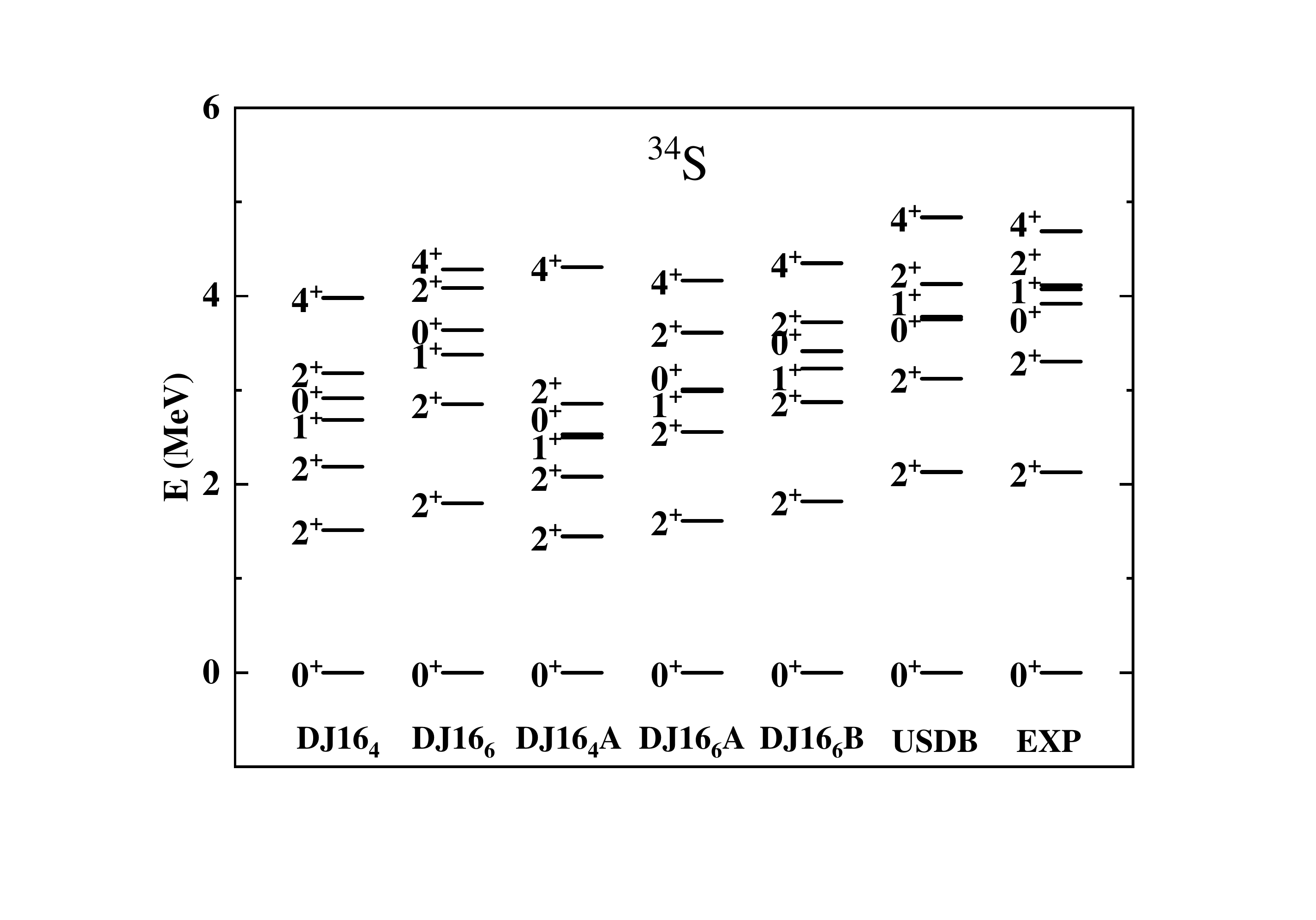}\hspace{-2cm}
  \includegraphics[width=.55\textwidth]{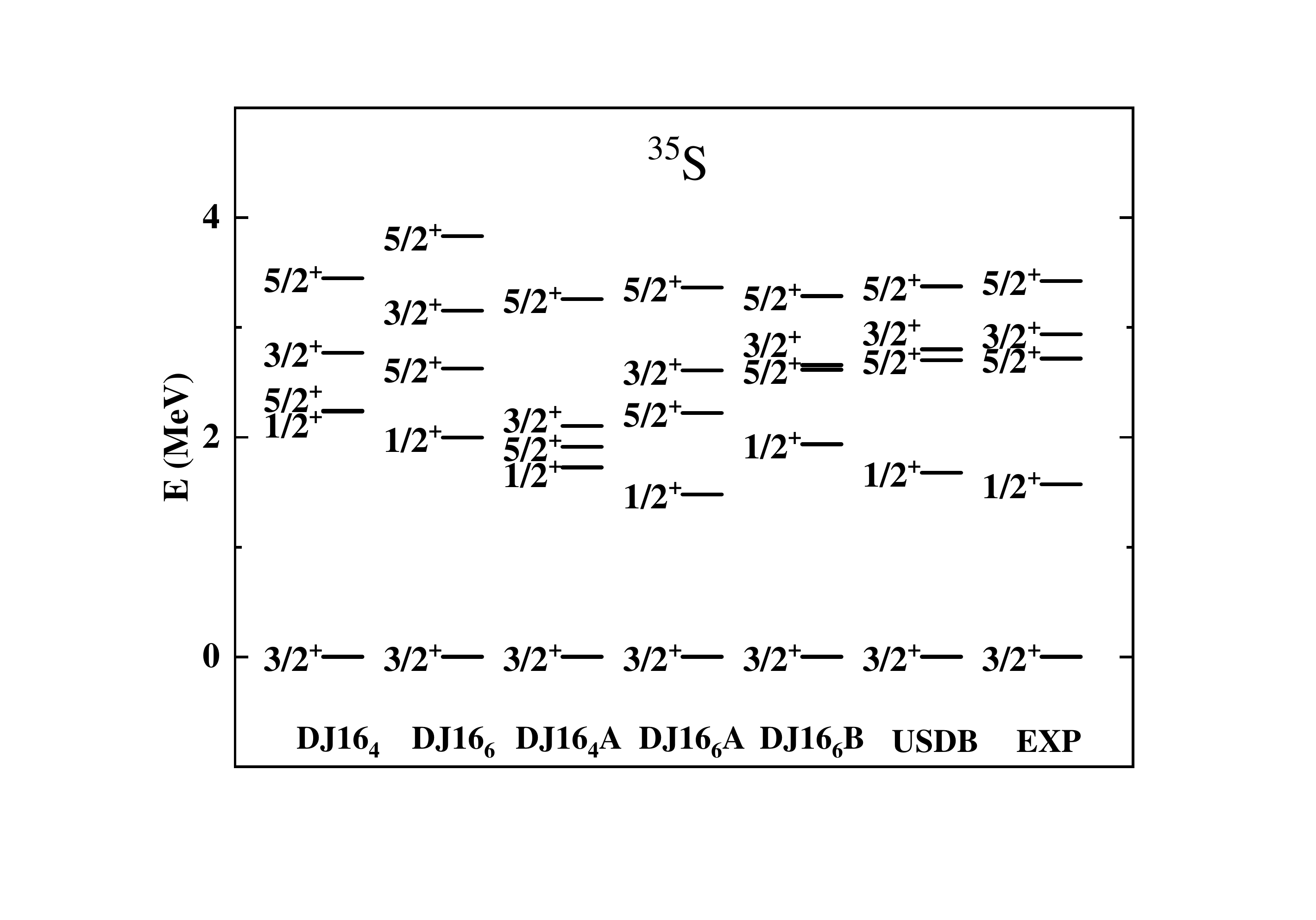}\\[-1.2cm]
  \includegraphics[width=.55\textwidth]{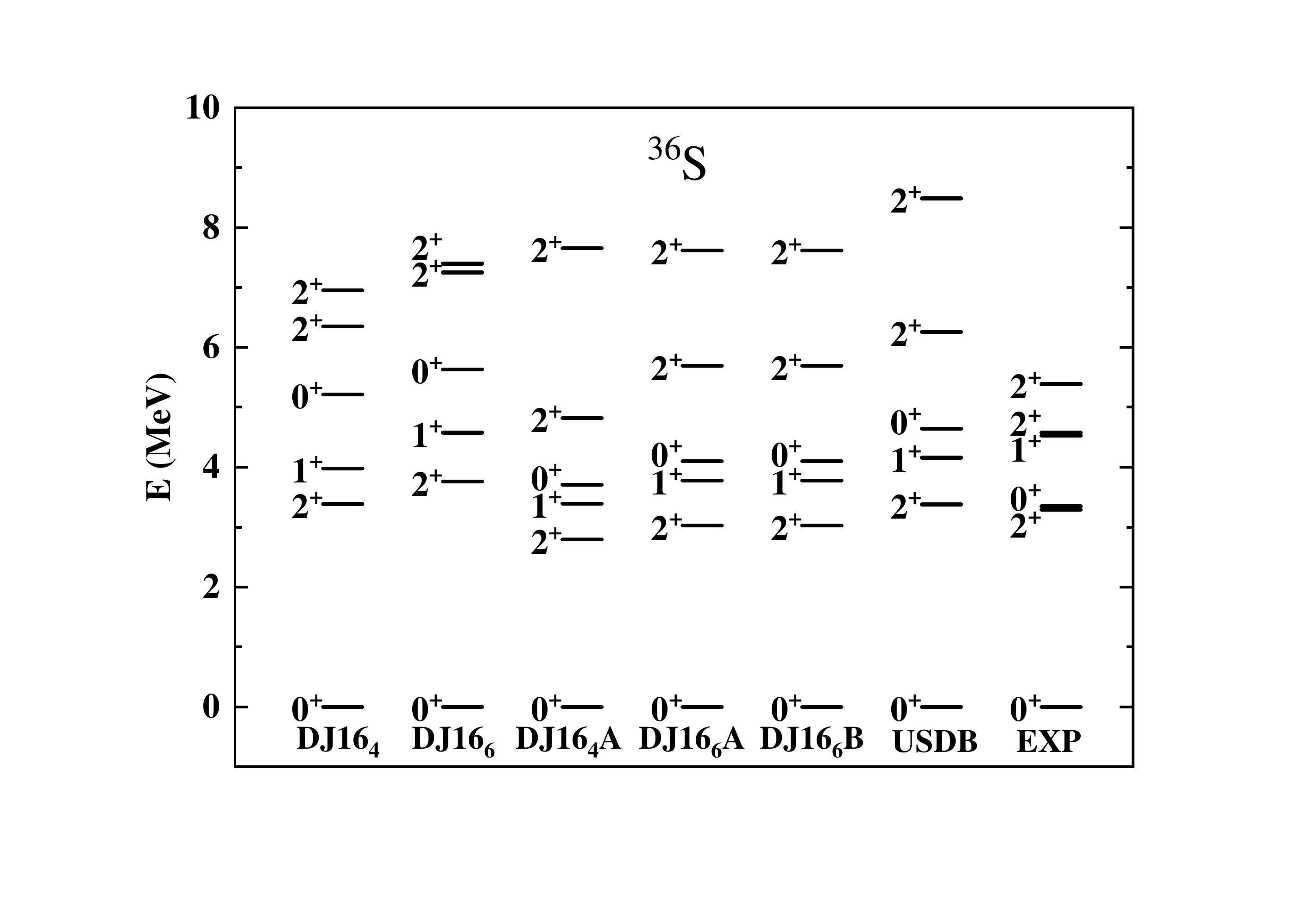}\\[-5mm]
  \caption{Theoretical and experimental low-energy spectra of $^{32-36}$S
(six or seven lowest positive-parity states are presented).}
\end{figure}

\begin{figure}
 
  \includegraphics[width=.55\textwidth]{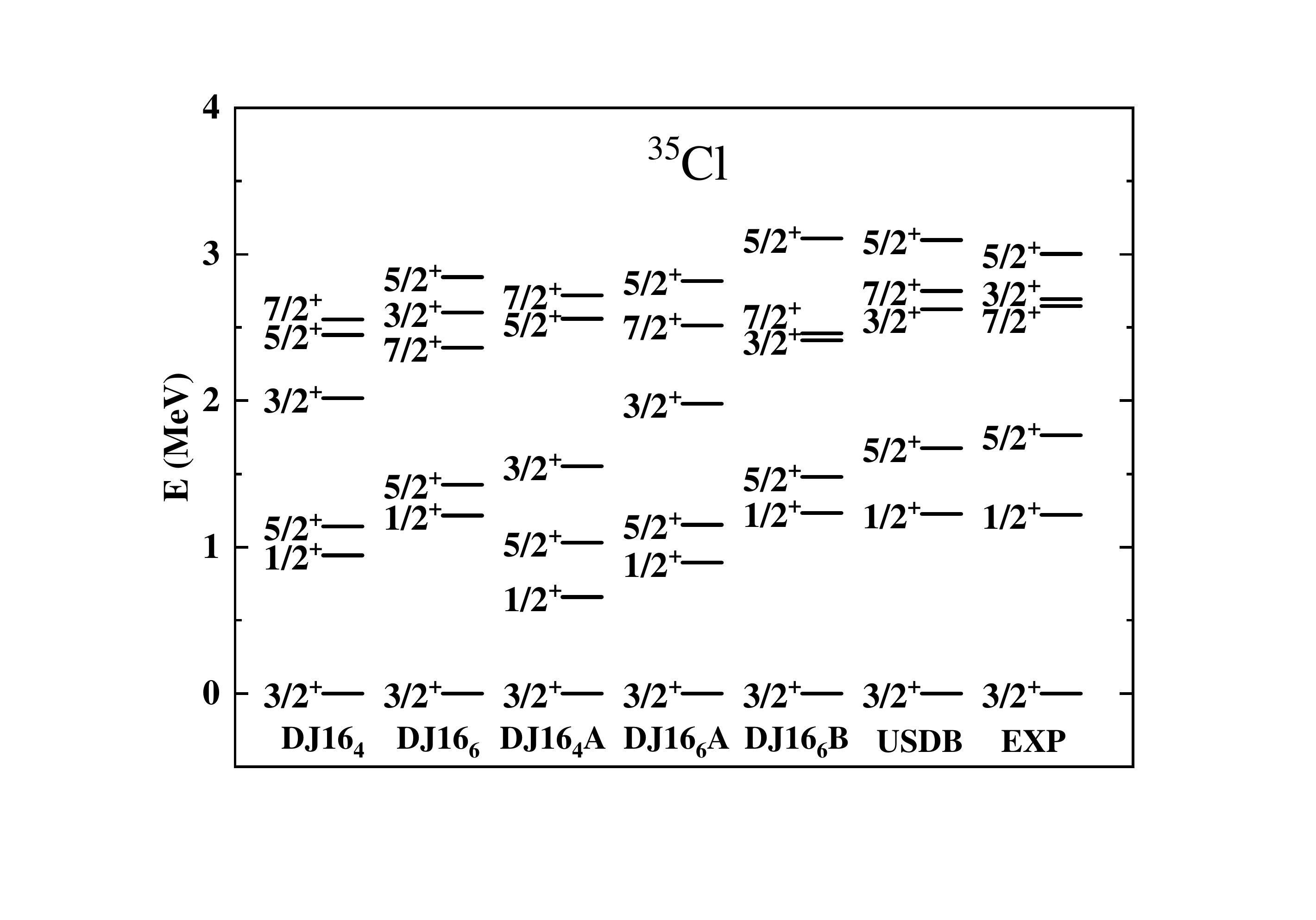}\hspace{-2cm}
  \includegraphics[width=.55\textwidth]{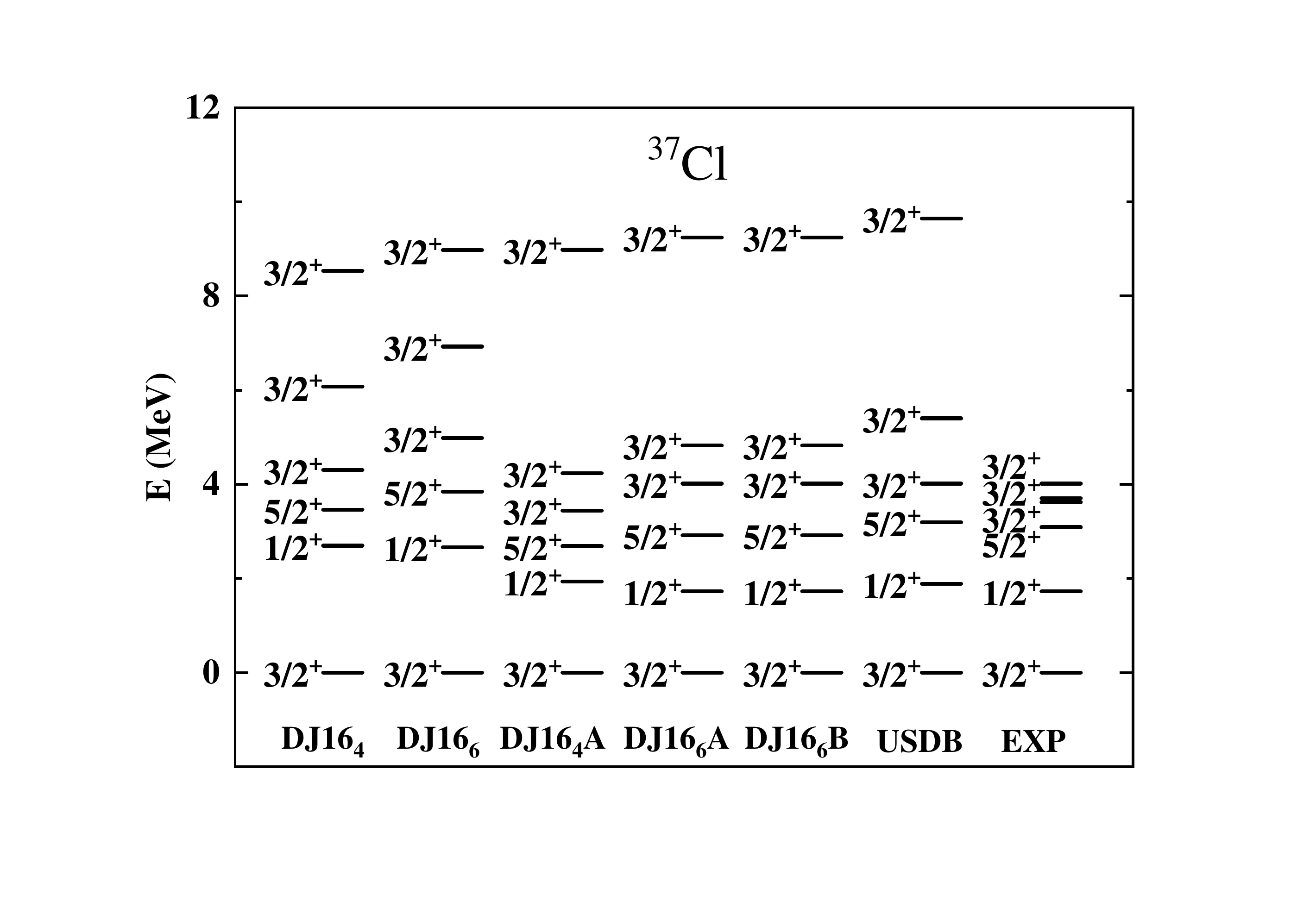}\\[-5mm]
  \caption{ Theoretical and experimental low-energy spectra of odd-$A$ $^{35-37}$Cl 
(six lowest positive-parity states are presented).}
\end{figure}

\begin{figure}
  \includegraphics[width=.55\textwidth]{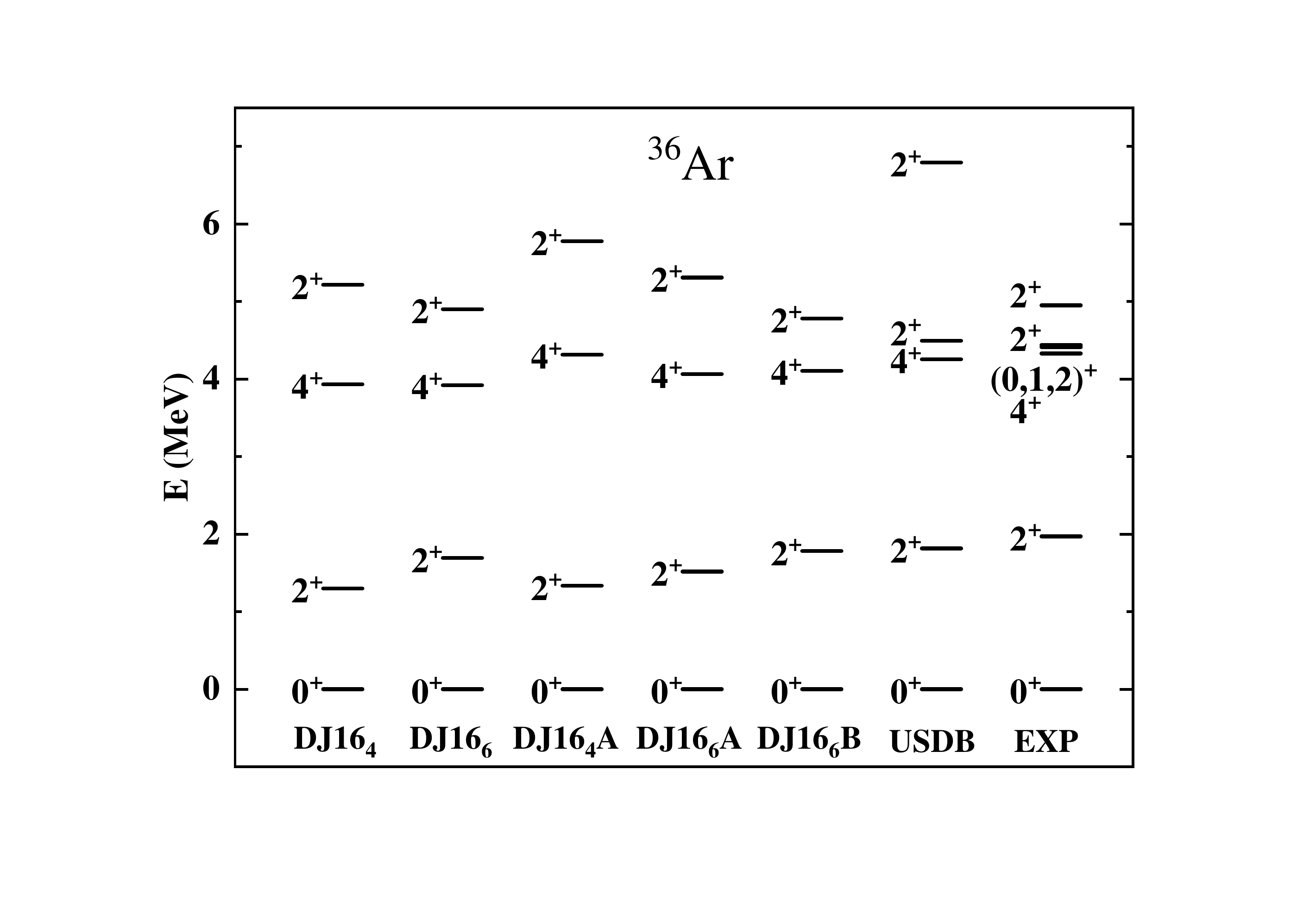}\hspace{-2cm}
  \includegraphics[width=.55\textwidth]{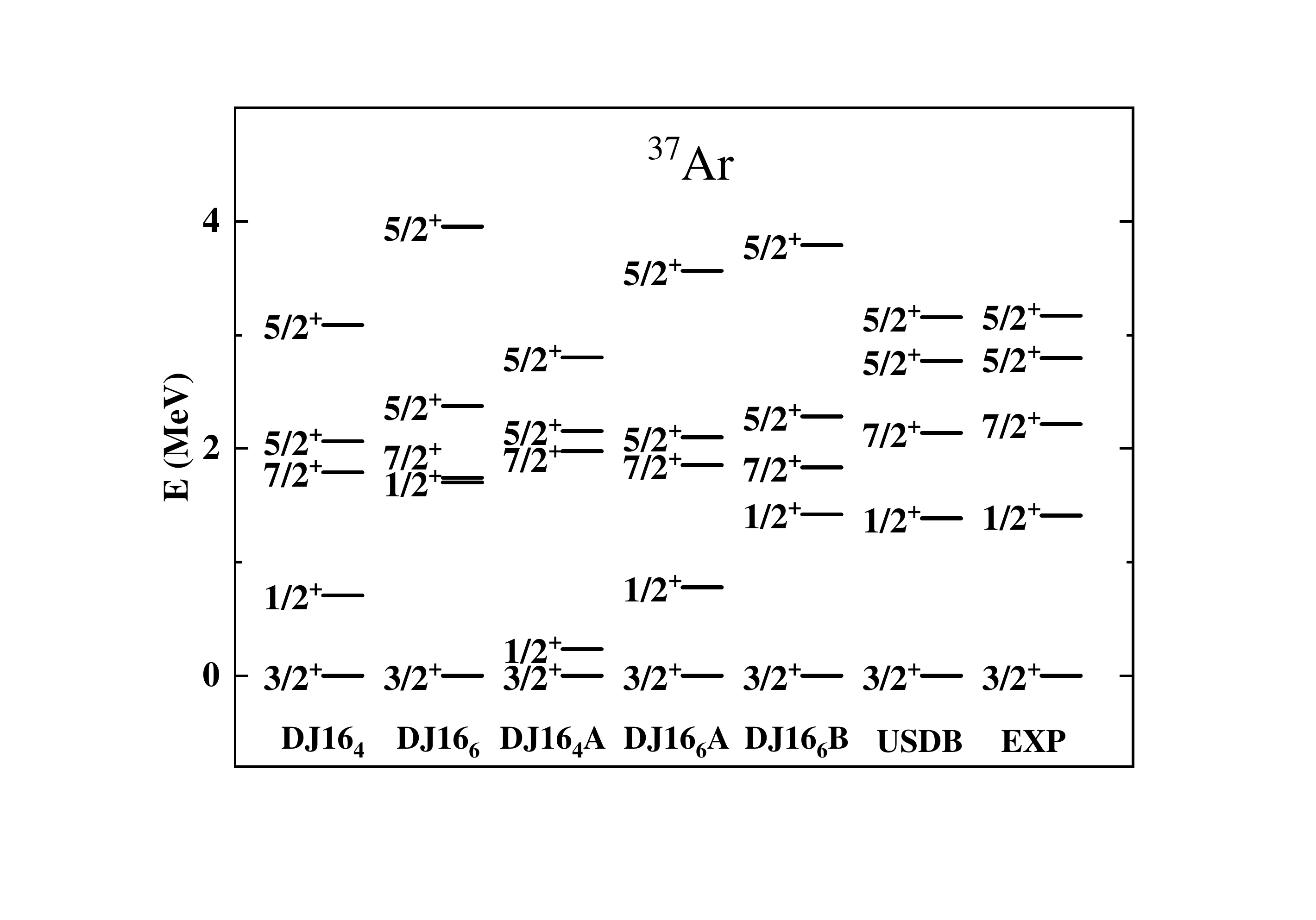}\\[-1.2cm]
  \includegraphics[width=.55\textwidth]{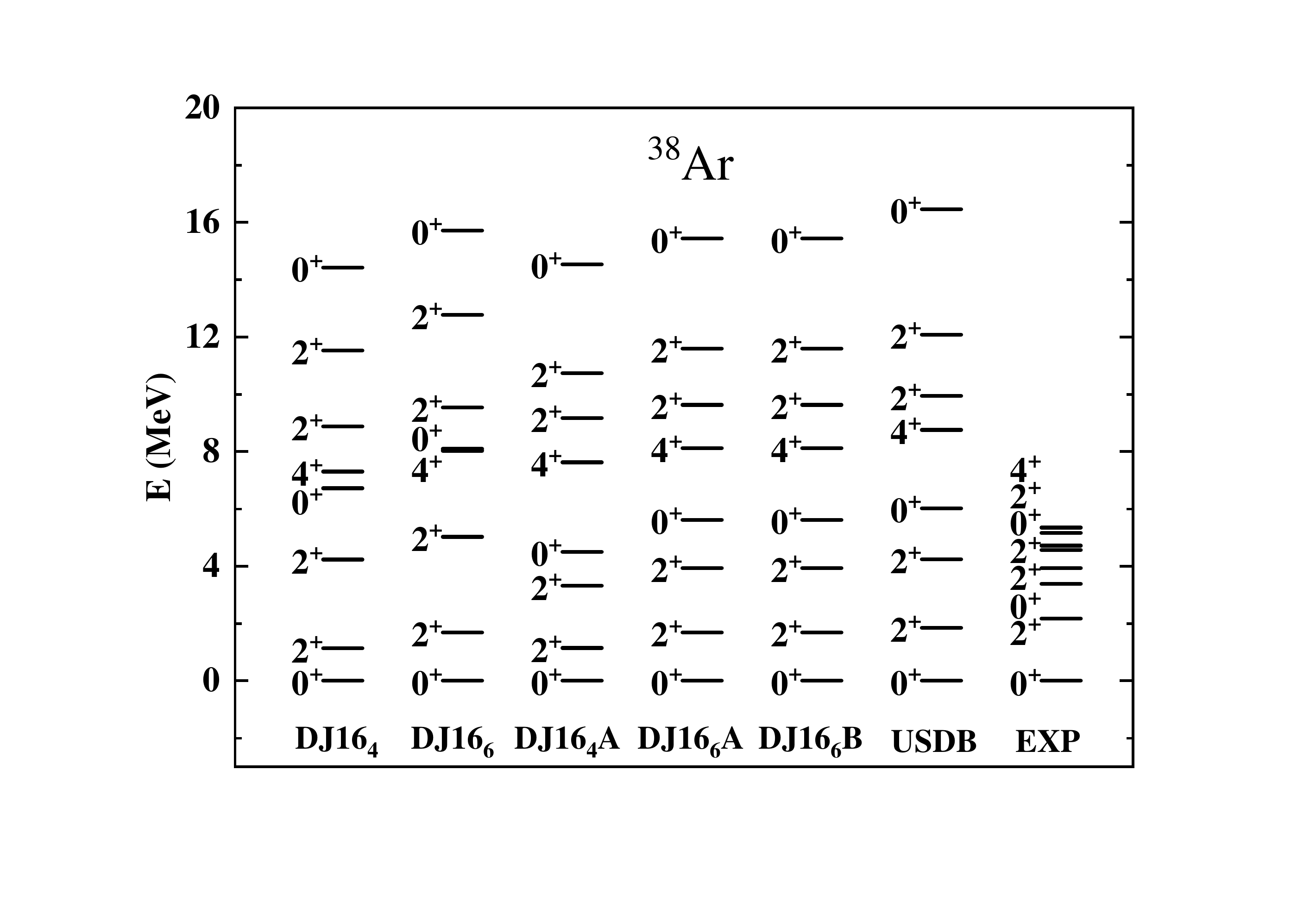}
  \caption{Theoretical and experimental low-energy spectra of $^{36-38}$Ar
   (lowest positive-parity states are presented -- five for $^{36,37}$Ar, eight for $^{38}$Ar.}
\end{figure}

\end{document}